# DISCOVERING EAST AFRICA'S INDUSTRIAL OPPORTUNITIES

César A. Hidalgo PhD[1,2],

[1] Macro Connections, The MIT Media Lab, Massachusetts Institute of Technology
[2] Center for International Development, Harvard University

**ABSTRACT**

What are East Africa's industrial opportunities? In this article we explore this question by using the Product Space to study the productive structure of five south-east African countries: Kenya, Mozambique, Rwanda, Tanzania and Zambia. The Product Space is a network connecting products that tend to be exported by the same sets of countries. Since countries are more likely to develop products that are close by in the Product Space to the ones that they already produce, the Product Space can be used to help anticipate a country's industrial opportunities.

Our results suggest that the most natural avenue for future product diversification for these five south-east African nations resides in the agricultural sector, since all of these nations appear to have productive structures that are pre-adapted to the production of many agricultural products that none of them are currently exporting.

We conclude this paper by exploring the potential benefits of further regional economic integration by doing an exercise in which we pull together the productive structures of these five countries. This exercise shows that the products that become more accessible in the combined economy are once again predominantly agricultural. These results suggest that while diversification into all sectors should remain an important long-term goal of the region, the path towards increased diversification in the near future may well lie in a more empowered and diverse agricultural sector.

## Acknowledgments

This work was supported by the German Marshall Fund of the United States as well as the Empowerment Lab and Growth Lab at the Center for International Development at Harvard University. We acknowledge comments that helped shape this work from Serdar Altay, Katrin Kuhlmann, Kathryn Ritterspach and Emily Batt. We also acknowledge Jennifer Gala and Nicola Lightner for providing logistical support for this project. All mistakes are my own.

# INTRODUCTION TO THE PRODUCT SPACE

Economic diversity matters because the ability of a country to support an adequate level of income depends on its ability to produce a diverse set of goods and services (Hidalgo and Hausmann, 2009). Yet the production of a diverse set of products and services is not easy to obtain, since the ability of a country or a region to produce a product or provide a service depends on the existence of a large number of individual economic activities.

Traditional approaches to the understanding of economic growth and development, however, have tended to assume away this diversity by assuming instead that, the ability of a country or a region to produce different products, results not from the availability of a large number of inputs, but from the relative cost of a few aggregate and amorphous factors such as “labor” and “capital” (Acemoglu, 2009; Aghion and Howitt, 2009).

The intuition about economic development that emerges from these aggregate production models is that the mix of products that a country exports has little or no effect on that country’s economic future. Since in this world the production of a product is the reflection of aggregate factors such as capital, land and labor, making the particular mix of products that a country exports irrelevant, since this is a world in which the production of goods can be seamlessly accommodated to reflect any change in the relative price of factors.

Yet, if the productive structure of a country is constrained by product-specific factors (capabilities), and thus cannot be seamlessly shifted in response to changes in factor prices, then any theory of economic development which does not include an adequate description of a country’s detailed and specific productive structure might be assuming away the most relevant part of the development puzzle.

Recent research, however, has begun to provide an alternative approach to more traditional conceptualizations of the economic growth process by demonstrating the value, difficulties and challenges that emerge from both, the process of economic *self-discovery*[1] (Hausmann and Rodrik, 2003) and the structure of economic diversity (Hidalgo and Hausmann, 2009). Contrary to factor-based models of growth, these works assume that the production of a good results from the local convergence of a large number of non-tradable inputs, or capabilities (Hidalgo and Hausmann, 2009), which are mostly product- or sector-specific and are unlikely to be accumulated in the absence of the products that demand them. Ultimately, the "fixed" costs required to initiate production are not spatially fixed, since they depend on the local presence and absence of other industries. For illustration purposes, consider the cost of starting the production of a new type of regional jet in two different locations, one in which no mechanical machine has ever been built and another location in which a diverse number of firms producing different types of machines are currently located. While in both cases, an attempt to start producing jets will be costly, the success of the enterprise, and the chances that it will eventually produce airplanes, are likely to be higher at the location where other machines were previously being produced.

An important consequence of assuming that products depend on a large number of capabilities, instead of a few factors, is that in such a world the ability of a country to produce a particular mix of products will depend naturally on the mix of products that the country is already making. This is because in such a case, the mix of products that a country makes is a reflection of the capabilities that the country has and the products accessible for future production are most likely to be those requiring a similar mix of capabilities. This makes the ability of a country to produce a product in the future dependent on the products it made in the past.

[1] The process by which countries discover the cost of producing a good locally.

Evidence of this process can be found in the structure and dynamics of the Product Space (Hidalgo et al., 2007), the network connecting products that tend to be co-exported by the same countries. Research exploring the structure and dynamics of the Product Space has demonstrated that countries are more likely to "discover" and subsequently produce and export a product that is close by in the Product Space to the products that it already exports. This shows that (i) the co-export of product pairs by a significant number of countries suggests that the production of those products requires similar capabilities, and (ii) that the process of economic diversification is path-dependant and cannot be properly understood without taking into consideration its historical nature.

Evidence of industrial path dependencies is not limited to the evolution of national economies. At the regional level, Neffke et al. (2009) used longitudinal data on Swedish manufacturing plants to show that plants producing products that are unrelated to other goods being produced in the region are more likely to exit, and plants that enter the region tend to produce more related varieties. At the city level, Glaeser et al. (1992) have shown that urban variety, but not specialization, encourages employment growth in cities. Whereas at the firm level, Teece et al. (1994) have shown that firms tend to remain coherent as they diversify, meaning that firms that operate on a larger scope tend to produce related varieties, e.g. varieties that are likely to be coproduced by many firms.

At the country level, there are important policy considerations that become natural when one walks away from the notion that products result from the combination of a few aggregate factors. For example, in a world in which products result from the combination of just a few factors, the specific products that a country makes do not matter. This is because the process of factor accumulation is indifferent to differences between the physical capital accumulated from the export of natural resources or from high-tech devices. Such ambiguity, however, is incompatible with recent evidence showing that future economic growth depends strongly on the mix of products that countries export (Hausmann et al., 2007; Hidalgo and Hausmann, 2009). This suggests that abstractions of the production process based on aggregate factors of production might be more a reflection of the way economists think about the world than of the world as it actually is.

These theories also affect the way that we interpret our world's economic history. For example, consider the case of Korea and Peru. In 1970, Korea's income was half of Peru's, but by 2003 Korea had four times Peru's income. How did this happen? If we look at aggregate factors we will find this question to be even more puzzling, since in 1970 Peru had almost 2.5 times more land per worker than Korea, four times the capital per worker and a similar level of human capital, measured in average years of education.

The paradox disappears, however, if instead of looking at factors, we look at the mix of products that were exported by each of these countries. This clarifies matters because the productive structure of Korea was orders of magnitude more sophisticated than that of Peru, a fact that was clearly visible from the mix of products that Korea was exporting. In fact, a recent paper estimating the number of capabilities available in countries during a 42-year period shows that, in a ranking of 85 different countries, Korea ranked 23rd in 1963, whereas Peru ranked 62nd, indicating that Korea's productive structure was more sophisticated than that of Peru's all along (Hidalgo 2010). Seen from this perspective, the reversal of economic fates experienced by these countries was more expected than surprising.

In this paper we bring the question of economic development to south-east Africa by presenting an exercise in which we look at the position in the Product Space of five countries: Kenya (KEN), Mozambique (MOZ), Rwanda (RWA), Tanzania (TZA) and Zambia (ZMB). Additionally, we study the collective regional position of these countries in the Product Space in an attempt to identify the opportunities that would be unveiled if the productive structures of these countries were to be combined. Since the capacity of a country to discover a good depends on the goods that it is already exporting, the industrial opportunities that appear as natural next steps for these countries will be different depending on whether we examine them as independent units or as a single integrated economy. In this paper we use the Product Space to help us understand whether there are rungs in the development ladder that would become within reach, were these economies to be integrated. To do so, we compare the position in the Product Space, and the opportunities associated with these positions, for each individual country and for the combined economy. This exercise shows that most of the self-discovery opportunities available in the region, before and after integration, are concentrated in the agricultural sector, followed by the garments cluster. This motivates the need for both national governments and their development partners to develop adequate incentive structures that can help catalyze the self-discovery of the agricultural industries that are within reach for these economies, but have not yet been exploited.

## DATA AND METHODS

We construct the Product Space using the BACI (Base pour l'Analyse du Commerce International) dataset at the HS-6 level of aggregation (Gaulier and Zignano, 2009). For the years considered (2003-2005) the data connected 5109 different product categories to 232 countries. This represents a more than fivefold increase in the number of product categories used in previous versions of the Product Space, which were constructed using SITC-4 data (Hidalgo et al., 2007).

We construct the Product Space following (Hidalgo et al., 2007) by calculating the frequency with which pairs of products are co-exported across all countries We call this quantity *proximity* and interpret it as an estimate of the relative number of capabilities shared by a pair of products.

We visualize the Product Space by taking the average proximity between pairs of products for the years 2003-2005 and consider only proximities larger than 0.45. A proximity of 0.45 means that at least 45% of the countries that export one product export the other. This threshold was chosen because it represents the percolation threshold of the weighted network. In the visualization, products that do not have any links to any other products at a proximity of 0.45 or higher are located on the right hand side of each figure arranged in a grid.

## RESULTS

### THE STRUCTURE OF THE PRODUCT SPACE

In agreement with previous constructions (Hidalgo et al., 2007), we find that the Product Space is characterized by a core of densely interconnected products, most of which are highly sophisticated (Figure 1) and include hundreds of different varieties of chemicals and machinery, and by a periphery in which poorly connected products are located. Here we measure *sophistication* using an estimate of the number of individual economic activities, or capabilities, that go into a product, following (Hidalgo and Hausmann, 2009). Besides the central cluster, there are other densely connected clusters in the Product Space including the garment sector, the textile sector and the electronics sector.

At this level of disaggregation we are also able to observe a clear gradient of sophistication in the textile cluster, which is visualized as an increase in node sizes from the less sophisticated varieties of textiles (e.g. cotton yarn or plain woven fabrics of cotton), located on the upper left corner of the textile cluster, to the most sophisticated textile varieties (e.g. denim or fabrics mixing different materials and colors) which are located towards the center and the bottom right corners of the textile cluster. We also note that textiles are not linked to garments, which is illustrative of a fact that this research has found to be true in general: input-output relationships do not explain proximity—i.e., products do not tend to be strongly connected to other varieties up or down the value chain (Hausmann et al., 2008). This indicates that proximity in the Product Space is driven by channels other than input-output connections.

Finally, we note the existence of a relatively sparse cluster of agricultural goods such as fruits, vegetables, leguminoses and animal products, which also includes diverse kinds of meats and fish. The fact that agricultural products cluster together, albeit weakly, suggests the existence of inter-industry spillovers in the agricultural sectors. Most raw materials and minerals, however, including oil, are extremely peripheral in the Product Space and have no connections to other products at the 0.45 proximity threshold (and are thus located together with the other disconnected products at the right of the visualization). This lack of connectedness suggests a lack of productive spillovers in these sectors, suggesting a difference in development benefits between income from raw materials and minerals versus agricultural products or other goods.

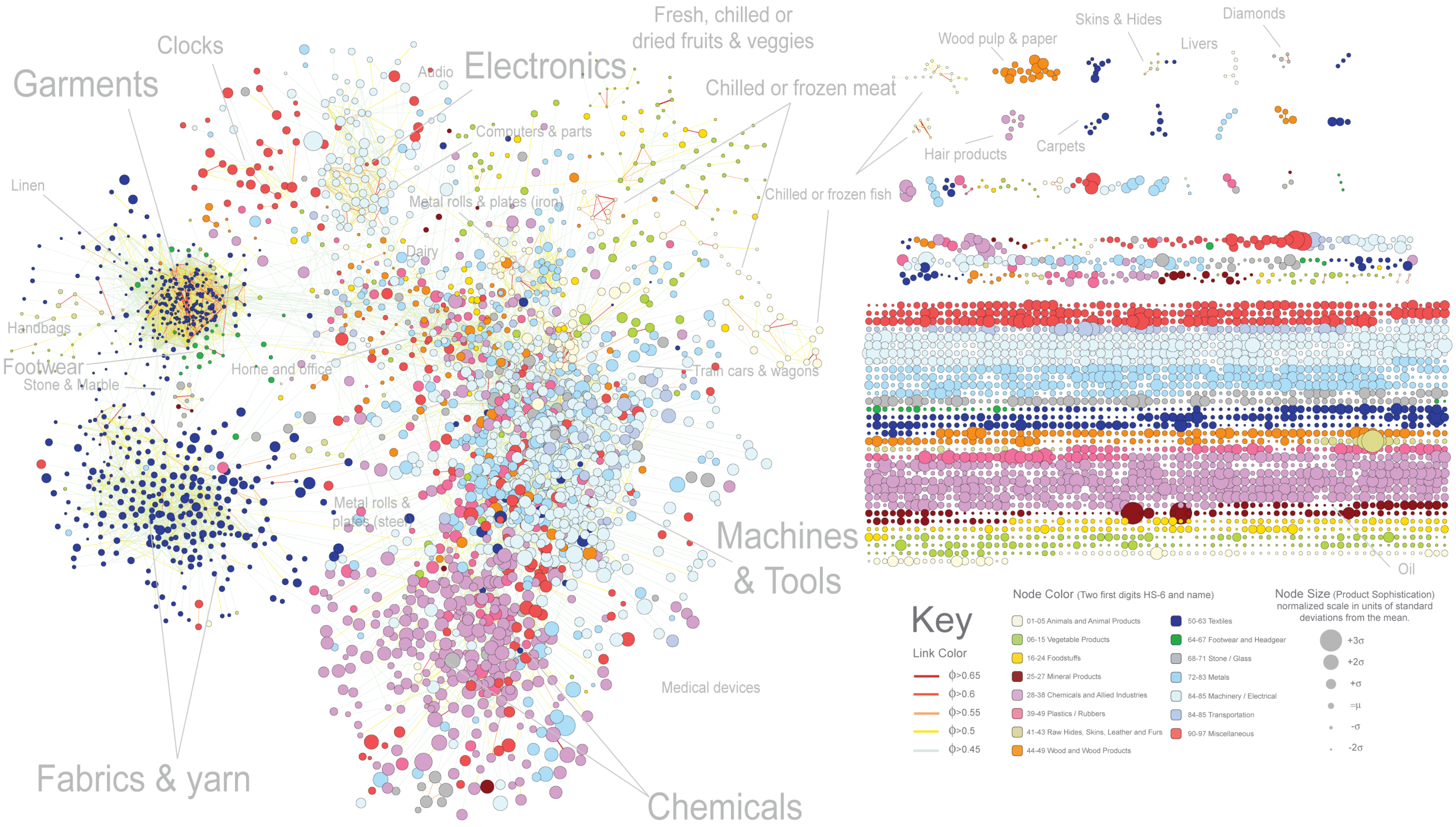

Figure 1 **The 2003-2005 Product Space.** See key for node color, size and link color specification

## SECTION II: OVERLAYING EXPORT STRUCTURE: TWO VIEWS OF THE PRODUCT SPACE AND ITS APPLICATION IN SOUTH-EAST AFRICA

In this section we use the Product Space to examine the productive structures of KEN, MOZ, RWA, TZA and ZMB. For this, we use *two different views of the Product Space:* the RCA view and the exports value view.

In the *RCA view*, we highlight the products for which these countries have Revealed Comparative Advantage (RCA). RCA is a measure indicating how "good" a country is at making a product given the size of its economy and the size of that product's market (Balassa, 1986).

For each of the five countries under review we provide an RCA view in which products that are exported with an RCA≥1, i.e. products that these countries are relatively good at making, are denoted by black squares with a green outer glow. Products that were exported with an RCA≥0.1, i.e. products in which these countries have a more marginal participation, are presented with squares colored according to the root categories of the HS-6 classification. Finally, products that are exported with an RCA<0.1, which can be considered to be products that are effectively not being exported at all by these countries, are not highlighted and are represented by circles, just like in the original map presented in Figure 1. In summary, the black squares in the RCA view show the products that each country is currently relatively good at making and exporting.

The *Export Value View*, on the other hand, uses the size of nodes to represent the U.S. dollar value of a country's exports of each product (see figure key for details). Additionally, products whose exports have increased, in nominal terms (U.S. dollars), between 2000 and 2005 are colored green, whereas products whose exports have decreased in nominal terms are colored red.

From these illustrations we can see a variety of things. First, it is easy to see that the position of these five countries in the Product Space is peripheral, meaning that all of these countries export only goods that are not connected to many other products. Moreover, these goods tend to be of relatively low sophistication, indicating that these are goods that require a fairly small and simple set of activities to be produced. From these five countries, however, Kenya appears to be an outlier, with a productive structure that is comparatively more diverse and better-connected than that of Mozambique, Rwanda, Tanzania or Zambia.

The overall simplicity of these countries' productive structures, and the contrast between the productive structure of Kenya and those of Mozambique, Rwanda, Tanzania and Zambia, can be easily seen by looking at the RCA view of the Product Space: Kenya's exports populate a much wider variety of nodes and expand a set of relatively more sophisticated industries (Figure 2-6). One of the most striking differences is the remarkable variety of products in the agricultural cluster that Kenya is exporting with RCA>1, compared to the four other countries considered.

Furthermore, inspection of the *Volume of Exports* view of the Product Space shows that most of the export value comes from a very small number of industries, such as aluminum for Mozambique, copper for Zambia, coffee for Rwanda and gold for Tanzania.[2] Kenya, on the other hand, has a productive structure that is relatively sophisticated compared to the others and has substantial participation in other sectors such as garments, while also exporting a substantially larger variety of agricultural goods including tea (14.6% of exports), fresh cut flowers (12.3%), coffee (not roasted) (4.91%), beans (3.74%), peas (1.85%) and pineapples (1.82%). Yet, upon inspection of the *Volume of Exports* view, we can also see that Kenya's participation in the most sophisticated products of its export basket is quite minimal (for instance, exposure meters represent only 0.001% of Kenyan exports, hair lacquers 0.006% and glucose syrup 0.057%).

[2] For numerical values and a longer list of products see APPENDIX:

Table 1, Table 5, Table 9, Table 13 and Table 17 in the Appendix.

In the Appendix we present two tables for each country showing the top 30 products exported by each of them in terms of RCA and in terms of export value.

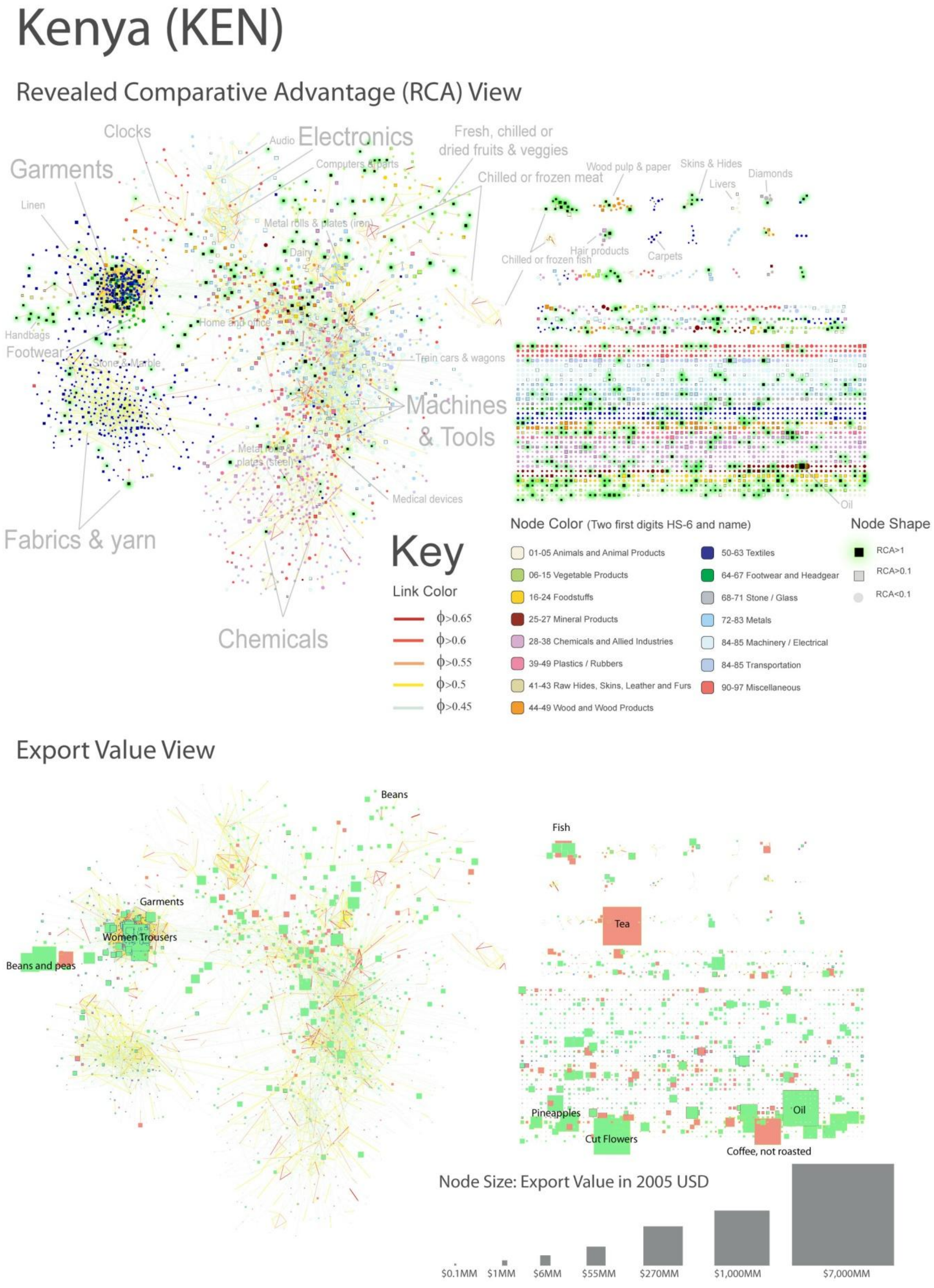

Figure 2 RCA and export value view of the Products space for Kenya

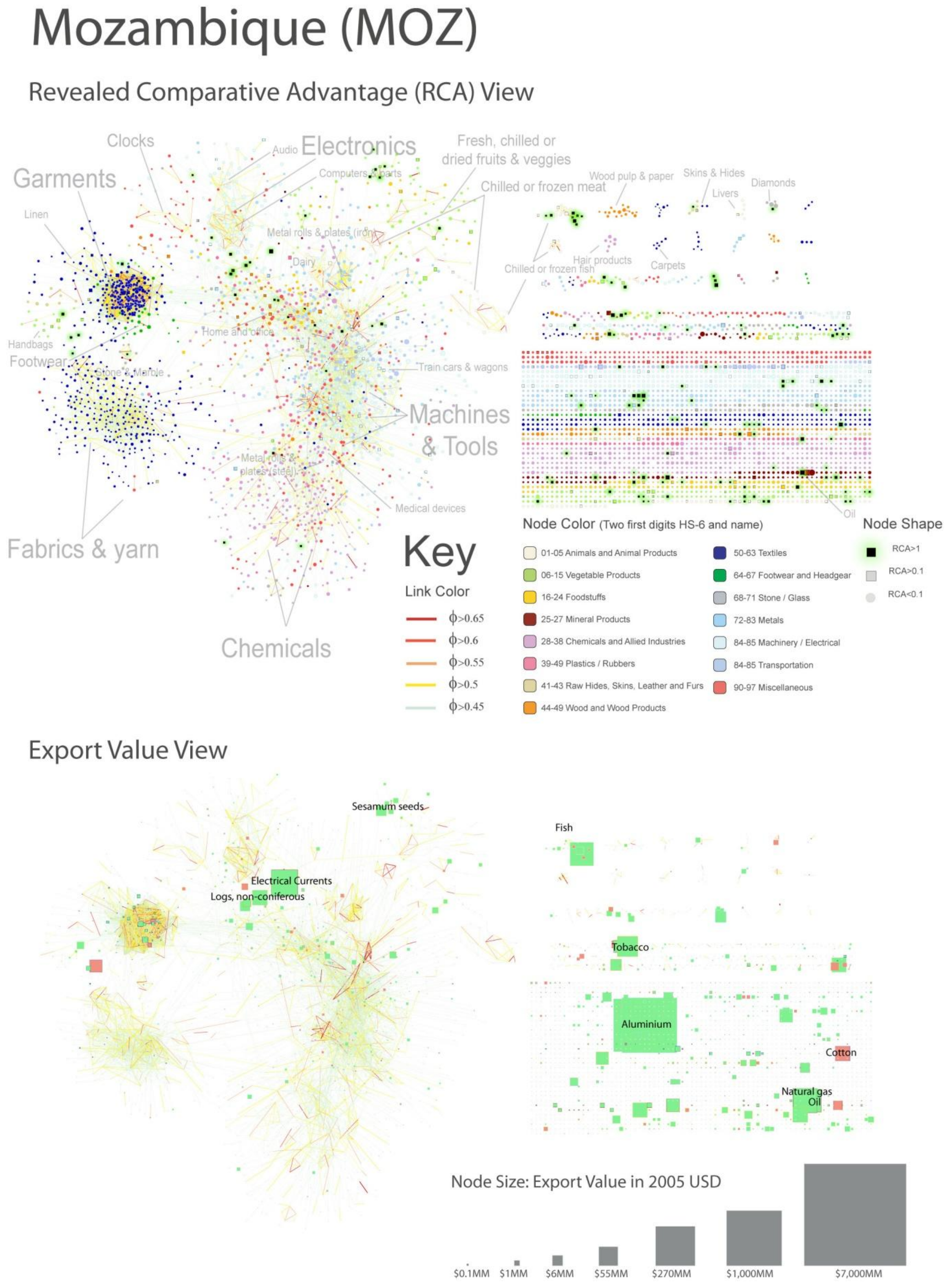

Figure 3 RCA and export value view of the Products space for Mozambique

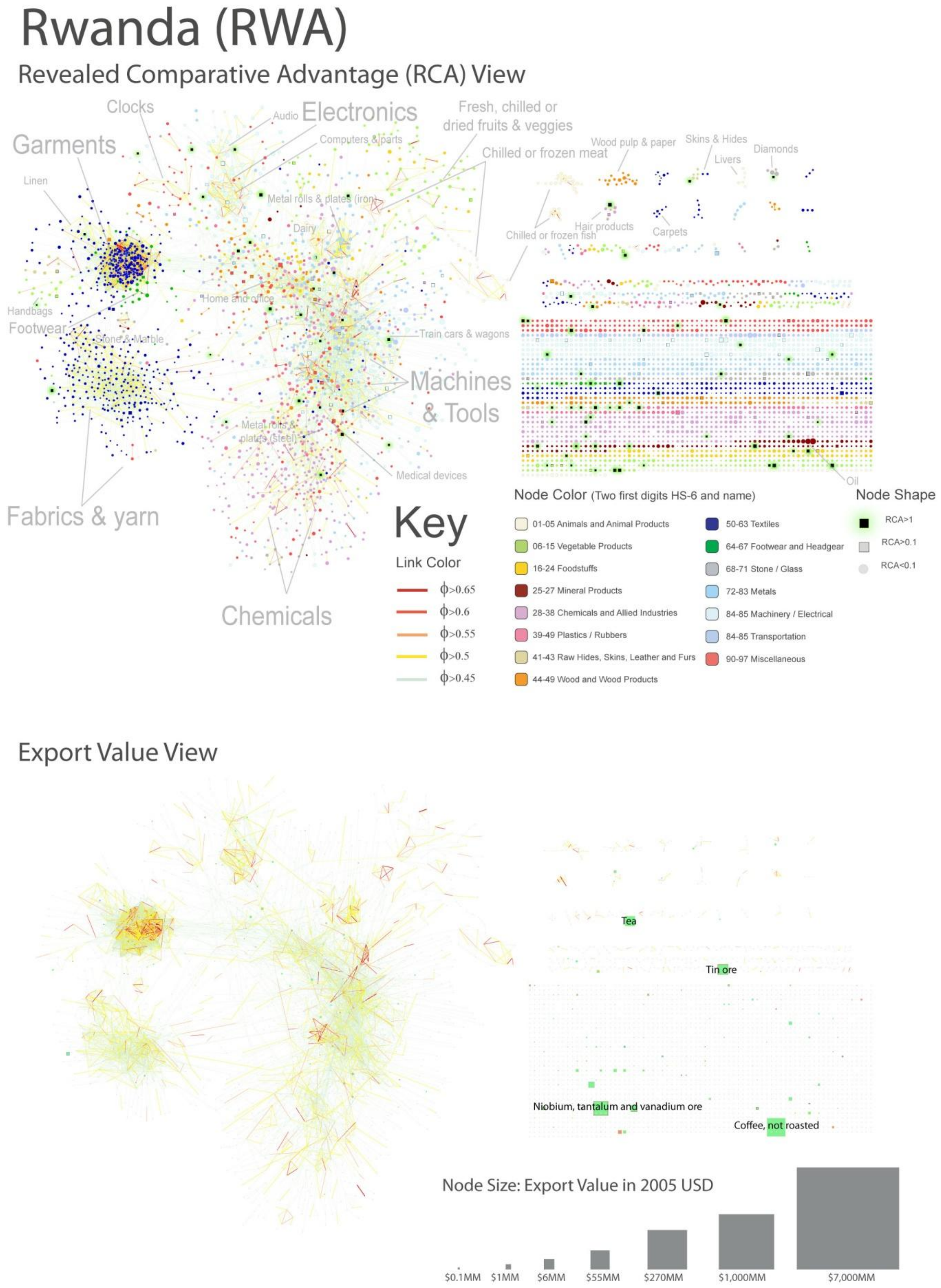

Figure 4 RCA and export value view of the Products space for Rwanda

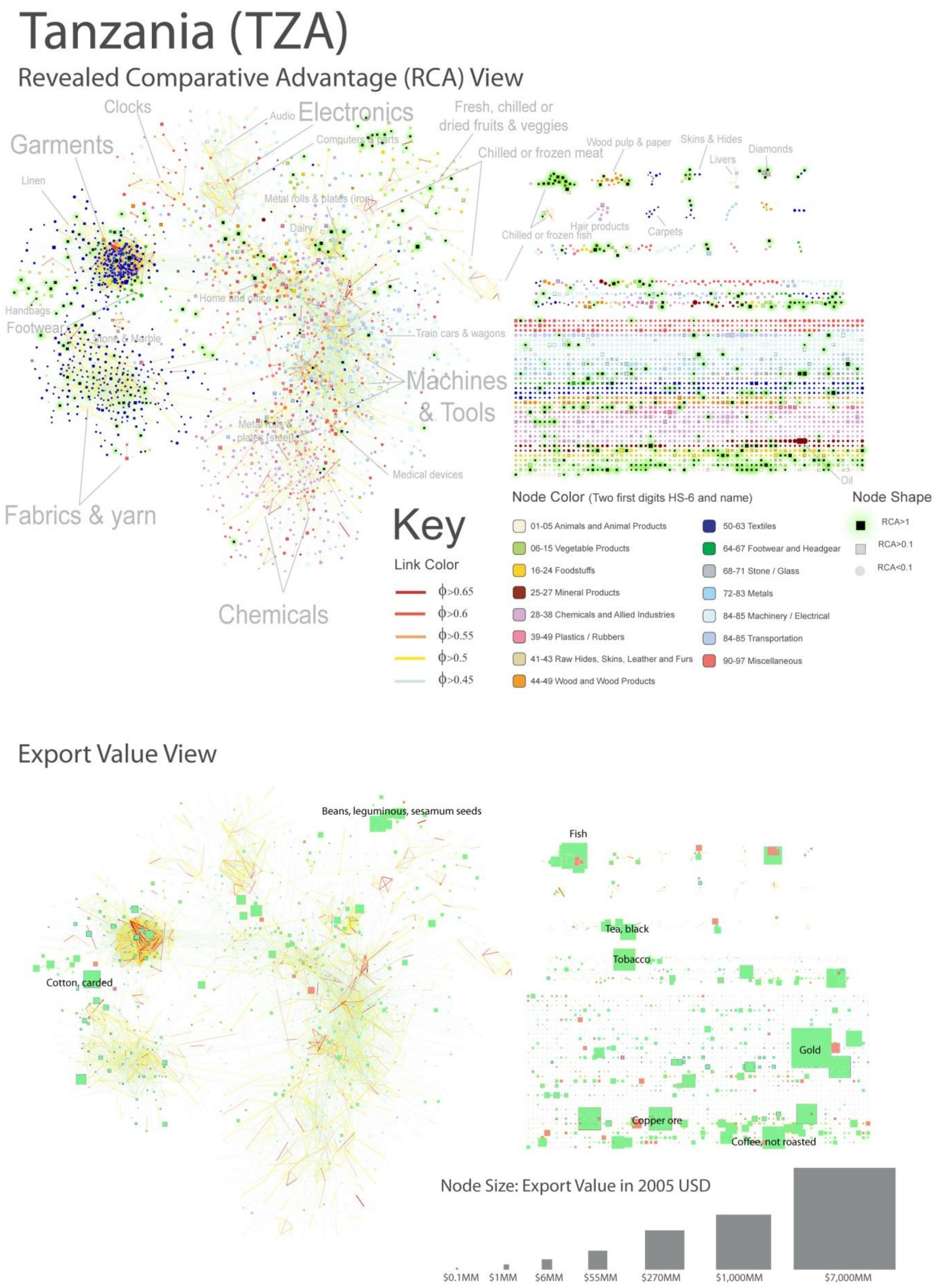

Figure 5 RCA and export value view of the Products space for Tanzania

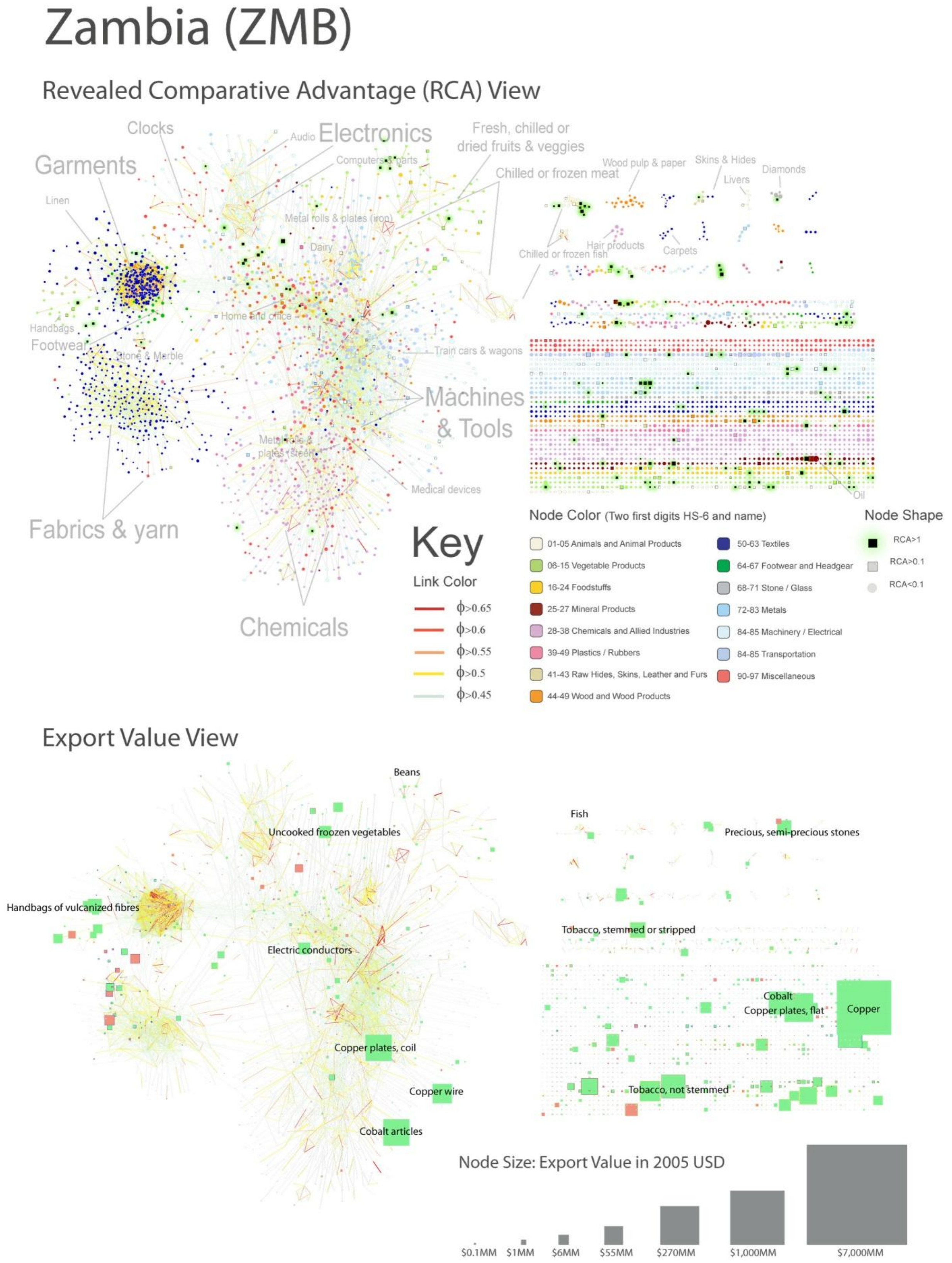

Figure 6 RCA and export value view of the Products space for Zambia

## AVENUES FOR DIVERSIFICATION: THE *OPPORTUNITIES VIEW* OF THE PRODUCT SPACE

Since one important aspect of the process of economic development is the channel by which countries move from the products that they currently produce to products that are nearby in the Product Space (Hausmann and Klinger, 2006; Hidalgo et al., 2007), we study the industrial opportunities of these five countries by introducing a third view of the Product Space: The *Opportunities View*.[3] The Opportunities View of the Product Space was constructed by coloring products not currently being exported by each of these countries according to how "close" or "far" these products are to the products that they are currently producing. Products that are relatively "close" to a country's productive structure are painted in red, whereas products that are relatively far are presented in blue (see chart for details). The distance between a product and a country's productive structure was determined by using *density,* a measure introduced for this purpose in (Hidalgo et al., 2007).

The Opportunities View of the Product Space shows the products that, given the country's current productive structure, could be most easily added to the country's export basket. We note that the Opportunities View presented in these figures shows the *relative* distance between a country's productive structure and the products that they are not yet exporting. Hence, it shows products which are relatively close to a country's productive structure, but that might still be far in absolute terms.

The Opportunities View of the Product Space shows that given the current productive structure of these countries, most of their potential export opportunities are located in the agriculture and garment clusters. In the appendix we present tables showing for each one of these countries the 30 products that they are currently not exporting, but that are closest to them given their current export structure. The findings presented in these tables have to be interpreted with caution, since they consider only the structure of the Product Space and not any other geographic constraints to production. Nonetheless, they show that for most of these countries the products that are currently closest to them, and can be thought of as the natural edge of their production frontiers, are several types of fish, leguminoses such as chick-peas, and other agricultural goods such as sesame seeds, ginger, mangoes, guavas and pineapples, among others.

For example, in the case of Kenya we find an assortment of agricultural and garment products on its production frontier. These include arrowroot, bananas, palm nuts and tuna, but also women's bathrobes and men's cotton underpants. In the case of Mozambique, however, most of the opportunities we found are agricultural and include different types of beans and legumes, as well as rice and coffee. Rwanda's industrial opportunities, on the other hand, include sugar cane, guavas and ginger. Those of Tanzania include sugar cane, sesame seeds, pineapples and green tea. Many of these products also appear in the opportunity set of Zambia, which includes sesame seeds, guavas, green tea, citrus fruits, beans, onions, and pineapples, among other products (see Appendix for full tables).

Each product not being exported by a country represents an individual mystery. While the fact that Tanzania does not export fresh guavas or pineapples, or Mozambique does not export rice, evidences a lack of some of the capabilities needed to make this happen, the fact that these are products that appear to be close by in the Product Space indicates that there is a possibility for these exports to emerge, given these countries' current productive structures.

---

[3] (Figure 7, Figure 8 and Figure 9) together with ten tables in the Appendix (Table 3, Table 4, Table 7, Table 8, Table 11, Table 12, Table 15, Table 16, Table 19 and Table 20).

The fact that the products found in our analysis are of relatively low sophistication, however, is a general feature of developing economies. This can be shown by the relationship between the sophistication of products and the distance between that product and a country's current productive structure. Figure 10 shows this by plotting the sophistication of these products (measured in units of standard deviations from the world's average) as a function of the *density* (ω) in the Product Space that each country has around each of these products. Density measures how far is the productive structure of a country from a particular product. The sophistication-density diagram for these five countries shows that these countries have higher densities—are closer—to products of relatively low sophistication. This can be seen from the negative slope of the distribution of points defined by the products not exported by these countries. This negative relationship indicates that density is highest for products in which sophistication tends to be low.

Figure 10 once again shows how Kenya stands out from the rest of the region. In absolute terms, Kenya is substantially closer than its geographical neighbor to most products, a reflection of its superior export structure.

Figure 10 also shows that Tanzania appears relatively well-positioned in the Product Space compared to Rwanda, Mozambique and Zambia, which have unsophisticated productive structures and are extremely far from all products.

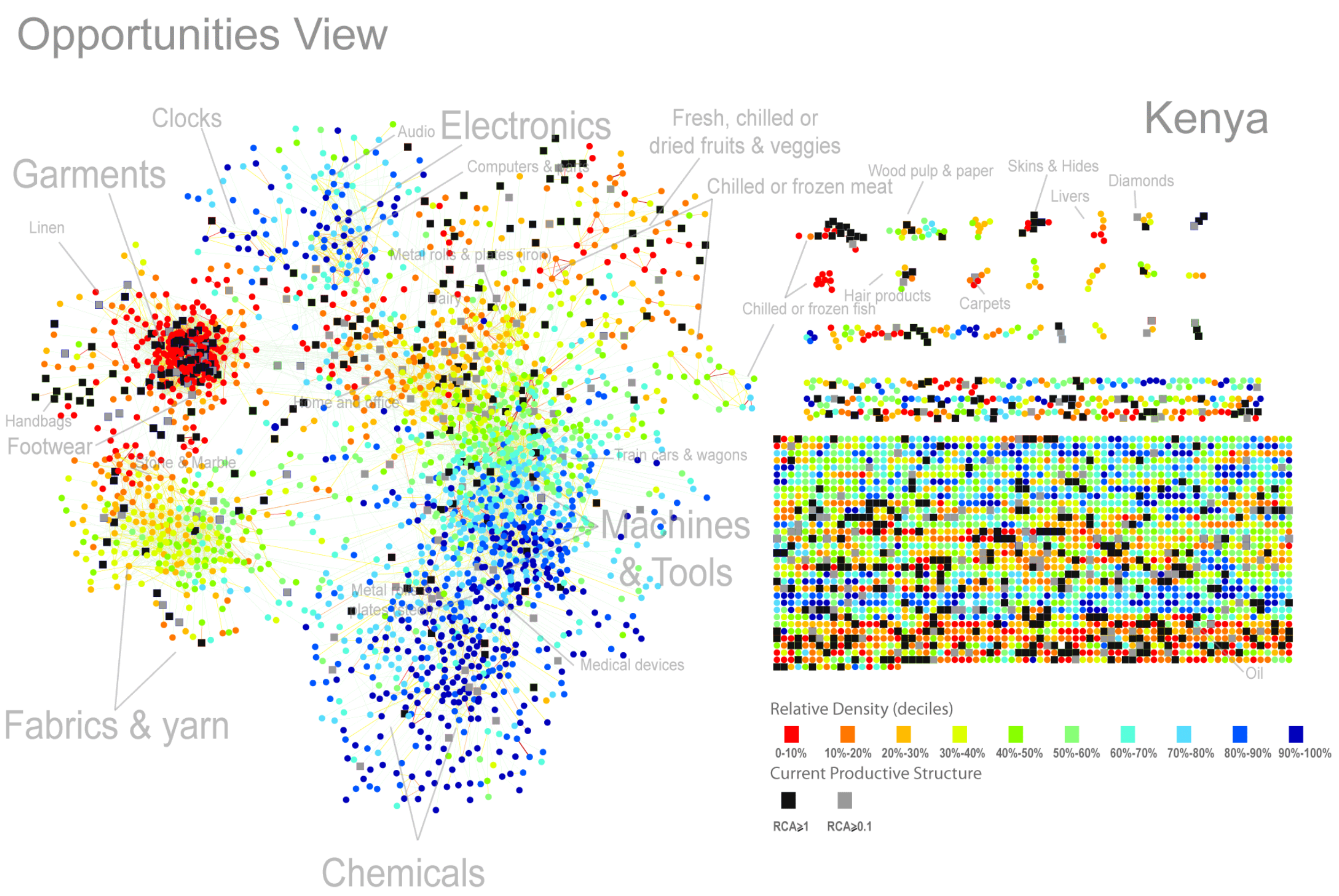

Figure 7 Kenya's industrial opportunities. Products not being exported by Kenya are color coded according to "density" deciles. Density is a measure of the similarity between a country's productive structure and a product.

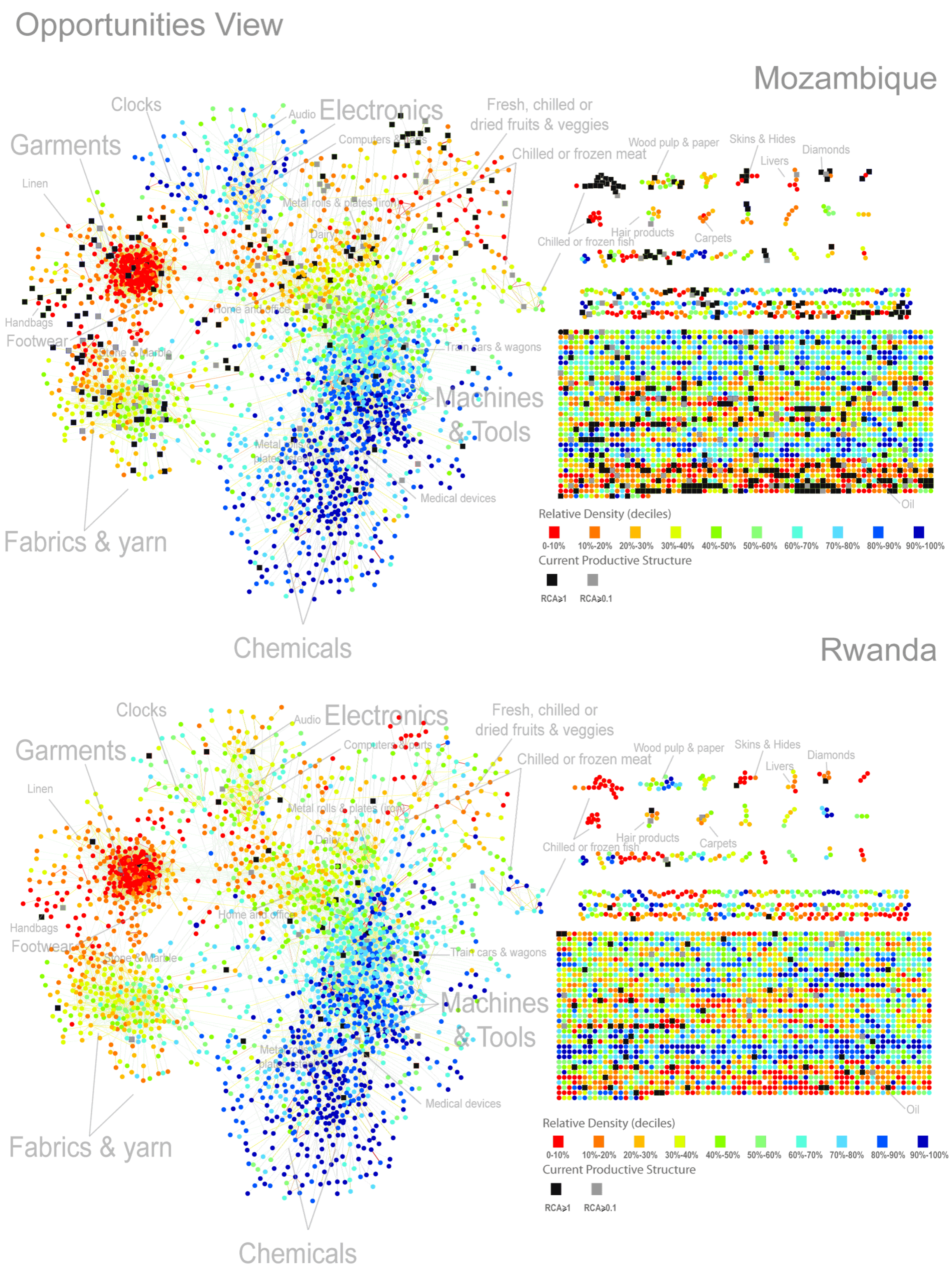

Figure 8 Mozambique's and Rwanda's industrial opportunities. Products not being exported by Mozambique and Rwanda are color coded, in the respective figures, according to "density" deciles. Density is a measure of the similarity between a country's productive structure and a product.

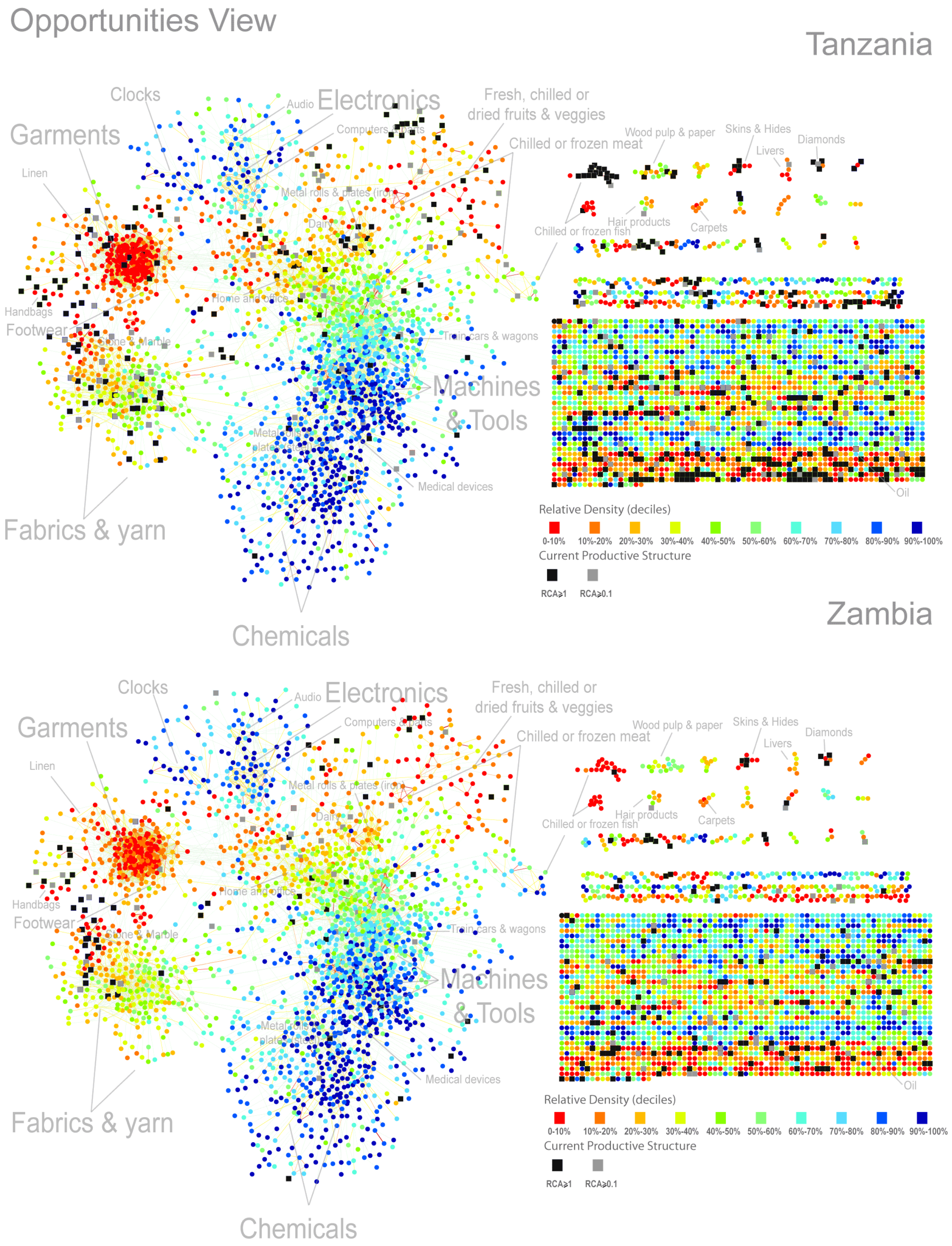

Figure 9 Tanzania's and Zambia's industrial opportunities. Products not being exported by Tanzania and Zambia are color coded, in their respective figures, according to "density" deciles. Density is a measure of the similarity between a country's productive structure and a product.

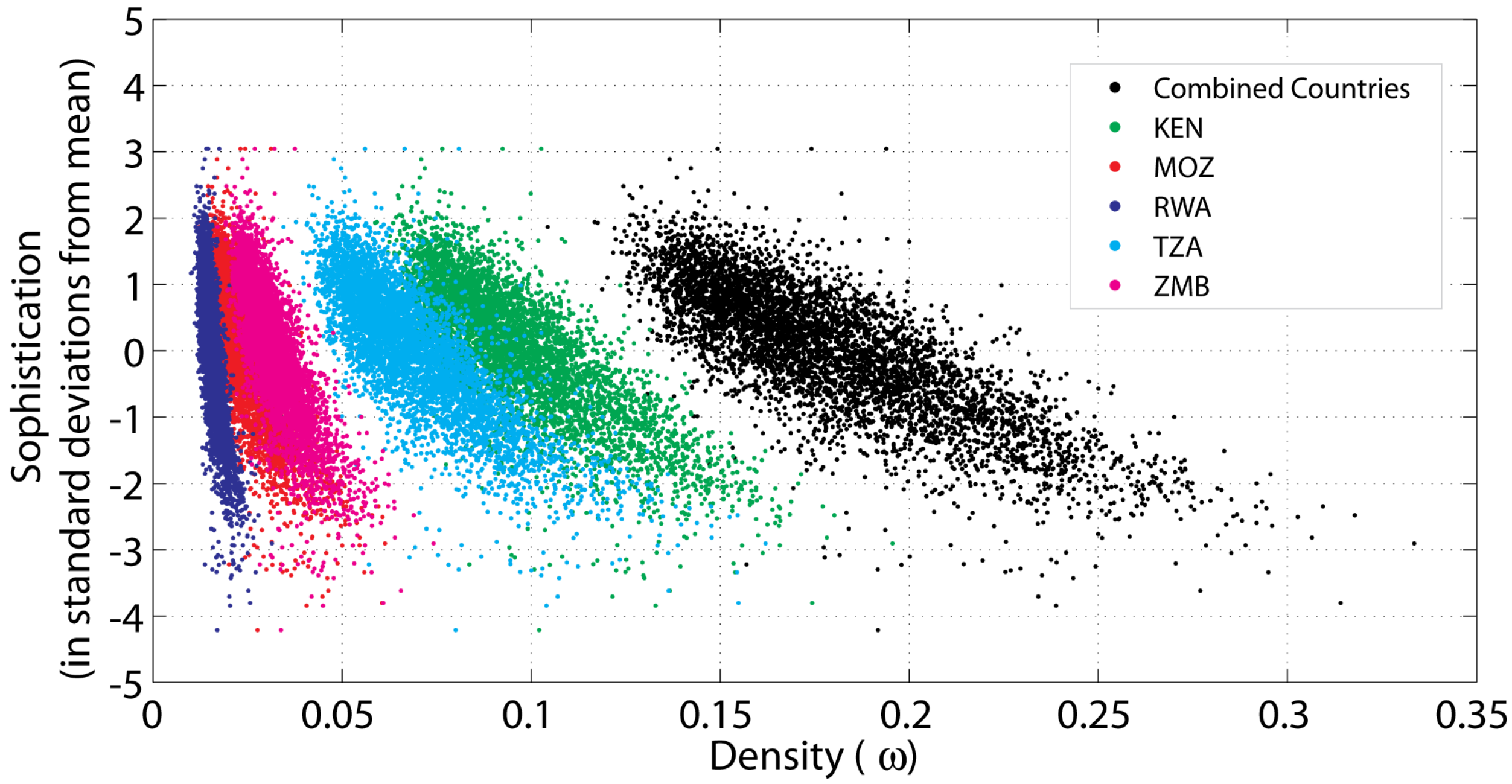

Figure 10 Diagram showing the relative "distance" (measured as density) and sophistication of all products not currently being exported by these countries, as well as by a hypothetical economic block constructed by combining the productive structures of all five. Note that density is a measure that incorporates information about a country's current productive structure, where sophistication is a characteristic of a product.

Finally, Figure 10 shows the relationship between product density and sophistication for a hypothetical combination of these five economies. This integration was done considering a simple "best case scenario" in which the RCA (Revealed Comparative Advantage) of the combined region is equal to the maximum RCA among the five countries for each product. Another scenario that could be considered could be to combine the exports of all of these countries and recalculate RCAs. For brevity, simplicity and illustration purposes, however, here we concentrate on the scenario in which every product needs only to have an RCA≥1 in any of the considered countries to be considered part of the region's combined productive structure.

The large increase in density that emerges after pulling together these five economies suggests that complementarities between their productive structures do exist, despite the large degree of similarity exhibited by their economies.[4] This fact has compelling implications for regional integration and its benefits, even among countries with unsophisticated and similar economic profiles. The diagram, however, shows clearly that the products that are closest to these countries' combined productive structures are once again among the least sophisticated products. These results nonetheless suggest that the combination of these countries' productive structures could result in the potential enhancement of their current productive capacities, since larger product densities can be interpreted as deeper markets for all the inputs required in the production of these products. Moreover, in a forthcoming work, we find that for a large number of products RCA coevolves with density, meaning that the level of RCA that a country can achieve depends on the number of related varieties that it is already exporting. In the case of these countries, this would suggest that the combination of these countries' productive structures could result not only in the creation of new export opportunities, but also in an increase in the competitiveness and export volume of existing industries.

[4] Data on the similarity on productive structures not shown.

Since increases in density are associated with increases in the probability of "discovering" a product, it is crucial to ascertain the set of products that see the largest increases in density after we combine these countries' productive structures. In the Appendix, Table 21 and Table 22 show the ranking of products that see the largest increase in density after we combine these countries' productive structures. Inspection of these tables reveals that products that see the largest increase in density are, once again, mostly agricultural. We quantified this by looking at the fraction of the top 150 products, ranked by their absolute increase in density, that belong, respectively, to the agricultural and foodstuffs category (HS-6[5] code < 280000 which includes the three coarse categories, Animals and Animal Products, Vegetable Products, and Foodstuffs); Garments and Textiles (HS-6 code between 500000 & 680000, including the coarse categories Textiles, Footwear and Headgear); and all other products.

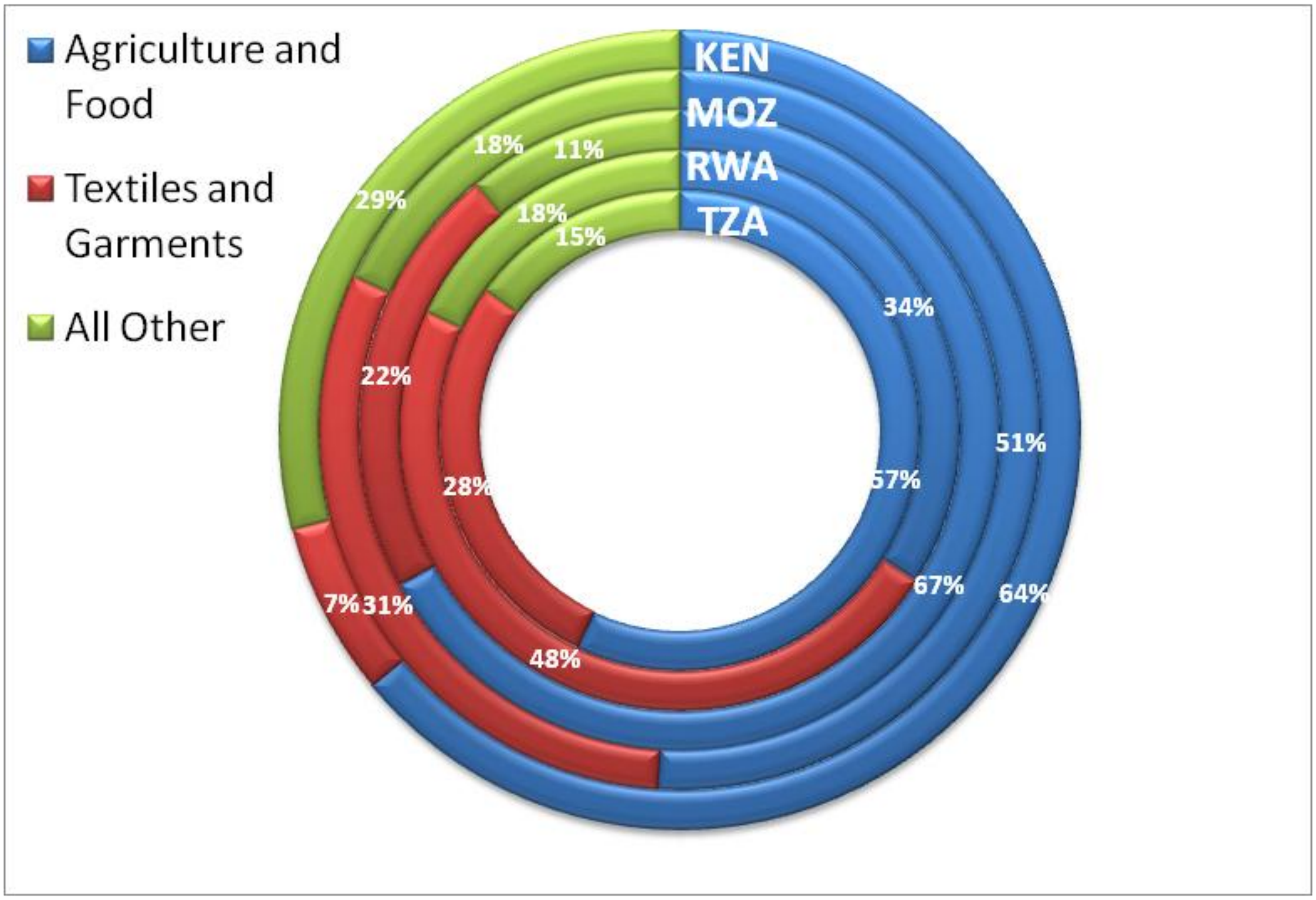

Figure 11 Decomposition of the top 150 products that experienced the largest absolute increase in density after regional integration

Figure 11 shows that, were these countries' economies to integrate, the largest increase in density would be expected in the agriculture and food sector, followed by textiles and garments. Spillovers, however, in other categories would be relatively minor. For all countries except

[5] The Harmonized Commodity Description and Coding System (HS) is the internationally standardized set of numbers for classifying traded products, developed and maintained by the World Customs Organization.

Rwanda, the agricultural sector represents more than half of the top 150 export opportunities that experience the greatest increase in density after integrating the economies. This shows that there are plenty of diversification opportunities for farmers in south-east Africa.

## CONCLUSION

In this paper we used the Product Space to study the productive structures of Kenya, Mozambique, Rwanda, Tanzania and Zambia. We found that the productive structures of these five countries are characterized by products of low sophistication that tend to be peripherally located in the Product Space--that is, goods that do not require a large number of inputs, or capabilities, to produce and that do not share a large number of required capabilities with many other products. Despite these limitations, we found plenty of diversification opportunities for these economies, located mostly in the agricultural sector. By doing an exercise in which we artificially remove national borders and look at the relative increases in density of products that these countries do not currently export, we found that most of the short-term benefits of economic integration for these south-east African nations are likely to be found in the agricultural and foodstuffs sectors, followed by textiles and garments.

The fact that many of these opportunities have not been seized points to a variety of problems that could be affecting the development of these five countries. Exporting agricultural goods requires the orchestration of a variety of economic and institutional capabilities that are probably lacking in the region. Agricultural exports do not only require proper land, climate and sufficient knowledge of a crop's agronomy and life cycle, but also require the existence of critical pieces of infrastructure, such as cold storage chains, together with the electricity needed to power them and the roads needed to move the fresh produce quickly. Delays can be devastating for agricultural exports, and could also emerge from poor flight frequency or irregular customs practices. Moreover, the need to comply with a variety of regulations and standards (e.g. sanitary and phytosanitary[6]) in order to bring products to most developed markets increases the difficulty of the coordination problem faced by producers contemplating starting export activities in these countries.

The existence of these opportunities, together with the problems that are linked to them, points to the need to explore the institutional and policy considerations that could help catalyze the export of agricultural products and overcome simultaneously both infrastructural and regulatory problems. To help untie the invisible hand in south-east Africa, it will be crucial to think about policy solutions that could integrate infrastructure with institutions and incentives, as well as financial and insurance products that could help hedge the risks implied in the development of new agricultural products. In the optimistic scenario in which these problems could be resolved, the productive structure of these countries offers the opportunity for a diverse agricultural sector to help drive prosperity.

[6] i.e. Animal and plant health and safety rules.

APPENDIX:

Table 1 Kenya's top 30 exports sorted in decreasing total exports volume (Exports in thousands of US dollars, sophistication in standard deviations of world's average. See text for explanation for the variables density and sophistication)

| RANK | Product Code (HS-6) | Product Name | RCA 2000 | RCA 2005 | Exports 2000 | Exports 2005 | Export Share 2005 | Density | Sophistication |
|---|---|---|---|---|---|---|---|---|---|
| 1 | 90240 | Tea, black (fermented or partly) in packages > 3 kg | 988 | 835 | 471342 | 416653 | 14.60% | 0.180 | -2.480 |
| 2 | 60310 | Cut flowers and flower buds for bouquets, etc., fresh | 114 | 217 | 133478 | 350677 | 12.29% | 0.171 | -1.860 |
| 3 | 271000 | Oils petroleum, bituminous, distillates, except crude | 4 | 3 | 214092 | 331394 | 11.61% | 0.153 | -1.243 |
| 4 | 620462 | Womens, girls trousers & shorts, of cotton, not knit | 6 | 28 | 17204 | 145831 | 5.11% | 0.143 | -1.859 |
| 5 | 90111 | Coffee, not roasted, not decaffeinated | 66 | 54 | 166567 | 140117 | 4.91% | 0.196 | -2.903 |
| 6 | 70820 | Beans, shelled or unshelled, fresh or chilled | 563 | 627 | 54429 | 106633 | 3.74% | 0.171 | -2.000 |
| 7 | 70810 | Peas, shelled or unshelled, fresh or chilled | 132 | 801 | 5619 | 52656 | 1.85% | 0.149 | -1.718 |
| 8 | 200820 | Pineapples, otherwise prepared or preserved | 209 | 219 | 39613 | 51903 | 1.82% | 0.131 | -1.986 |
| 9 | 160414 | Tuna, skipjack, bonito, prepared/preserved, not mince | 23 | 42 | 13056 | 39919 | 1.40% | 0.149 | -2.529 |
| 10 | 620342 | Mens, boys trousers & shorts, of cotton, not knit | 4 | 8 | 14873 | 39393 | 1.38% | 0.145 | -1.739 |
| 11 | 30420 | Fish fillets, frozen | 32 | 18 | 45030 | 39281 | 1.38% | 0.157 | -1.776 |
| 12 | 252329 | Portland cement, other than white cement | 18 | 27 | 17412 | 39157 | 1.37% | 0.143 | -1.349 |
| 13 | 60210 | Cuttings and slips, not rooted | 229 | 351 | 15275 | 35466 | 1.24% | 0.159 | -1.617 |
| 14 | 283620 | Disodium carbonate | 41 | 71 | 14002 | 33603 | 1.18% | 0.100 | -0.220 |
| 15 | 200559 | Beans nes, prepared or preserved, not frozen/vinegar | 230 | 441 | 9786 | 29255 | 1.03% | 0.122 | -1.011 |
| 16 | 70990 | Vegetables, fresh or chilled nes | 99 | 46 | 34276 | 25528 | 0.89% | 0.165 | -2.076 |
| 17 | 80440 | Avocados, fresh or dried | 108 | 89 | 12992 | 23353 | 0.82% | 0.133 | -0.665 |
| 18 | 30410 | Fish fillet or meat, fresh or chilled, not liver, roe | 1 | 28 | 420 | 21673 | 0.76% | 0.160 | -2.063 |
| 19 | 721049 | Flat rolled iron or non-alloy steel, coated with zinc, width >600mm, ne | 0 | 5 | 825 | 18996 | 0.67% | 0.098 | -0.073 |
| 20 | 80290 | Nuts edible, fresh or dried, nes | 33 | 85 | 4090 | 18180 | 0.64% | 0.147 | -2.022 |
| 21 | 210120 | Tea and mate extracts, essences and concentrates | 91 | 122 | 6016 | 16821 | 0.59% | 0.115 | -0.274 |
| 22 | 300490 | Medicaments nes, in dosage | 1 | 0 | 12800 | 15547 | 0.54% | 0.097 | 0.496 |
| 23 | 610462 | Womens, girls trousers & shorts, of cotton, knit | 0 | 20 | 150 | 14583 | 0.51% | 0.146 | -2.233 |
| 24 | 252922 | Fluorspar, >97% calcium fluoride | 215 | 169 | 11867 | 13669 | 0.48% | 0.118 | -1.546 |
| 25 | 200940 | Pineapple juice, not fermented or spirited | 118 | 104 | 9774 | 13462 | 0.47% | 0.139 | -1.430 |
| 26 | 410110 | Bovine skins, whole, raw | 48 | 56 | 5924 | 13386 | 0.47% | 0.109 | -1.772 |
| 27 | 130214 | Pyrethrum, roots containing rotenone, extracts | 1485 | 1647 | 11608 | 13258 | 0.46% | 0.166 | -2.640 |
| 28 | 720918 | | 16 | 13 | 9375 | 13241 | 0.46% | 0.114 | -0.816 |
| 29 | 611020 | Pullovers, cardigans etc of cotton, knit | 0 | 3 | 17 | 12455 | 0.44% | 0.155 | -2.245 |
| 30 | 121190 | Plants & parts, pharmacy, perfume, insecticide use ne | 13 | 38 | 3719 | 12451 | 0.44% | 0.161 | -2.209 |

| RANK | Product Code (HS-6) | Product Name | RCA 2000 | RCA 2005 | Exports 2000 | Exports 2005 | Export Share 2005 | Density | Sophistication |
|---|---|---|---|---|---|---|---|---|---|
| 1 | 130214 | Pyrethrum, roots containing rotenone, extracts | 1485 | 1647 | 11608 | 13258 | 0.46% | 0.166 | -2.640 |
| 2 | 410519 | Sheep or lamb skin leather, tanned or retanned, nes | 12 | 905 | 901 | 1624 | 0.06% | 0.121 | -1.226 |
| 3 | 90240 | Tea, black (fermented or partly) in packages > 3 kg | 988 | 835 | 471342 | 416653 | 14.60% | 0.180 | -2.480 |
| 4 | 70810 | Peas, shelled or unshelled, fresh or chilled | 132 | 801 | 5619 | 52656 | 1.85% | 0.149 | -1.718 |
| 5 | 530490 | Sisal and Agave, processed but not spun, tow & waste | 608 | 765 | 1884 | 3981 | 0.14% | 0.126 | -1.809 |
| 6 | 70820 | Beans, shelled or unshelled, fresh or chilled | 563 | 627 | 54429 | 106633 | 3.74% | 0.171 | -2.000 |
| 7 | 530410 | Sisal and Agave, raw | 780 | 577 | 8225 | 9223 | 0.32% | 0.162 | -2.620 |
| 8 | 200559 | Beans nes, prepared or preserved, not frozen/vinegar | 230 | 441 | 9786 | 29255 | 1.03% | 0.122 | -1.011 |
| 9 | 60210 | Cuttings and slips, not rooted | 229 | 351 | 15275 | 35466 | 1.24% | 0.159 | -1.617 |
| 10 | 200820 | Pineapples, otherwise prepared or preserved | 209 | 219 | 39613 | 51903 | 1.82% | 0.131 | -1.986 |
| 11 | 60310 | Cut flowers and flower buds for bouquets, etc., fresh | 114 | 217 | 133478 | 350677 | 12.29% | 0.171 | -1.860 |
| 12 | 410612 | Goat or kid skin leather, otherwise pre-tanned | 126 | 190 | 4308 | 9487 | 0.33% | 0.143 | -2.761 |
| 13 | 252922 | Fluorspar, >97% calcium fluoride | 215 | 169 | 11867 | 13669 | 0.48% | 0.118 | -1.546 |
| 14 | 410310 | Goat or kid hides and skins, raw, nes | 49 | 166 | 525 | 2672 | 0.09% | 0.152 | -2.175 |
| 15 | 150430 | Marine mammal fats, oils, etc. not chemically modifie | 0 | 138 | 0 | 334 | 0.01% | 0.109 | -0.141 |
| 16 | 30377 | Sea bass, frozen, whole | 7 | 137 | 112 | 1316 | 0.05% | 0.123 | -1.324 |
| 17 | 252230 | Hydraulic lime | 102 | 125 | 569 | 799 | 0.03% | 0.122 | -0.709 |
| 18 | 210120 | Tea and mate extracts, essences and concentrates | 91 | 122 | 6016 | 16821 | 0.59% | 0.115 | -0.274 |
| 19 | 271410 | Bituminous or oil shale and tar sands | 1 | 118 | 11 | 1050 | 0.04% | 0.114 | -0.055 |
| 20 | 200940 | Pineapple juice, not fermented or spirited | 118 | 104 | 9774 | 13462 | 0.47% | 0.139 | -1.430 |
| 21 | 30569 | Fish nes, salted or in brine, not dried or smoked | 35 | 102 | 1981 | 7560 | 0.26% | 0.150 | -1.588 |
| 22 | 721041 | Flat rolled iron or non-alloy steel, coat/zinc, corrugated, w >600m, ne | 56 | 91 | 4110 | 8449 | 0.30% | 0.121 | -1.763 |
| 23 | 80440 | Avocados, fresh or dried | 108 | 89 | 12992 | 23353 | 0.82% | 0.133 | -0.665 |
| 24 | 80290 | Nuts edible, fresh or dried, nes | 33 | 85 | 4090 | 18180 | 0.64% | 0.147 | -2.022 |
| 25 | 210230 | Baking powders, prepared | 37 | 81 | 758 | 2000 | 0.07% | 0.118 | -0.759 |
| 26 | 110720 | Malt, roasted | 38 | 73 | 1011 | 1910 | 0.07% | 0.101 | 0.345 |
| 27 | 410512 | Sheep or lamb skin leather, otherwise pre-tanned | 20 | 73 | 1828 | 5262 | 0.18% | 0.130 | -2.232 |
| 28 | 283620 | Disodium carbonate | 41 | 71 | 14002 | 33603 | 1.18% | 0.100 | -0.220 |
| 29 | 410110 | Bovine skins, whole, raw | 48 | 56 | 5924 | 13386 | 0.47% | 0.109 | -1.772 |
| 30 | 90111 | Coffee, not roasted, not decaffeinated | 66 | 54 | 166567 | 140117 | 4.91% | 0.196 | -2.903 |

Table 2 Kenya's top 30 exports sorted in decreasing order of RCA (Exports in thousands of US dollars, sophistication in standard deviations of world's average. See text for explanation for the variables density and sophistication)

| RANK | Product Code (HS-6) | Product Name | RCA 2000 | RCA 2005 | Exports 2000 | Exports 2005 | Export Share 2005 | Density | Sophistication |
|---|---|---|---|---|---|---|---|---|---|
| 1 | 71490 | Arrowroot, salep, etc fresh or dried and sago pith | 0.16 | 0.03 | 9.2 | 1.9 | 0.00% | 0.16590895 | -2.563694138 |
| 2 | 30232 | Tuna(yellowfin) fresh or chilled, whole | 0.00 | 0.00 | 0.0 | 0.0 | 0.00% | 0.15499774 | -2.081891758 |
| 3 | 180200 | Cocoa shells, husks, skins and waste | 0.00 | 0.00 | 0.0 | 0.0 | 0.00% | 0.15394123 | -3.338703232 |
| 4 | 80300 | Bananas, including plantains, fresh or dried | 0.07 | 0.01 | 114.3 | 18.6 | 0.00% | 0.15330262 | -1.841768924 |
| 5 | 30239 | Tuna nes, fresh or chilled, whole | 0.00 | 0.00 | 0.5 | 0.0 | 0.00% | 0.15165378 | -2.040132957 |
| 6 | 410520 | Sheep or lamb skin leather, nes | 0.02 | 0.01 | 7.6 | 3.0 | 0.00% | 0.15023272 | -1.446402982 |
| 7 | 110100 | Wheat or meslin flour | 5.65 | 0.03 | 2768.1 | 19.2 | 0.00% | 0.14922306 | -1.695554745 |
| 8 | 30342 | Tunas(yellowfin) frozen, whole | 0.00 | 0.00 | 0.0 | 0.0 | 0.00% | 0.14786909 | -2.423667889 |
| 9 | 120799 | Oil seeds and oleaginous fruits, nes | 0.58 | 0.01 | 26.5 | 1.0 | 0.00% | 0.1475061 | -1.919604527 |
| 10 | 30219 | Salmonidae, not trout or salmon,fresh or chilled whol | 0.12 | 0.00 | 4.3 | 0.0 | 0.00% | 0.14713105 | -2.017015783 |
| 11 | 30375 | Dogfish and other sharks, frozen, whole | 0.00 | 0.00 | 0.0 | 0.0 | 0.00% | 0.14693217 | -2.096914029 |
| 12 | 410320 | Reptile skins, raw | 3.30 | 0.02 | 99.7 | 0.8 | 0.00% | 0.1469282 | -2.121548998 |
| 13 | 610891 | Womens, girls bathrobe, dressing gowns, of knit cotto | 0.00 | 0.01 | 0.0 | 0.9 | 0.00% | 0.14633862 | -1.975256983 |
| 14 | 630532 |  | 0.51 | 0.00 | 80.8 | 0.0 | 0.00% | 0.14504271 | -1.282987134 |
| 15 | 30329 | Salmonidae, nes,frozen, whole | 0.00 | 0.00 | 0.0 | 0.0 | 0.00% | 0.14482213 | -1.970041997 |
| 16 | 30549 | Smoked fish & fillets other than herrings or salmon | 0.00 | 0.00 | 0.0 | 0.0 | 0.00% | 0.14438387 | -1.910575597 |
| 17 | 520291 | Garnetted stock of cotton | 0.00 | 0.01 | 0.0 | 0.1 | 0.00% | 0.14425613 | -1.630367428 |
| 18 | 611010 | Pullovers, cardigans etc of wool or hair, knit | 0.02 | 0.01 | 30.0 | 9.5 | 0.00% | 0.14408363 | -1.325757798 |
| 19 | 620891 | Womens, girls panties, bathrobes etc, cotton, not kni | 0.01 | 0.03 | 1.1 | 5.1 | 0.00% | 0.14393918 | -1.589114559 |
| 20 | 30623 | Shrimps and prawns, not frozen | 7.21 | 0.05 | 848.6 | 7.4 | 0.00% | 0.1439253 | -1.879168931 |
| 21 | 120720 | Cotton seeds | 0.03 | 0.00 | 2.1 | 0.1 | 0.00% | 0.14283361 | -2.022269687 |
| 22 | 240210 | Cigars, cheroots and cigarillos, containing tobacco | 0.08 | 0.06 | 23.7 | 24.0 | 0.00% | 0.14274385 | -1.412700166 |
| 23 | 121292 | Sugar cane | 9.44 | 0.00 | 15.3 | 0.0 | 0.00% | 0.14211084 | -1.252553414 |
| 24 | 120710 | Palm nuts and kernels | 3.23 | 0.00 | 26.6 | 0.0 | 0.00% | 0.14161669 | -2.472276222 |
| 25 | 170390 | Molasses, except cane molasses | 0.05 | 0.10 | 2.1 | 5.4 | 0.00% | 0.14114649 | -0.639559126 |
| 26 | 80711 |  | 0.17 | 0.03 | 19.8 | 5.4 | 0.00% | 0.14111187 | -1.406278728 |
| 27 | 621210 | Brassieres and parts thereof | 0.00 | 0.00 | 0.0 | 0.5 | 0.00% | 0.14098032 | -1.279601285 |
| 28 | 610711 | Mens, boys underpants or briefs, of cotton, knit | 0.02 | 0.02 | 13.9 | 15.9 | 0.00% | 0.14098001 | -1.466562403 |
| 29 | 90300 | Mate | 2.13 | 0.00 | 38.2 | 0.0 | 0.00% | 0.14067208 | -2.464298073 |
| 30 | 410140 | Equine hides and skins, raw | 2.37 | 0.00 | 49.7 | 0.0 | 0.00% | 0.14054148 | -0.568923317 |

Table 3 Kenya's top 30 export opportunities (products currently with an RCA<0.1, sorted by density). No sophistication filter was applied (Exports in thousands of US dollars, sophistication in standard deviations of world's average. See text for explanation for the variables density and sophistication)

| RANK | Product Code (HS-6) | Product Name | RCA 2000 | RCA 2005 | Exports 2000 | Exports 2005 | Export Share 2005 | Density | Sophistication |
|---|---|---|---|---|---|---|---|---|---|
| 1 | 230620 | Linseed oil-cake and other solid residues | 0.00 | 0.00 | 0.0 | 0.0 | 0.00% | 0.12460324 | 0.271428765 |
| 2 | 470200 | Chemical wood pulp, dissolving grades | 0.00 | 0.00 | 0.0 | 1.7 | 0.00% | 0.12361753 | 0.987360637 |
| 3 | 470100 | Mechanical wood pulp | 0.00 | 0.00 | 0.0 | 0.0 | 0.00% | 0.12317334 | 0.175729892 |
| 4 | 310420 | Potassium chloride, in packs >10 kg | 0.01 | 0.00 | 9.4 | 0.0 | 0.00% | 0.12105994 | 0.116769423 |
| 5 | 440392 | Logs, Beech (Fagus spp) | 0.00 | 0.00 | 0.0 | 0.0 | 0.00% | 0.11952658 | 0.226284116 |
| 6 | 940390 | Furniture parts nes | 0.06 | 0.02 | 94.2 | 38.2 | 0.00% | 0.11649989 | 0.201454559 |
| 7 | 150100 | Lard, other pig fat and poultry fat, rendered | 0.05 | 0.00 | 2.7 | 0.0 | 0.00% | 0.11582289 | 0.674889611 |
| 8 | 210330 | Mustard flour or meal and prepared mustard | 0.02 | 0.03 | 0.8 | 1.3 | 0.00% | 0.11548559 | 0.260103685 |
| 9 | 360690 | Pyrophoric alloys, firelighters, etc | 0.00 | 0.00 | 0.0 | 0.0 | 0.00% | 0.11519792 | 0.321399222 |
| 10 | 721491 | | 0.00 | 0.00 | 0.0 | 0.1 | 0.00% | 0.11518552 | 0.60429272 |
| 11 | 731413 | | 0.00 | 0.00 | 0.0 | 0.0 | 0.00% | 0.11341655 | 0.685124993 |
| 12 | 780411 | Lead foil of a thickness <2mm | 0.00 | 0.02 | 0.0 | 0.2 | 0.00% | 0.11318912 | 0.458350965 |
| 13 | 281512 | Sodium hydroxide (caustic soda) in aqueous solution | 0.01 | 0.01 | 2.9 | 9.5 | 0.00% | 0.11294351 | 0.057575448 |
| 14 | 330290 | Mixed odoriferous substances - industrial use nes | 0.01 | 0.00 | 5.0 | 5.2 | 0.00% | 0.11260801 | 0.443134105 |
| 15 | 280120 | Iodine | 0.01 | 0.02 | 0.7 | 1.9 | 0.00% | 0.11243564 | 0.054500941 |
| 16 | 280450 | Boron, tellurium | 0.07 | 0.00 | 0.4 | 0.0 | 0.00% | 0.11226624 | 0.000327362 |
| 17 | 392520 | Plastic doors and windows and frames thereof | 0.04 | 0.00 | 10.0 | 0.1 | 0.00% | 0.11215493 | 0.286256448 |
| 18 | 870540 | Mobile concrete mixers | 0.04 | 0.00 | 3.0 | 0.0 | 0.00% | 0.11214246 | 0.058742982 |
| 19 | 731210 | Stranded steel wire/cable/etc, no electric insulation | 0.02 | 0.03 | 14.5 | 35.8 | 0.00% | 0.1120754 | 0.166311784 |
| 20 | 841850 | Refrigerator/freezer chests/cabinets/showcases | 0.05 | 0.03 | 24.1 | 28.3 | 0.00% | 0.11207012 | 0.004647238 |
| 21 | 841840 | Freezers of the upright type, < 900 litre capacity | 0.00 | 0.01 | 0.7 | 3.5 | 0.00% | 0.11143735 | 0.086180032 |
| 22 | 310221 | Ammonium sulphate, in packs >10 kg | 0.78 | 0.05 | 124.9 | 14.4 | 0.00% | 0.11128818 | 0.147047473 |
| 23 | 701110 | Glass envelopes (bulbs & tubes) for electric lighting | 0.06 | 0.00 | 5.2 | 0.4 | 0.00% | 0.11118911 | 0.560549111 |
| 24 | 790112 | Zinc, not alloyed, unwrought, <99% pure | 0.28 | 0.10 | 120.7 | 42.2 | 0.00% | 0.11034333 | 0.010523826 |
| 25 | 441840 | Shuttering for concrete constructional work, of wood | 0.39 | 0.00 | 20.2 | 0.0 | 0.00% | 0.11015878 | 0.134749448 |
| 26 | 310240 | Ammonium nitrate limestone etc mixes, pack >10 kg | 0.08 | 0.03 | 16.8 | 9.9 | 0.00% | 0.10981403 | 0.471660853 |
| 27 | 200120 | Onions prepared or preserved by vinegar | 0.08 | 0.00 | 0.4 | 0.0 | 0.00% | 0.10965131 | 0.030994589 |
| 28 | 441600 | Wooden casks, barrels, vats, tubs, etc. | 0.01 | 0.00 | 0.9 | 0.1 | 0.00% | 0.10953135 | 0.031383767 |
| 29 | 271114 | Ethylene, propylene, butylene, butadiene, liquefied | 0.00 | 0.00 | 0.0 | 0.2 | 0.00% | 0.10947978 | 0.256873508 |
| 30 | 721550 | | 0.06 | 0.00 | 8.8 | 0.0 | 0.00% | 0.10938025 | 0.122256833 |

Table 4 Kenya's top 30 export opportunities: products currently with an RCA<0.1, sorted by density and sophistication larger than the World's average (Exports in thousands of US dollars, sophistication in standard deviations of world's average. See text for explanation for the variables density and sophisticati on)

| RANK | Product Code (HS-6) | Product Name | RCA 2000 | RCA 2005 | Exports 2000 | Exports 2005 | Export Share 2005 | Density | Sophistication |
|---|---|---|---|---|---|---|---|---|---|
| 1 | 760110 | Aluminium unwrought, not alloyed | 25.05 | 172.79 | 29037.7 | 1146047.4 | 35.26% | 0.0261182 | -1.063062639 |
| 2 | 760200 | Waste or scrap, aluminium | 1.39 | 420.73 | 454.9 | 1021251.2 | 31.42% | 0.04014 | -1.186743412 |
| 3 | 271600 | Electrical energy | 65.25 | 16.66 | 58743.7 | 141800.4 | 4.36% | 0.0350581 | -0.380989255 |
| 4 | 271111 | Natural gas, liquefied | 0 | 8 | 1 | 110743 | 3.41% | 0.0460342 | -2.50948164 |
| 5 | 760120 | Aluminium unwrought, alloyed | 64 | 20 | 60155 | 110703 | 3.41% | 0.0267918 | -0.44956242 |
| 6 | 30613 | Shrimps and prawns, frozen | 127.91 | 34.07 | 101335.4 | 102134.5 | 3.14% | 0.0546445 | -2.934308359 |
| 7 | 271000 | Oils petroleum, bituminous, distillates, except crude | 0.79 | 0.80 | 10873.3 | 93370.3 | 2.87% | 0.0468735 | -1.243368812 |
| 8 | 240120 | Tobacco, unmanufactured, stemmed or stripped | 24 | 39 | 9386 | 67668 | 2.08% | 0.0513471 | -2.276013751 |
| 9 | 170111 | Raw sugar, cane | 29.35 | 14.06 | 10347.7 | 28701.0 | 0.88% | 0.0420806 | -2.82012353 |
| 10 | 440399 | Logs, non-coniferous nes | 88.79 | 27.15 | 16627.0 | 26975.7 | 0.83% | 0.037359 | -2.445578611 |
| 11 | 240110 | Tobacco, unmanufactured, not stemmed or stripped | 24 | 42 | 3434 | 26560 | 0.82% | 0.0402222 | -2.43444812 |
| 12 | 520100 | Cotton, not carded or combed | 52.69 | 8.56 | 31067.1 | 26232.0 | 0.81% | 0.0310062 | -2.429388805 |
| 13 | 80131 |  | 991 | 169 | 22149 | 25042 | 0.77% | 0.0605552 | -3.802525587 |
| 14 | 840710 | Aircraft engines, spark-ignition | 0.08 | 41.40 | 5.5 | 19385.2 | 0.60% | 0.0266769 | -1.180127385 |
| 15 | 260111 | Iron ore, concentrate, not iron pyrites,unagglomerate | 0 | 3 | 0 | 18674 | 0.57% | 0.039209 | -1.440253969 |
| 16 | 520300 | Cotton, carded or combed | 958.56 | 223.76 | 18901.6 | 16493.1 | 0.51% | 0.039571 | -1.893335011 |
| 17 | 490700 | Documents of title (bonds etc), unused stamps etc | 0 | 34 | 3 | 16470 | 0.51% | 0.032 | -1.162 |
| 18 | 270119 | Coal except anthracite or bituminous, not agglomerate | 12 | 5 | 3562 | 14469 | 0.45% | 0.0329599 | -1.297075378 |
| 19 | 440320 | Logs, poles, coniferous not treated or painted | 6.79 | 6.38 | 2195.8 | 11652.5 | 0.36% | 0.0374154 | -0.232401089 |
| 20 | 720449 | Ferrous waste or scrap, nes | 3 | 2 | 941 | 10952 | 0.34% | 0.0434412 | -1.364792352 |
| 21 | 120740 | Sesamum seeds | 17.58 | 34.63 | 816.8 | 9166.1 | 0.28% | 0.0546572 | -2.81658201 |
| 22 | 731100 | Containers for compressed/liquefied gas, iron or stee | 1.19 | 16.42 | 94.7 | 8929.7 | 0.27% | 0.0325016 | -0.599785133 |
| 23 | 392310 | Boxes, cases, crates etc. of plastic | 0 | 4 | 3 | 8380 | 0.26% | 0.0267797 | -0.247734703 |
| 24 | 871130 | Motorcycles, spark ignition engine of 250-500 cc | 0.01 | 16.88 | 0.5 | 7263.9 | 0.22% | 0.0203834 | 0.566308946 |
| 25 | 880240 | Fixed wing aircraft, unladen weight > 15,000 kg | 0 | 0 | 0 | 6423 | 0.20% | 0.0329302 | -0.245749895 |
| 26 | 251611 | Granite, crude or roughly trimmed | 150.70 | 20.11 | 8161.5 | 6205.1 | 0.19% | 0.0354978 | -1.086452238 |
| 27 | 720429 | Waste or scrap, of alloy steel, other than stainless | 7.73 | 14.97 | 317.0 | 5354.6 | 0.16% | 0.0443898 | -1.624374087 |
| 28 | 80132 |  | 194.59 | 11.38 | 13749.0 | 4830.8 | 0.15% | 0.0500789 | -2.936215331 |
| 29 | 440799 | Lumber, non-coniferous nes | 11.62 | 3.60 | 3297.2 | 4725.8 | 0.15% | 0.0390725 | -2.229001047 |
| 30 | 310560 | Fertilizers containing phosphorus & potassium, <=10kg | 0 | 60 | 0 | 4599 | 0.14% | 0.0293164 | -0.291672901 |

Table 5 Mozambique's top 30 exports sorted in decreasing total exports volume (Exports in thousands of US dollars, sophistication in standard deviations of world's average. See text for explanation for the variables density and sophistication)

| RANK | Product Code (HS-6) | Product Name | RCA 2000 | RCA 2005 | Exports 2000 | Exports 2005 | Export Share 2005 | Density | Sophistication |
|---|---|---|---|---|---|---|---|---|---|
| 1 | 760200 | Waste or scrap, aluminium | 1.39 | 420.73 | 454.9 | 1021251.2 | 31.42% | 0.04014 | -1.186743412 |
| 2 | 520300 | Cotton, carded or combed | 958.56 | 223.76 | 18901.6 | 16493.1 | 0.51% | 0.039571 | -1.893335011 |
| 3 | 530390 | Jute and other bast fibres, not spun, nes, tow, waste | 0.00 | 202.56 | 0.0 | 992.0 | 0.03% | 0.028468 | -1.655741834 |
| 4 | 760110 | Aluminium unwrought, not alloyed | 25.05 | 172.79 | 29037.7 | 1146047.4 | 35.26% | 0.0261182 | -1.063062639 |
| 5 | 80131 | | 991 | 169 | 22149 | 25042 | 0.77% | 0.0605552 | -3.802525587 |
| 6 | 120730 | Castor oil seeds | 0.00 | 127.57 | 0.0 | 301.4 | 0.01% | 0.0431732 | -2.457526376 |
| 7 | 560729 | Twine nes, cordage, ropes and cables, of sisal | 193.00 | 105.55 | 631.7 | 1337.3 | 0.04% | 0.0371803 | -1.799115014 |
| 8 | 530490 | Sisal and Agave, processed but not spun, tow & waste | 240 | 104 | 203 | 618 | 0.02% | 0.035069 | -1.809039054 |
| 9 | 282630 | Sodium hexafluoroaluminate (synthetic cryolite) | 0.00 | 63.23 | 0.0 | 1267.0 | 0.04% | 0.0260161 | 0.127160476 |
| 10 | 310560 | Fertilizers containing phosphorus & potassium, <=10kg | 0 | 60 | 0 | 4599 | 0.14% | 0.0293164 | -0.291672901 |
| 11 | 240110 | Tobacco, unmanufactured, not stemmed or stripped | 24 | 42 | 3434 | 26560 | 0.82% | 0.0402222 | -2.43444812 |
| 12 | 840710 | Aircraft engines, spark-ignition | 0.08 | 41.40 | 5.5 | 19385.2 | 0.60% | 0.0266769 | -1.180127385 |
| 13 | 681130 | Tubes, pipes etc of asbestos or cellulose fibre cemen | 0 | 41 | 0 | 416 | 0.01% | 0.0339002 | -0.330512866 |
| 14 | 240120 | Tobacco, unmanufactured, stemmed or stripped | 24 | 39 | 9386 | 67668 | 2.08% | 0.0513471 | -2.276013751 |
| 15 | 120740 | Sesamum seeds | 17.58 | 34.63 | 816.8 | 9166.1 | 0.28% | 0.0546572 | -2.81658201 |
| 16 | 30613 | Shrimps and prawns, frozen | 127.91 | 34.07 | 101335.4 | 102134.5 | 3.14% | 0.0546445 | -2.934308359 |
| 17 | 490700 | Documents of title (bonds etc), unused stamps etc | 0 | 34 | 3 | 16470 | 0.51% | 0.032 | -1.162 |
| 18 | 170310 | Cane molasses | 21.28 | 27.97 | 548.3 | 4289.7 | 0.13% | 0.0372113 | -2.415222726 |
| 19 | 440399 | Logs, non-coniferous nes | 88.79 | 27.15 | 16627.0 | 26975.7 | 0.83% | 0.037359 | -2.445578611 |
| 20 | 410320 | Reptile skins, raw | 4 | 25 | 35 | 1181 | 0.04% | 0.0481133 | -2.121548998 |
| 21 | 120720 | Cotton seeds | 39.43 | 22.40 | 722.0 | 1776.5 | 0.05% | 0.0506567 | -2.022269687 |
| 22 | 30219 | Salmonidae, not trout or salmon,fresh or chilled whol | 0.28 | 21.47 | 2.7 | 1444.7 | 0.04% | 0.0483971 | -2.017015783 |
| 23 | 110220 | Maize (corn) flour | 0.00 | 21.05 | 0.0 | 1096.3 | 0.03% | 0.0417157 | -1.450645022 |
| 24 | 230610 | Cotton seed oil-cake and other solid residues | 94.13 | 20.62 | 546.1 | 513.5 | 0.02% | 0.0422035 | -2.431334695 |
| 25 | 251611 | Granite, crude or roughly trimmed | 150.70 | 20.11 | 8161.5 | 6205.1 | 0.19% | 0.0354978 | -1.086452238 |
| 26 | 760120 | Aluminium unwrought, alloyed | 64 | 20 | 60155 | 110703 | 3.41% | 0.0267918 | -0.44956242 |
| 27 | 30559 | Dried fish, other than cod, not smoked | 31.79 | 17.10 | 1810.7 | 3991.2 | 0.12% | 0.0514922 | -2.497884136 |
| 28 | 871130 | Motorcycles, spark ignition engine of 250-500 cc | 0.01 | 16.88 | 0.5 | 7263.9 | 0.22% | 0.0203834 | 0.566308946 |
| 29 | 271600 | Electrical energy | 65.25 | 16.66 | 58743.7 | 141800.4 | 4.36% | 0.0350581 | -0.380989255 |
| 30 | 230230 | Wheat bran, sharps, other residues | 66.70 | 16.43 | 1181.9 | 1850.9 | 0.06% | 0.0395094 | -2.07087802 |

Table 6 Mozambique's top 30 exports sorted in decreasing order of RCA (Exports in thousands of US dollars, sophistication in standard deviations of world's average. See text for explanation for the variables density and sophistication)

| RANK | Product Code (HS-6) | Product Name | RCA 2000 | RCA 2005 | Exports 2000 | Exports 2005 | Export Share 2005 | Density | Sophistication |
|---|---|---|---|---|---|---|---|---|---|
| 1 | 90111 | Coffee, not roasted, not decaffeinated | 0 | 0 | 0 | 0 | 0.00% | 0.0613812 | -2.9031352 |
| 2 | 180200 | Cocoa shells, husks, skins and waste | 0 | 0 | 0 | 0 | 0.00% | 0.0558853 | -3.338703232 |
| 3 | 71490 | Arrowroot, salep, etc fresh or dried and sago pith | 0 | 0 | 0 | 0 | 0.00% | 0.0508766 | -2.563694138 |
| 4 | 130214 | Pyrethrum, roots containing rotenone, extracts | 0.00 | 0.00 | 0.0 | 0.0 | 0.00% | 0.049673 | -2.639856275 |
| 5 | 70820 | Beans, shelled or unshelled, fresh or chilled | 13 | 0 | 340 | 0 | 0.00% | 0.0490969 | -1.999853033 |
| 6 | 30410 | Fish fillet or meat, fresh or chilled, not liver, roe | 0.01 | 0.00 | 0.9 | 0.0 | 0.00% | 0.048143 | -2.06344472 |
| 7 | 30329 | Salmonidae, nes,frozen, whole | 36.75 | 0.00 | 564.0 | 0.0 | 0.00% | 0.0474844 | -1.970041997 |
| 8 | 100640 | Rice, broken | 1.36 | 0.00 | 58.1 | 0.0 | 0.00% | 0.0472192 | -2.195376067 |
| 9 | 440420 | Poles, piles etc, non-coniferous, pointed, not sawn | 0.04 | 0.00 | 0.1 | 0.0 | 0.00% | 0.0467838 | -1.659283354 |
| 10 | 30490 | Fish meat & mince, except liver, roe & fillets, froze | 0.79 | 0.00 | 72.0 | 1.1 | 0.00% | 0.046715 | -1.979576859 |
| 11 | 120999 | Seed, fruits and spores for sowing, nes | 0.00 | 0.06 | 0.0 | 5.1 | 0.00% | 0.0466654 | -1.77300117 |
| 12 | 30110 | Ornamental fish, live | 0 | 0 | 0 | 1 | 0.00% | 0.0465418 | -2.539720772 |
| 13 | 260500 | Cobalt ores and concentrates | 0 | 0 | 0 | 0 | 0.00% | 0.0463857 | -3.619028154 |
| 14 | 90220 | Tea, green (unfermented) in packages > 3 kg | 0 | 0 | 0 | 0 | 0.00% | 0.0463743 | -2.780154948 |
| 15 | 440726 |  | 0 | 0 | 0 | 0 | 0.00% | 0.0463271 | -1.62289521 |
| 16 | 30420 | Fish fillets, frozen | 0.00 | 0.00 | 0.0 | 0.0 | 0.00% | 0.0462504 | -1.775608662 |
| 17 | 71320 | Chickpeas, dried, shelled | 0.00 | 0.00 | 0.0 | 0.0 | 0.00% | 0.0460165 | -2.25974611 |
| 18 | 91030 | Turmeric (curcuma) | 0.00 | 0.00 | 0.0 | 0.0 | 0.00% | 0.0458273 | -3.048765612 |
| 19 | 90190 |  | 0.00 | 0.00 | 0.0 | 0.0 | 0.00% | 0.0455661 | -2.415689739 |
| 20 | 410229 | Sheep or lamb skins, raw, except pickled, no wool | 0.00 | 0.00 | 0.0 | 0.0 | 0.00% | 0.0453491 | -1.829081721 |
| 21 | 70890 | Legumes except peas & beans, fresh or chilled | 0.00 | 0.00 | 0.0 | 0.0 | 0.00% | 0.0449028 | -1.973427846 |
| 22 | 520291 | Garnetted stock of cotton | 0.00 | 0.00 | 0.0 | 0.0 | 0.00% | 0.04469 | -1.630367428 |
| 23 | 30232 | Tuna(yellowfin) fresh or chilled, whole | 0.00 | 0.00 | 0.0 | 0.0 | 0.00% | 0.0446777 | -2.081891758 |
| 24 | 261610 | Silver ores and concentrates | 0.00 | 0.00 | 0.0 | 0.0 | 0.00% | 0.0446421 | -1.716492522 |
| 25 | 260300 | Copper ores and concentrates | 0.00 | 0.00 | 0.0 | 0.0 | 0.00% | 0.0444872 | -2.29017983 |
| 26 | 30339 | Flatfish except halibut, plaice or sole, frozen, whol | 0 | 0 | 0 | 0 | 0.00% | 0.0443845 | -2.145094267 |
| 27 | 140420 | Cotton linters | 0.00 | 0.00 | 0.0 | 0.0 | 0.00% | 0.0443697 | -1.934081949 |
| 28 | 81090 | Fruits, fresh nes | 0.22 | 0.00 | 10.6 | 0.0 | 0.00% | 0.0441943 | -1.649203644 |
| 29 | 410140 | Equine hides and skins, raw | 0.00 | 0.00 | 0.0 | 0.0 | 0.00% | 0.0441542 | -0.568923317 |
| 30 | 30549 | Smoked fish & fillets other than herrings or salmon | 1.33 | 0.09 | 16.3 | 7.7 | 0.00% | 0.0436408 | -1.910575597 |

Table 7 Mozambique's top 30 export opportunities (products currently with an RCA<0.1, sorted by density). No sophistication filter was applied (Exports in thousands of US dollars, sophistication in standard deviations of world's average. See text for explanation for the variables density and sophistication)

| RANK | Product Code (HS-6) | Product Name | RCA 2000 | RCA 2005 | Exports 2000 | Exports 2005 | Export Share 2005 | Density | Sophistication |
|---|---|---|---|---|---|---|---|---|---|
| 1 | 230620 | Linseed oil-cake and other solid residues | 0.28 | 0.00 | 1.3 | 0.0 | 0.00% | 0.0393061 | 0.271428765 |
| 2 | 440392 | Logs, Beech (Fagus spp) | 0.05 | 0.00 | 2.1 | 0.0 | 0.00% | 0.039078 | 0.226284116 |
| 3 | 720441 | Waste from the mechanical working of iron or steel ne | 0.00 | 0.00 | 0.0 | 0.0 | 0.00% | 0.0369676 | 0.343348862 |
| 4 | 470100 | Mechanical wood pulp | 0.00 | 0.00 | 0.0 | 0.0 | 0.00% | 0.0361449 | 0.175729892 |
| 5 | 470200 | Chemical wood pulp, dissolving grades | 0.00 | 0.00 | 0.0 | 0.0 | 0.00% | 0.0361095 | 0.987360637 |
| 6 | 252330 | Aluminous cement | 0.00 | 0.00 | 0.0 | 0.0 | 0.00% | 0.0351538 | 0.137862872 |
| 7 | 210420 | Homogenised composite food preparations | 0.00 | 0.00 | 0.0 | 0.0 | 0.00% | 0.0346671 | 0.113539246 |
| 8 | 340211 | Anionic surface-active agents | 0.00 | 0.00 | 0.1 | 0.0 | 0.00% | 0.0343677 | 0.319531168 |
| 9 | 710900 | Base metals or silver, clad with gold, semi-manuf | 0.00 | 0.00 | 0.0 | 0.0 | 0.00% | 0.0341821 | 0.200754038 |
| 10 | 310420 | Potassium chloride, in packs >10 kg | 0 | 0 | 16 | 0 | 0.00% | 0.034 | 0.117 |
| 11 | 721050 | Flat rolled iron or non-alloy steel, coated chromium/oxides, w> 600 mm | 0 | 0 | 0 | 0 | 0.00% | 0.0340552 | 0.532333705 |
| 12 | 760310 | Powders, aluminium, of non-lamellar structure | 0 | 0 | 0 | 0 | 0.00% | 0.033321 | 0.596353488 |
| 13 | 260120 | Roasted iron pyrites | 0.00 | 0.00 | 0.0 | 0.0 | 0.00% | 0.0332822 | 0.33860089 |
| 14 | 780411 | Lead foil of a thickness <2mm | 0.20 | 0.00 | 0.3 | 0.0 | 0.00% | 0.0329394 | 0.458350965 |
| 15 | 170211 | | 0.00 | 0.00 | 0.0 | 0.0 | 0.00% | 0.0326548 | 0.896487571 |
| 16 | 441019 | | 0 | 0 | 0 | 0 | 0.00% | 0.0326269 | 0.335798809 |
| 17 | 360690 | Pyrophoric alloys, firelighters, etc | 0.00 | 0.00 | 0.0 | 0.0 | 0.00% | 0.0323689 | 0.321399222 |
| 18 | 310250 | Sodium nitrate, in packs >10 kg | 0.00 | 0.00 | 0.0 | 0.0 | 0.00% | 0.0322911 | 0.134905119 |
| 19 | 760429 | Bars, rods and other profiles, aluminium alloyed | 0.00 | 0.00 | 0.0 | 1.8 | 0.00% | 0.0320093 | 0.031461603 |
| 20 | 280120 | Iodine | 0.00 | 0.00 | 0.0 | 0.0 | 0.00% | 0.031999 | 0.054500941 |
| 21 | 440391 | Logs, Oak (Quercus spp) | 0.00 | 0.00 | 0.0 | 0.0 | 0.00% | 0.0318578 | 0.108830192 |
| 22 | 281512 | Sodium hydroxide (caustic soda) in aqueous solution | 0.00 | 0.00 | 0.0 | 0.0 | 0.00% | 0.0318572 | 0.057575448 |
| 23 | 150100 | Lard, other pig fat and poultry fat, rendered | 0 | 0 | 0 | 0 | 0.00% | 0.031726 | 0.674889611 |
| 24 | 731413 | | 0.00 | 0.00 | 0.0 | 0.0 | 0.00% | 0.0317125 | 0.685124993 |
| 25 | 721491 | | 0 | 0 | 1 | 0 | 0.00% | 0.0316998 | 0.60429272 |
| 26 | 280450 | Boron, tellurium | 0.00 | 0.00 | 0.0 | 0.0 | 0.00% | 0.0315519 | 0.000327362 |
| 27 | 210330 | Mustard flour or meal and prepared mustard | 0 | 0 | 0 | 0 | 0.00% | 0.0313111 | 0.260103685 |
| 28 | 262050 | Ash or residues containing mainly vanadium | 0 | 0 | 0 | 0 | 0.00% | 0.0312621 | 0.540350772 |
| 29 | 253020 | Kieserite, epsomite (natural magnesium sulphates) | 0.00 | 0.00 | 0.0 | 0.0 | 0.00% | 0.0312164 | 3.04599557 |
| 30 | 760521 | Wire, aluminium alloy, t > 7mm | 0 | 0 | 0 | 0 | 0.00% | 0.03121 | 0.127355065 |

Table 8 Mozambique's top 30 export opportunities: products currently with an RCA<0.1, sorted by density and sophistication larger than the World's average (Exports in thousands of US dollars, sophistication in standard deviations of world's average. See text for explanation for the variables density and sophistication)

| RANK | Product Code (HS-6) | Product Name | RCA 2000 | RCA 2005 | Exports 2000 | Exports 2005 | Export Share 2005 | Density | Sophistication |
|---|---|---|---|---|---|---|---|---|---|
| 1 | 90111 | Coffee, not roasted, not decaffeinated | 441.420314 | 521.979012 | 31385.24006 | 44265.46832 | 47.83% | 0.0314505 | -2.9031352 |
| 2 | 261590 | Niobium, tantalum and vanadium ores and concentrates | 3732.50743 | 6318.24308 | 9909.444169 | 21613.942 | 23.35% | 0.0235412 | -2.50878112 |
| 3 | 260900 | Tin ores and concentrates | 2289.47427 | 4808.94492 | 3373.093367 | 8464.817339 | 9.15% | 0.0208389 | -3.384431648 |
| 4 | 90240 | Tea, black (fermented or partly) in packages > 3 kg | 139.879389 | 473.358789 | 1872.194797 | 7663.408951 | 8.28% | 0.0278906 | -2.479865194 |
| 5 | 261100 | Tungsten ores and concentrates | 1548.9115 | 2181.79157 | 333.2573809 | 2094.625118 | 2.26% | 0.0230205 | -2.542717443 |
| 6 | 310530 | Diammonium phosphate, in packs >10 kg | 0 | 76.1549061 | 0 | 1848.254436 | 2.00% | 0.0177185 | -0.859717128 |
| 7 | 60310 | Cut flowers and flower buds for bouquets, etc., fresh | 20.8606549 | 8.66538872 | 685.1654281 | 455 | 0.49% | 0.0269937 | -1.859826784 |
| 8 | 282590 | Metal bases, oxides, hydroxides, peroxides, nes | 0 | 50.3830248 | 0 | 363.905 | 0.39% | 0.0163625 | 1.109523615 |
| 9 | 810390 | Tantalum and articles thereof nes | 0 | 156.220679 | 0 | 319.853 | 0.35% | 0.0151484 | 0.329182783 |
| 10 | 410310 | Goat or kid hides and skins, raw, nes | 45.3741458 | 539.217128 | 13.654 | 281.1980483 | 0.30% | 0.0233977 | -2.174516125 |
| 11 | 262090 | Ash or residues containing metals, metal compounds ne | 0 | 22.8166069 | 0 | 258.0322061 | 0.28% | 0.0193265 | -0.167836457 |
| 12 | 60290 | | 8.66909509 | 5.85386452 | 204.2129602 | 250.265 | 0.27% | 0.0182555 | -1.083416649 |
| 13 | 551311 | Woven plain >85% polyester+cotton, <170g/m2 unbl/blch | 0.71953305 | 33.4790218 | 6.252 | 241.447 | 0.26% | 0.0170294 | -1.530426514 |
| 14 | 410612 | Goat or kid skin leather, otherwise pre-tanned | 0 | 121.331643 | 0 | 196.526 | 0.21% | 0.0227006 | -2.760696047 |
| 15 | 220210 | Beverage waters, sweetened or flavoured | 0 | 3.922635 | 0 | 154.439998 | 0.17% | 0.0190999 | -0.828543969 |
| 16 | 410210 | Sheep or lamb skins, raw, wool on, except Persian etc | 17.3518312 | 27.1799028 | 65.02494674 | 144.5783082 | 0.16% | 0.0214604 | -1.532138898 |
| 17 | 330499 | Beauty, makeup and suntan preparations nes | 0.24099709 | 1.27658937 | 12.488 | 140.127 | 0.15% | 0.0182458 | -0.296031694 |
| 18 | 460210 | Basketwork, wickerwork products of vegetable material | 3.11424052 | 10.7980159 | 19.14899076 | 120.9009092 | 0.13% | 0.0222021 | -2.607437746 |
| 19 | 410110 | Bovine skins, whole, raw | 7.55034105 | 15.020622 | 26.00471884 | 117.1578253 | 0.13% | 0.0173656 | -1.772222814 |
| 20 | 252329 | Portland cement, other than white cement | 0 | 2.44555682 | 0 | 113.9507842 | 0.12% | 0.0202547 | -1.34926415 |
| 21 | 80300 | Bananas, including plantains, fresh or dried | 2.78094839 | 1.38096624 | 125.6454876 | 97.73889666 | 0.11% | 0.0260931 | -1.841768924 |
| 22 | 842940 | Tamping machines and road rollers, self- propelled | 0 | 6.92419362 | 0 | 95.229 | 0.10% | 0.0207594 | -0.367212353 |
| 23 | 293921 | Quinine, salts, in bulk | 107.078948 | 231.961165 | 51.6 | 87.942 | 0.10% | 0.0157019 | -1.314977567 |
| 24 | 410422 | Bovine leather, otherwise pre-tanned except whole ski | 0 | 3.59846984 | 0 | 87.03255921 | 0.09% | 0.0209569 | -1.433404436 |
| 25 | 442010 | Statuettes and other ornaments of wood | 10.559544 | 15.2657409 | 53.48977873 | 86.39178358 | 0.09% | 0.0207953 | -3.045107339 |
| 26 | 90121 | Coffee, roasted, not decaffeinated | 0 | 3.98562872 | 0 | 86.28964315 | 0.09% | 0.0186342 | -0.53222383 |
| 27 | 843149 | Parts of cranes, work-trucks, shovels, constr machine | 0.16606874 | 0.61737135 | 9.839 | 84 | 0.09% | 0.0146318 | 0.971832435 |
| 28 | 261790 | ores and concentrates nes | 0 | 75.9108082 | 0 | 80.88203368 | 0.09% | 0.0199537 | -1.253098263 |
| 29 | 950691 | Physical exercise, gymnasium and athletics equipment | 0.09607913 | 1.87522327 | 1.917 | 76.978 | 0.08% | 0.0187774 | 0.391451265 |
| 30 | 282810 | Commercial and other calcium hypochlorite | 0 | 52.6389843 | 0 | 69.0113358 | 0.07% | 0.0161032 | -1.292677667 |

Table 9 Rwanda's top 30 exports sorted in decreasing total exports volume (Exports in thousands of US dollars, sophistication in standard deviations of world's average. See text for explanation for the variables density and sophistication)

| RANK | Product Code (HS-6) | Product Name | RCA 2000 | RCA 2005 | Exports 2000 | Exports 2005 | Export Share 2005 | Density | Sophistication |
|---|---|---|---|---|---|---|---|---|---|
| 1 | 261590 | Niobium, tantalum and vanadium ores and concentrates | 3732.50743 | 6318.24308 | 9909.444169 | 21613.942 | 23.35% | 0.0235412 | -2.50878112 |
| 2 | 260900 | Tin ores and concentrates | 2289.47427 | 4808.94492 | 3373.093367 | 8464.817339 | 9.15% | 0.0208389 | -3.384431648 |
| 3 | 261100 | Tungsten ores and concentrates | 1548.9115 | 2181.79157 | 333.2573809 | 2094.625118 | 2.26% | 0.0230205 | -2.542717443 |
| 4 | 410310 | Goat or kid hides and skins, raw, nes | 45.3741458 | 539.217128 | 13.654 | 281.1980483 | 0.30% | 0.0233977 | -2.174516125 |
| 5 | 90111 | Coffee, not roasted, not decaffeinated | 441.420314 | 521.979012 | 31385.24006 | 44265.46832 | 47.83% | 0.0314505 | -2.9031352 |
| 6 | 90240 | Tea, black (fermented or partly) in packages > 3 kg | 139.879389 | 473.358789 | 1872.194797 | 7663.408951 | 8.28% | 0.0278906 | -2.479865194 |
| 7 | 293921 | Quinine, salts, in bulk | 107.078948 | 231.961165 | 51.6 | 87.942 | 0.10% | 0.0157019 | -1.314977567 |
| 8 | 110620 | Flour or meal of sago, starchy roots or tubers | 0 | 201.76558 | 0 | 32.77575258 | 0.04% | 0.0232183 | -2.197944641 |
| 9 | 810390 | Tantalum and articles thereof nes | 0 | 156.220679 | 0 | 319.853 | 0.35% | 0.0151484 | 0.329182783 |
| 10 | 410612 | Goat or kid skin leather, otherwise pre-tanned | 0 | 121.331643 | 0 | 196.526 | 0.21% | 0.0227006 | -2.760696047 |
| 11 | 310530 | Diammonium phosphate, in packs >10 kg | 0 | 76.1549061 | 0 | 1848.254436 | 2.00% | 0.0177185 | -0.859717128 |
| 12 | 261790 | ores and concentrates nes | 0 | 75.9108082 | 0 | 80.88203368 | 0.09% | 0.0199537 | -1.253098263 |
| 13 | 282810 | Commercial and other calcium hypochlorite | 0 | 52.6389843 | 0 | 69.0113358 | 0.07% | 0.0161032 | -1.292677667 |
| 14 | 282590 | Metal bases, oxides, hydroxides, peroxides, nes | 0 | 50.3830248 | 0 | 363.905 | 0.39% | 0.0163625 | 1.109523615 |
| 15 | 551311 | Woven plain >85% polyester+cotton, <170g/m2 unbl/blch | 0.71953305 | 33.4790218 | 6.252 | 241.447 | 0.26% | 0.0170294 | -1.530426514 |
| 16 | 410210 | Sheep or lamb skins, raw, wool on, except Persian etc | 17.3518312 | 27.1799028 | 65.02494674 | 144.5783082 | 0.16% | 0.0214604 | -1.532138898 |
| 17 | 262090 | Ash or residues containing metals, metal compounds ne | 0 | 22.8166069 | 0 | 258.0322061 | 0.28% | 0.0193265 | -0.167836457 |
| 18 | 710310 | Precious, semi-precious stones unworked, partly worke | 0.41860155 | 16.5540614 | 1 | 53.065 | 0.06% | 0.0219526 | -2.370389419 |
| 19 | 442010 | Statuettes and other ornaments of wood | 10.559544 | 15.2657409 | 53.48977873 | 86.39178358 | 0.09% | 0.0207953 | -3.045107339 |
| 20 | 410110 | Bovine skins, whole, raw | 7.55034105 | 15.020622 | 26.00471884 | 117.1578253 | 0.13% | 0.0173656 | -1.772222814 |
| 21 | 840120 | Machinery & apparatus for isotopic separation & parts | 0 | 11.9827296 | 0 | 18.403 | 0.02% | 0.0169681 | 0.142182748 |
| 22 | 110819 | Starches except wheat, maize, potato, manioc | 0 | 11.5883999 | 0 | 9.39987536 | 0.01% | 0.020413 | -1.483180304 |
| 23 | 320120 | Wattle tanning extract | 0 | 11.4096599 | 0 | 7.516766613 | 0.01% | 0.0227492 | -2.100650138 |
| 24 | 460210 | Basketwork, wickerwork products of vegetable material | 3.11424052 | 10.7980159 | 19.14899076 | 120.9009092 | 0.13% | 0.0222021 | -2.607437746 |
| 25 | 60310 | Cut flowers and flower buds for bouquets, etc., fresh | 20.8606549 | 8.66538872 | 685.1654281 | 455 | 0.49% | 0.0269937 | -1.859826784 |
| 26 | 410620 | Goat or kid skin leather, nes | 0 | 7.25173457 | 0 | 47.605 | 0.05% | 0.0227001 | -2.305435608 |
| 27 | 842940 | Tamping machines and road rollers, self- propelled | 0 | 6.92419362 | 0 | 95.229 | 0.10% | 0.0207594 | -0.367212353 |
| 28 | 721041 | Flat rolled iron or non-alloy steel, coat/zinc, corrugated, w >600m, ne | 0 | 6.21091961 | 0 | 18.66680501 | 0.02% | 0.0189646 | -1.762882541 |
| 29 | 60290 | | 8.66909509 | 5.85386452 | 204.2129602 | 250.265 | 0.27% | 0.0182555 | -1.083416649 |
| 30 | 830220 | Castors of base metal | 0 | 5.80887683 | 0 | 22.63201362 | 0.02% | 0.0150612 | 0.439047735 |

Table 10 Rwanda's top 30 exports sorted in decreasing order of RCA (Exports in thousands of US dollars, sophistication in standard deviations of world's average. See text for explanation for the variables density and sophistication)

| RANK | Product Code (HS-6) | Product Name | RCA 2000 | RCA 2005 | Exports 2000 | Exports 2005 | Export Share 2005 | Density | Sophistication |
|---|---|---|---|---|---|---|---|---|---|
| 1 | 120792 | Shea nuts (karite nuts) | 0 | 0 | 0 | 0 | 0.00% | 0.0339892 | -0.997291555 |
| 2 | 140210 | Kapok | 0 | 0 | 0 | 0 | 0.00% | 0.0274315 | -1.73209856 |
| 3 | 180200 | Cocoa shells, husks, skins and waste | 0 | 0 | 0 | 0 | 0.00% | 0.0271777 | -3.338703232 |
| 4 | 71490 | Arrowroot, salep, etc fresh or dried and sago pith | 0 | 0 | 0 | 0 | 0.00% | 0.0268164 | -2.563694138 |
| 5 | 120740 | Sesamum seeds | 0 | 0 | 0 | 0 | 0.00% | 0.0265097 | -2.81658201 |
| 6 | 121292 | Sugar cane | 0 | 0 | 0 | 0 | 0.00% | 0.0264891 | -1.252553414 |
| 7 | 80131 | | 0 | 0 | 0 | 0 | 0.00% | 0.02573 | -3.802525587 |
| 8 | 80450 | Guavas, mangoes and mangosteens, fresh or dried | 0.25050269 | 0 | 1.113826714 | 0 | 0.00% | 0.0254722 | -2.467333662 |
| 9 | 260500 | Cobalt ores and concentrates | 0 | 0 | 0 | 0 | 0.00% | 0.0252754 | -3.619028154 |
| 10 | 80119 | | 0 | 0 | 0 | 0 | 0.00% | 0.0250781 | -1.982807036 |
| 11 | 91030 | Turmeric (curcuma) | 0 | 0 | 0 | 0 | 0.00% | 0.0250484 | -3.048765612 |
| 12 | 90220 | Tea, green (unfermented) in packages > 3 kg | 0 | 0 | 0 | 0 | 0.00% | 0.0250292 | -2.780154948 |
| 13 | 91010 | Ginger | 0 | 0 | 0 | 0 | 0.00% | 0.0249543 | -2.802026752 |
| 14 | 30110 | Ornamental fish, live | 0 | 0 | 0 | 0 | 0.00% | 0.0248312 | -2.539720772 |
| 15 | 30611 | Rock lobster and other sea crawfish, frozen | 0 | 0 | 0 | 0 | 0.00% | 0.0247892 | -2.344820423 |
| 16 | 71331 | Urd,mung,black or green gram beans dried shelled | 0 | 0 | 0 | 0 | 0.00% | 0.0245613 | -2.343613971 |
| 17 | 71339 | Beans dried, shelled, nes | 0 | 0 | 0 | 0 | 0.00% | 0.0245212 | -2.374047692 |
| 18 | 151190 | Palm oil or fractions simply refined | 0 | 0 | 0 | 0 | 0.00% | 0.0243161 | -2.087457004 |
| 19 | 30612 | Lobsters (Homarus) frozen | 0 | 0 | 0 | 0 | 0.00% | 0.0242244 | -2.507652504 |
| 20 | 240120 | Tobacco, unmanufactured, stemmed or stripped | 0 | 0 | 0 | 0 | 0.00% | 0.0239559 | -2.276013751 |
| 21 | 30559 | Dried fish, other than cod, not smoked | 0 | 0 | 0 | 0 | 0.00% | 0.0239158 | -2.497884136 |
| 22 | 90411 | Pepper of the genus Piper, whole | 0 | 0 | 0 | 0 | 0.00% | 0.0238299 | -2.135676159 |
| 23 | 30613 | Shrimps and prawns, frozen | 0 | 0 | 0 | 0 | 0.00% | 0.0238165 | -2.934308359 |
| 24 | 121190 | Plants & parts, pharmacy, perfume, insecticide use ne | 0 | 0 | 0 | 0 | 0.00% | 0.0237921 | -2.209191886 |
| 25 | 80430 | Pineapples, fresh or dried | 0 | 0 | 0 | 0 | 0.00% | 0.023791 | -2.25231281 |
| 26 | 71390 | Leguminous vegetables dried, shelled | 0 | 0 | 0 | 0 | 0.00% | 0.0237862 | -2.597825049 |
| 27 | 70820 | Beans, shelled or unshelled, fresh or chilled | 0 | 0 | 0 | 0 | 0.00% | 0.0237429 | -1.999853033 |
| 28 | 121299 | Vegetable products nes for human consumption | 0 | 0 | 0 | 0 | 0.00% | 0.023661 | -2.395491401 |
| 29 | 10120 | Asses, mules and hinnies, live | 0 | 0 | 0 | 0 | 0.00% | 0.0236262 | -2.268813957 |
| 30 | 30232 | Tuna(yellowfin) fresh or chilled, whole | 0 | 0 | 0 | 0 | 0.00% | 0.0236166 | -2.081891758 |

Table 11 Rwanda's top 30 export opportunities (products currently with an RCA<0.1, sorted by density). No sophistication filter was ap plied (Exports in thousands of US dollars, sophistication in standard deviations of world's average. See text for explanation for the variables density and sophistication)

| RANK | Product Code (HS-6) | Product Name | RCA 2000 | RCA 2005 | Exports 2000 | Exports 2005 | Export Share 2005 | Density | Sophistication |
|---|---|---|---|---|---|---|---|---|---|
| 1 | 470200 | Chemical wood pulp, dissolving grades | 0 | 0 | 0 | 0 | 0.00% | 0.020241 | 0.987360637 |
| 2 | 470100 | Mechanical wood pulp | 0 | 0 | 0 | 0 | 0.00% | 0.0198766 | 0.175729892 |
| 3 | 710900 | Base metals or silver, clad with gold, semi-manuf | 0 | 0 | 0 | 0 | 0.00% | 0.0190258 | 0.200754038 |
| 4 | 440392 | Logs, Beech (Fagus spp) | 0 | 0 | 0 | 0 | 0.00% | 0.0189117 | 0.226284116 |
| 5 | 340211 | Anionic surface-active agents | 0 | 0 | 0 | 0 | 0.00% | 0.0186213 | 0.319531168 |
| 6 | 330690 | Oral & dental hygiene preparations, except dentifrice | 0 | 0 | 0 | 0 | 0.00% | 0.0186189 | 0.016205825 |
| 7 | 230620 | Linseed oil-cake and other solid residues | 0 | 0 | 0 | 0 | 0.00% | 0.0184561 | 0.271428765 |
| 8 | 721050 | Flat rolled iron or non-alloy steel, coated chromium/oxides, w> 600 mm | 0 | 0 | 0 | 0 | 0.00% | 0.0184319 | 0.532333705 |
| 9 | 310420 | Potassium chloride, in packs >10 kg | 0 | 0 | 0 | 0 | 0.00% | 0.0182493 | 0.116769423 |
| 10 | 710420 | Synthetic precious, semi-precious stone, rough shaped | 0 | 0 | 0 | 0 | 0.00% | 0.0182383 | 0.562767426 |
| 11 | 482190 | Paper labels of all kinds, not printed | 0 | 0 | 0 | 0 | 0.00% | 0.0181863 | 0.118326135 |
| 12 | 701110 | Glass envelopes (bulbs & tubes) for electric lighting | 0 | 0 | 0 | 0 | 0.00% | 0.0181519 | 0.560549111 |
| 13 | 330790 | Perfumery, cosmetic or toilet preparations, nes | 0 | 0 | 0 | 0 | 0.00% | 0.0180799 | 0.068628103 |
| 14 | 210420 | Homogenised composite food preparations | 0 | 0 | 0 | 0 | 0.00% | 0.0179866 | 0.113539246 |
| 15 | 252330 | Aluminous cement | 0 | 0 | 0 | 0 | 0.00% | 0.017984 | 0.137862872 |
| 16 | 850490 | Parts of electrical transformers and inductors | 0 | 0 | 0 | 0 | 0.00% | 0.0179227 | 0.289564461 |
| 17 | 851840 | Audio-frequency electric amplifiers | 0 | 0 | 0 | 0 | 0.00% | 0.0178029 | 0.044148806 |
| 18 | 711011 | Platinum unwrought or in powder form | 0 | 0 | 0 | 0 | 0.00% | 0.0177916 | 1.023748781 |
| 19 | 960810 | Ball point pens | 0 | 0 | 0 | 0 | 0.00% | 0.0177625 | 0.011847031 |
| 20 | 720441 | Waste from the mechanical working of iron or steel ne | 0 | 0 | 0 | 0 | 0.00% | 0.0177442 | 0.343348862 |
| 21 | 440391 | Logs, Oak (Quercus spp) | 0 | 0 | 0 | 0 | 0.00% | 0.0177152 | 0.108830192 |
| 22 | 391731 | Plastic tube, pipe or hose, flexible, mbp > 27.6 MPa | 0 | 0 | 0 | 0 | 0.00% | 0.0177065 | 0.126965887 |
| 23 | 940390 | Furniture parts nes | 0 | 0 | 0 | 0 | 0.00% | 0.0176953 | 0.201454559 |
| 24 | 721070 | Flat rolled iron or non-alloy steel, painted/plastic coated,width>600mm | 0 | 0 | 0 | 0 | 0.00% | 0.0176806 | 0.156971512 |
| 25 | 851910 | Coin or disc-operated record-players | 0 | 0 | 0 | 0 | 0.00% | 0.0176795 | 0.107818329 |
| 26 | 400211 | Styrene-butadiene rubber (SBR/XSBR) latex | 0 | 0 | 0 | 0 | 0.00% | 0.0175891 | 0.920538772 |
| 27 | 731812 | Screws, wood, iron or steel, except coach screws | 0 | 0 | 0 | 0 | 0.00% | 0.0175882 | 0.524316638 |
| 28 | 680410 | Stones for milling, grinding or pulping | 0 | 0 | 0 | 0 | 0.00% | 0.0175779 | 0.504040464 |
| 29 | 441019 | | 0 | 0 | 0 | 0 | 0.00% | 0.0175701 | 0.335798809 |
| 30 | 253020 | Kieserite, epsomite (natural magnesium sulphates) | 0 | 0 | 0 | 0 | 0.00% | 0.0175147 | 3.04599557 |

Table 12 Rwanda's top 30 export opportunities: products currently with an RCA<0.1, sorted by density and sophistication larger than the World's average (Exports in thousands of US dollars, sophistication in standard deviations of world's average. See text for explanation for the variables density and sophistication)

| RANK | Product Code (HS-6) | Product Name | RCA 2000 | RCA 2005 | Exports 2000 | Exports 2005 | Export Share 2005 | Density | Sophistication |
|---|---|---|---|---|---|---|---|---|---|
| 1 | 710812 | Gold in unwrought forms non-monetary | 19.4725986 | 63.5997611 | 74146.556 | 465258.4137 | 22.31% | 0.0708663 | -2.07336876 |
| 2 | 30410 | Fish fillet or meat, fresh or chilled, not liver, roe | 436.7338 | 239.780402 | 96804.08541 | 136719.9242 | 6.56% | 0.1230575 | -2.06344472 |
| 3 | 710813 | Gold, semi-manufactured forms, non-monetary | 49.4439679 | 35.1570794 | 39502.466 | 95192.69397 | 4.57% | 0.1081749 | -1.310540937 |
| 4 | 260300 | Copper ores and concentrates | 2.41541497 | 27.7521909 | 2185.276817 | 94275.94288 | 4.52% | 0.1189498 | -2.29017983 |
| 5 | 261690 | Precious metal ores and concentrates except silver | 0 | 355.003979 | 0 | 90838.97854 | 4.36% | 0.1083677 | -1.519140352 |
| 6 | 90111 | Coffee, not roasted, not decaffeinated | 73.3934549 | 46.7711456 | 86999.36503 | 89352.34938 | 4.29% | 0.1619909 | -2.9031352 |
| 7 | 240120 | Tobacco, unmanufactured, stemmed or stripped | 60.9885918 | 76.439935 | 41412.92076 | 84280.08254 | 4.04% | 0.1345305 | -2.276013751 |
| 8 | 520100 | Cotton, not carded or combed | 37.4650261 | 39.7951063 | 37874.56166 | 78252.20928 | 3.75% | 0.0885193 | -2.429388805 |
| 9 | 30420 | Fish fillets, frozen | 67.189586 | 42.1303116 | 44735.7721 | 67518.47816 | 3.24% | 0.125353 | -1.775608662 |
| 10 | 271000 | Oils petroleum, bituminous, distillates, except crude | 0.38001149 | 0.85219286 | 8987.884238 | 64032.609 | 3.07% | 0.1163002 | -1.243368812 |
| 11 | 80131 |  | 1234.13722 | 589.459961 | 47274.7517 | 56067.3194 | 2.69% | 0.1548755 | -3.802525587 |
| 12 | 710310 | Precious, semi-precious stones unworked, partly worke | 464.711618 | 654.792423 | 18508.37576 | 47284.97863 | 2.27% | 0.1091191 | -2.370389419 |
| 13 | 520300 | Cotton, carded or combed | 166.599705 | 855.640201 | 5632.821897 | 40454.75888 | 1.94% | 0.1089218 | -1.893335011 |
| 14 | 740311 | Copper cathodes and sections of cathodes unwrought | 10.1406842 | 6.55414422 | 17007.88818 | 33374.12896 | 1.60% | 0.090325 | -1.754826556 |
| 15 | 90240 | Tea, black (fermented or partly) in packages > 3 kg | 122.693093 | 89.5326783 | 27378.08282 | 32653.45045 | 1.57% | 0.1546029 | -2.479865194 |
| 16 | 120740 | Sesamum seeds | 126.668794 | 190.604209 | 10091.23366 | 32358.89215 | 1.55% | 0.1477475 | -2.81658201 |
| 17 | 710210 | Diamonds, unsorted | 1.20052695 | 41.5654244 | 747.6579962 | 20696.18501 | 0.99% | 0.1010593 | -2.348634368 |
| 18 | 240130 | Tobacco refuse | 75.185921 | 387.326748 | 1907.303394 | 19767.5611 | 0.95% | 0.1230729 | -1.995805582 |
| 19 | 401511 | Rubber surgical gloves | 0 | 99.9492095 | 0 | 19722.65043 | 0.95% | 0.0854849 | -1.590009668 |
| 20 | 60310 | Cut flowers and flower buds for bouquets, etc., fresh | 21.7065138 | 14.3143916 | 11886.20437 | 16932.15317 | 0.81% | 0.136168 | -1.859826784 |
| 21 | 90700 | Cloves (whole fruit, cloves and stems) | 527.587958 | 626.175471 | 10231.19119 | 16353.1983 | 0.78% | 0.1228792 | -2.036941698 |
| 22 | 30613 | Shrimps and prawns, frozen | 8.16974711 | 7.52497473 | 11097.55895 | 14470.30306 | 0.69% | 0.1387776 | -2.934308359 |
| 23 | 170111 | Raw sugar, cane | 13.4888948 | 10.3477091 | 8153.495884 | 13551.80301 | 0.65% | 0.1112081 | -2.82012353 |
| 24 | 180100 | Cocoa beans, whole or broken, raw or roasted | 6.82621721 | 14.3511386 | 2365.566665 | 13201.01638 | 0.63% | 0.1040376 | -3.842727675 |
| 25 | 440399 | Logs, non-coniferous nes | 4.97421227 | 19.0617854 | 1597.191184 | 12149.71153 | 0.58% | 0.0918535 | -2.445578611 |
| 26 | 71320 | Chickpeas, dried, shelled | 33.3975352 | 112.256483 | 1501.997929 | 10700.27543 | 0.51% | 0.1311214 | -2.25974611 |
| 27 | 701092 |  | 28.7634593 | 8.7520771 | 5492.255454 | 9906.863609 | 0.48% | 0.076541 | -0.76339557 |
| 28 | 30759 | Octopus, frozen, dried, salted or in brine | 19.8046244 | 45.7805119 | 1946.671745 | 9419.017134 | 0.45% | 0.1173934 | -2.299286596 |
| 29 | 530410 | Sisal and Agave, raw | 1296.09488 | 797.109775 | 6391.747565 | 9305.071343 | 0.45% | 0.1396587 | -2.620436292 |
| 30 | 100590 | Maize except seed corn | 0.74053493 | 4.13470612 | 859.5143378 | 9045.028895 | 0.43% | 0.1052306 | -1.415541165 |

Table 13 Tanzania's top 30 exports sorted in decreasing total exports volume (Exports in thousands of US dollars, sophistication in standard deviations of world's average. See text for explanation for the variables density and sophistication)

| RANK | Product Code (HS-6) | ProductName | RCA 2000 | RCA 2005 | Exports 2000 | Exports 2005 | Export Share 2005 | Density | Sophistication |
|---|---|---|---|---|---|---|---|---|---|
| 1 | 430190 | Raw furskin pieces (e.g. heads, tails, paws) | 962.863405 | 1301.69673 | 420.751 | 1130.724 | 0.05% | 0.0754544 | -0.585268793 |
| 2 | 520300 | Cotton, carded or combed | 166.599705 | 855.640201 | 5632.821897 | 40454.75888 | 1.94% | 0.1089218 | -1.893335011 |
| 3 | 530410 | Sisal and Agave, raw | 1296.09488 | 797.109775 | 6391.747565 | 9305.071343 | 0.45% | 0.1396587 | -2.620436292 |
| 4 | 710310 | Precious, semi-precious stones unworked, partly worke | 464.711618 | 654.792423 | 18508.37576 | 47284.97863 | 2.27% | 0.1091191 | -2.370389419 |
| 5 | 90700 | Cloves (whole fruit, cloves and stems) | 527.587958 | 626.175471 | 10231.19119 | 16353.1983 | 0.78% | 0.1228792 | -2.036941698 |
| 6 | 80131 | | 1234.13722 | 589.459961 | 47274.7517 | 56067.3194 | 2.69% | 0.1548755 | -3.802525587 |
| 7 | 551634 | Woven fabric <85% artificial staple+wool/hair, printe | 0 | 568.639855 | 0 | 905.844 | 0.04% | 0.0722811 | -1.07224724 |
| 8 | 530490 | Sisal and Agave, processed but not spun, tow & waste | 705.185644 | 414.539635 | 1022.46142 | 1575.901974 | 0.08% | 0.1008714 | -1.809039054 |
| 9 | 120760 | Safflower seeds | 0.01663523 | 402.744977 | 0.059 | 1211.793775 | 0.06% | 0.0917529 | -0.374684571 |
| 10 | 240130 | Tobacco refuse | 75.185921 | 387.326748 | 1907.303394 | 19767.5611 | 0.95% | 0.1230729 | -1.995805582 |
| 11 | 261690 | Precious metal ores and concentrates except silver | 0 | 355.003979 | 0 | 90838.97854 | 4.36% | 0.1083677 | -1.519140352 |
| 12 | 560729 | Twine nes, cordage, ropes and cables, of sisal | 192.967422 | 297.709371 | 1082.943701 | 2419.428122 | 0.12% | 0.1021022 | -1.799115014 |
| 13 | 230610 | Cotton seed oil-cake and other solid residues | 45.1061178 | 255.319887 | 448.7048269 | 4078.127697 | 0.20% | 0.1157456 | -2.431334695 |
| 14 | 30410 | Fish fillet or meat, fresh or chilled, not liver, roe | 436.7338 | 239.780402 | 96804.08541 | 136719.9242 | 6.56% | 0.1230575 | -2.06344472 |
| 15 | 210130 | Chicory & other coffee substitutes, roasted & product | 93.077605 | 237.811046 | 696.616 | 3703.794 | 0.18% | 0.0848482 | -0.928990814 |
| 16 | 120740 | Sesamum seeds | 126.668794 | 190.604209 | 10091.23366 | 32358.89215 | 1.55% | 0.1477475 | -2.81658201 |
| 17 | 50710 | Ivory, unworked or simply prepared, powder and waste | 392.327647 | 174.737563 | 292.1387184 | 261.2474137 | 0.01% | 0.1161991 | -1.616940787 |
| 18 | 560721 | Binder or baler twine, of sisal or agave | 194.540671 | 172.118941 | 1852.944601 | 2357.286001 | 0.11% | 0.0821535 | -1.888587039 |
| 19 | 152190 | Beeswax, other insect waxes and spermaceti | 323.515215 | 167.443056 | 2228.906031 | 2196.159027 | 0.11% | 0.0890984 | -1.651888972 |
| 20 | 320120 | Wattle tanning extract | 202.931424 | 162.394186 | 1547.139411 | 2410.153974 | 0.12% | 0.11337 | -2.100650138 |
| 21 | 410310 | Goat or kid hides and skins, raw, nes | 54.1231436 | 147.445387 | 271.5314014 | 1732.190921 | 0.08% | 0.1160154 | -2.174516125 |
| 22 | 630491 | Textile furnishing articles nes, knit or crochet | 82.7063274 | 142.934734 | 1276.556 | 8570.336381 | 0.41% | 0.0920472 | -1.626903744 |
| 23 | 120730 | Castor oil seeds | 107.470203 | 126.501108 | 253.3946526 | 191.7416615 | 0.01% | 0.1230544 | -2.457526376 |
| 24 | 110210 | Rye flour | 0.05239224 | 112.44698 | 0.08351543 | 590.5217244 | 0.03% | 0.0806537 | 0.277032929 |
| 25 | 71320 | Chickpeas, dried, shelled | 33.3975352 | 112.256483 | 1501.997929 | 10700.27543 | 0.51% | 0.1311214 | -2.25974611 |
| 26 | 60210 | Cuttings and slips, not rooted | 77.3215339 | 108.092879 | 2411.614859 | 7970.199206 | 0.38% | 0.1167116 | -1.616512691 |
| 27 | 71390 | Leguminous vegetables dried, shelled | 5.08955107 | 105.931788 | 72.95370737 | 6906.243021 | 0.33% | 0.1438175 | -2.597825049 |
| 28 | 401511 | Rubber surgical gloves | 0 | 99.9492095 | 0 | 19722.65043 | 0.95% | 0.0854849 | -1.590009668 |
| 29 | 50900 | Sponges, natural, of animal origin | 0 | 98.5449687 | 0 | 342.117 | 0.02% | 0.0872049 | -1.299215857 |
| 30 | 71339 | Beans dried, shelled, nes | 2.66131884 | 91.5519283 | 72.07315554 | 6515.029259 | 0.31% | 0.1419105 | -2.374047692 |

Table 14 Tanzania's top 30 exports sorted in decreasing order of RCA (Exports in thousands of US dollars, sophistication in standard deviations of world's average. See text for explanation for the variables density and sophistication)

| RANK | Product Code (HS-6) | Product Name | RCA 2000 | RCA 2005 | Exports 2000 | Exports 2005 | Export Share 2005 | Density | Sophistication |
|---|---|---|---|---|---|---|---|---|---|
| 1 | 120792 | Shea nuts (karite nuts) | 0 | 0 | 0 | 0 | 0.00% | 0.0339892 | -0.997291555 |
| 2 | 140210 | Kapok | 0 | 0 | 0 | 0 | 0.00% | 0.0274315 | -1.73209856 |
| 3 | 180200 | Cocoa shells, husks, skins and waste | 0 | 0 | 0 | 0 | 0.00% | 0.0271777 | -3.338703232 |
| 4 | 71490 | Arrowroot, salep, etc fresh or dried and sago pith | 0 | 0 | 0 | 0 | 0.00% | 0.0268164 | -2.563694138 |
| 5 | 120740 | Sesamum seeds | 0 | 0 | 0 | 0 | 0.00% | 0.0265097 | -2.81658201 |
| 6 | 121292 | Sugar cane | 0 | 0 | 0 | 0 | 0.00% | 0.0264891 | -1.252553414 |
| 7 | 80131 |  | 0 | 0 | 0 | 0 | 0.00% | 0.02573 | -3.802525587 |
| 8 | 80450 | Guavas, mangoes and mangosteens, fresh or dried | 0.25050269 | 0 | 1.113826714 | 0 | 0.00% | 0.0254722 | -2.467333662 |
| 9 | 260500 | Cobalt ores and concentrates | 0 | 0 | 0 | 0 | 0.00% | 0.0252754 | -3.619028154 |
| 10 | 80119 |  | 0 | 0 | 0 | 0 | 0.00% | 0.0250781 | -1.982807036 |
| 11 | 91030 | Turmeric (curcuma) | 0 | 0 | 0 | 0 | 0.00% | 0.0250484 | -3.048765612 |
| 12 | 90220 | Tea, green (unfermented) in packages > 3 kg | 0 | 0 | 0 | 0 | 0.00% | 0.0250292 | -2.780154948 |
| 13 | 91010 | Ginger | 0 | 0 | 0 | 0 | 0.00% | 0.0249543 | -2.802026752 |
| 14 | 30110 | Ornamental fish, live | 0 | 0 | 0 | 0 | 0.00% | 0.0248312 | -2.539720772 |
| 15 | 30611 | Rock lobster and other sea crawfish, frozen | 0 | 0 | 0 | 0 | 0.00% | 0.0247892 | -2.344820423 |
| 16 | 71331 | Urd,mung,black or green gram beans dried shelled | 0 | 0 | 0 | 0 | 0.00% | 0.0245613 | -2.343613971 |
| 17 | 71339 | Beans dried, shelled, nes | 0 | 0 | 0 | 0 | 0.00% | 0.0245212 | -2.374047692 |
| 18 | 151190 | Palm oil or fractions simply refined | 0 | 0 | 0 | 0 | 0.00% | 0.0243161 | -2.087457004 |
| 19 | 30612 | Lobsters (Homarus) frozen | 0 | 0 | 0 | 0 | 0.00% | 0.0242244 | -2.507652504 |
| 20 | 240120 | Tobacco, unmanufactured, stemmed or stripped | 0 | 0 | 0 | 0 | 0.00% | 0.0239559 | -2.276013751 |
| 21 | 30559 | Dried fish, other than cod, not smoked | 0 | 0 | 0 | 0 | 0.00% | 0.0239158 | -2.497884136 |
| 22 | 90411 | Pepper of the genus Piper, whole | 0 | 0 | 0 | 0 | 0.00% | 0.0238299 | -2.135676159 |
| 23 | 30613 | Shrimps and prawns, frozen | 0 | 0 | 0 | 0 | 0.00% | 0.0238165 | -2.934308359 |
| 24 | 121190 | Plants & parts, pharmacy, perfume, insecticide use ne | 0 | 0 | 0 | 0 | 0.00% | 0.0237921 | -2.209191886 |
| 25 | 80430 | Pineapples, fresh or dried | 0 | 0 | 0 | 0 | 0.00% | 0.023791 | -2.25231281 |
| 26 | 71390 | Leguminous vegetables dried, shelled | 0 | 0 | 0 | 0 | 0.00% | 0.0237862 | -2.597825049 |
| 27 | 70820 | Beans, shelled or unshelled, fresh or chilled | 0 | 0 | 0 | 0 | 0.00% | 0.0237429 | -1.999853033 |
| 28 | 121299 | Vegetable products nes for human consumption | 0 | 0 | 0 | 0 | 0.00% | 0.023661 | -2.395491401 |
| 29 | 10120 | Asses, mules and hinnies, live | 0 | 0 | 0 | 0 | 0.00% | 0.0236262 | -2.268813957 |
| 30 | 30232 | Tuna(yellowfin) fresh or chilled, whole | 0 | 0 | 0 | 0 | 0.00% | 0.0236166 | -2.081891758 |

Table 15 Tanzania's top 30 export opportunities (products currently with an RCA<0.1, sorted by density). No sophistication filter was applied (Exports in thousands of US dollars, sophistication in standard deviations of world's average. See text for explanation for the variables density and sophistication)

| RANK | Product Code (HS-6) | ProductName | RCA 2000 | RCA 2005 | Exports 2000 | Exports 2005 | Export Share 2005 | Density | Sophistication |
|---|---|---|---|---|---|---|---|---|---|
| 1 | 110210 | Rye flour | 0.05239224 | 112.44698 | 0.08351543 | 590.5217244 | 0.03% | 0.0806537 | 0.277032929 |
| 2 | 320210 | Synthetic organic tanning substances | 0.86776865 | 22.3871588 | 30.578 | 1886.482 | 0.09% | 0.0890657 | 0.353817751 |
| 3 | 741991 | Articles of copper, cast/moulded/stamped, nfw | 0 | 10.5712155 | 0 | 726.667 | 0.03% | 0.0672939 | 0.253254152 |
| 4 | 711049 | Iridium, osmium and ruthenium, semi-manufactured | 0 | 8.66757642 | 0 | 51.065 | 0.00% | 0.0596304 | 1.035735464 |
| 5 | 780420 | Lead powders and flakes | 0 | 7.78180916 | 0 | 24.023 | 0.00% | 0.06884 | 0.219006487 |
| 6 | 844140 | Machines for moulding articles in pulp, paper, board | 0 | 7.50395876 | 0 | 242.418 | 0.01% | 0.0569679 | 1.005068237 |
| 7 | 721810 | Ingots and other primary forms, stainless steel | 0 | 5.76438973 | 0 | 300.225 | 0.01% | 0.0667813 | 0.430524737 |
| 8 | 591110 | Textile fabric for card clothing, technical use | 0 | 5.68087946 | 0 | 340.306 | 0.02% | 0.0648656 | 0.50906086 |
| 9 | 490600 | Plans and drawings for architectural etc use | 0 | 4.80893872 | 0 | 209.804 | 0.01% | 0.0715138 | 0.539689169 |
| 10 | 900669 | Photographic flashlight apparatus, nes | 0.00393436 | 4.01839769 | 0.04 | 42.657 | 0.00% | 0.0584033 | 0.426438368 |
| 11 | 722230 | Stainless steel bar or rod nes | 0 | 3.4952899 | 0 | 420.609 | 0.02% | 0.0639747 | 0.045394176 |
| 12 | 900640 | Instant print cameras | 0 | 3.48991315 | 0 | 69.984 | 0.00% | 0.0524681 | 0.770043635 |
| 13 | 330690 | Oral & dental hygiene preparations, except dentifrice | 0.29839086 | 3.38757775 | 14.192 | 455.448 | 0.02% | 0.0867139 | 0.016205825 |
| 14 | 480910 | Paper, carbon or similar width >36 cm | 0.00980646 | 3.31843476 | 0.040971163 | 35.595 | 0.00% | 0.067856 | 0.083416868 |
| 15 | 480439 | Paper, kraft, <150g/m2, uncoated, nes | 0 | 3.30169647 | 0 | 639.005 | 0.03% | 0.0751522 | 0.791176001 |
| 16 | 731412 | | 0 | 3.1253463 | 0 | 33.506 | 0.00% | 0.0779752 | 0.504585313 |
| 17 | 470693 | Semi-chemical pulps of other fibrous material | 14.9715269 | 2.98429841 | 94.5895799 | 46.16618875 | 0.00% | 0.0729809 | 0.119221245 |
| 18 | 130213 | Hop extract | 0 | 2.70528821 | 0 | 86.377 | 0.00% | 0.0609903 | 0.241228552 |
| 19 | 844520 | Textile yarn spinning machines | 0 | 2.61533373 | 0 | 417.619 | 0.02% | 0.0515779 | 0.756733747 |
| 20 | 722219 | | 0.0064112 | 2.43944397 | 0.198 | 178.81 | 0.01% | 0.0624646 | 0.784871317 |
| 21 | 720925 | | 0 | 2.27299291 | 0 | 64.12709566 | 0.00% | 0.079965 | 0.438580722 |
| 22 | 481710 | Envelopes of paper | 0.2340319 | 2.01632988 | 18.58786244 | 340.5232857 | 0.02% | 0.0742006 | 0.388804854 |
| 23 | 720917 | | 0 | 2.00173864 | 0 | 2964.375104 | 0.14% | 0.0785711 | 0.076217074 |
| 24 | 430180 | Raw furskins of other animals, whole | 15.6553079 | 1.93144846 | 154.634 | 65.8455557 | 0.00% | 0.0753825 | 0.022354838 |
| 25 | 846299 | Metal shaping presses, nes | 0 | 1.86268671 | 0 | 525.306 | 0.03% | 0.0641725 | 0.803668615 |
| 26 | 210330 | Mustard flour or meal and prepared mustard | 3.83692276 | 1.79706555 | 61.78465001 | 64.128 | 0.00% | 0.086364 | 0.260103685 |
| 27 | 283220 | Sulphites of metals other than sodium | 0 | 1.75439597 | 0 | 16.757 | 0.00% | 0.0732321 | 0.32622503 |
| 28 | 470620 | | 0 | 1.5309694 | 0 | 24.29 | 0.00% | 0.0738672 | 0.014065346 |
| 29 | 320490 | Synthetic organic products used as luminophores | 0.00188432 | 1.34930638 | 0.066 | 78.703 | 0.00% | 0.0630434 | 0.43807479 |
| 30 | 230330 | Brewing or distilling dregs and waste | 0 | 1.21061754 | 0 | 50.29 | 0.00% | 0.0782295 | 0.399662921 |

Table 16 Tanzania's top 30 export opportunities: products currently with an RCA<0.1, sorted by density and sophistication larger than the World's average (Exports in thousands of US dollars, sophistication in standard deviations of world's average. See text for explanation for the variables density and sophisticati on)

| RANK | Product Code (HS-6) | ProductName | RCA 2000 | RCA 2005 | Exports 2000 | Exports 2005 | Export Share 2005 | Density | Sophistication |
|---|---|---|---|---|---|---|---|---|---|
| 1 | 740311 | Copper cathodes and sections of cathodes unwrought | 146.87 | 179.74 | 218274.66 | 1160623.22 | 43.90% | 0.0449 | -1.7548 |
| 2 | 740919 | Plate, sheet, strip, refined copper, flat, t > 0.15mm | 623.48 | 754.25 | 49565.74 | 188332.05 | 7.12% | 0.0339 | -0.1000 |
| 3 | 740911 | Plate, sheet, strip, refined copper, coil, t > 0.15mm | 0.42 | 284.05 | 59.47 | 146676.20 | 5.55% | 0.0254 | 0.4970 |
| 4 | 810590 | Cobalt, articles thereof, nes | 1665.33 | 1018.91 | 44084.37 | 146580.67 | 5.54% | 0.0316 | -0.9089 |
| 5 | 520100 | Cotton, not carded or combed | 16.82 | 45.36 | 15064.99 | 113100.81 | 4.28% | 0.0427 | -2.4294 |
| 6 | 260300 | Copper ores and concentrates | 21.33 | 24.61 | 17099.50 | 105992.60 | 4.01% | 0.0563 | -2.2902 |
| 7 | 810510 | Cobalt, unwrought, matte, waste or scrap, powders | 494.40 | 230.00 | 86488.35 | 101319.13 | 3.83% | 0.0410 | -1.3302 |
| 8 | 240110 | Tobacco, unmanufactured, not stemmed or stripped | 41.76 | 134.43 | 9192.46 | 68413.66 | 2.59% | 0.0485 | -2.4344 |
| 9 | 740811 | Wire of refined copper > 6mm wide | 24.73 | 26.77 | 11622.02 | 61706.57 | 2.33% | 0.0350 | -0.2326 |
| 10 | 170111 | Raw sugar, cane | 38.29 | 26.76 | 20508.14 | 44446.28 | 1.68% | 0.0490 | -2.8201 |
| 11 | 261800 | Granulated slag (slag sand) from iron, steel industry | 0.00 | 409.21 | 0.00 | 36992.24 | 1.40% | 0.0430 | 0.3723 |
| 12 | 240120 | Tobacco, unmanufactured, stemmed or stripped | 18.71 | 22.19 | 11255.32 | 31030.37 | 1.17% | 0.0638 | -2.2760 |
| 13 | 710310 | Precious, semi-precious stones unworked, partly worke | 408.77 | 272.34 | 14425.79 | 24939.10 | 0.94% | 0.0524 | -2.3704 |
| 14 | 420229 | Handbags, of vulcanised fibre or paperboard | 0.00 | 268.61 | 0.09 | 22599.98 | 0.85% | 0.0367 | -1.0807 |
| 15 | 60310 | Cut flowers and flower buds for bouquets, etc., fresh | 35.16 | 11.26 | 17059.02 | 16895.66 | 0.64% | 0.0623 | -1.8598 |
| 16 | 740200 | Unrefined copper, copper anodes, electrolytic refinin | 70.12 | 18.14 | 13645.10 | 16849.42 | 0.64% | 0.0507 | -1.1023 |
| 17 | 71090 | Frozen vegetable mixtures, uncooked, boiled or steame | 10.47 | 124.39 | 430.46 | 15878.16 | 0.60% | 0.0426 | -0.7065 |
| 18 | 490700 | Documents of title (bonds etc), unused stamps etc | 11.48 | 39.77 | 1576.59 | 15686.61 | 0.59% | 0.0369 | -1.1621 |
| 19 | 854459 | Electric conductors, 80-1,000 volts, no connectors | 5.20 | 6.96 | 3317.57 | 15652.76 | 0.59% | 0.0347 | -0.1284 |
| 20 | 252329 | Portland cement, other than white cement | 16.47 | 11.03 | 6556.89 | 14680.22 | 0.56% | 0.0482 | -1.3493 |
| 21 | 440710 | Lumber, coniferous (softwood) thickness < 6 mm | 0.05 | 2.50 | 107.61 | 14247.60 | 0.54% | 0.0441 | -0.3679 |
| 22 | 90111 | Coffee, not roasted, not decaffeinated | 8.70 | 5.73 | 9134.40 | 13884.54 | 0.53% | 0.0742 | -2.9031 |
| 23 | 210690 | Food preparations nes | 0.10 | 3.04 | 108.71 | 13116.36 | 0.50% | 0.0393 | -0.4971 |
| 24 | 750210 | Nickel unwrought, not alloyed | 0.93 | 4.63 | 811.15 | 12475.47 | 0.47% | 0.0480 | -0.2976 |
| 25 | 100510 | Maize (corn) seed | 24.16 | 21.95 | 3143.33 | 9733.24 | 0.37% | 0.0466 | -0.6923 |
| 26 | 740710 | Bars, rods & profiles of refined copper | 20.06 | 21.93 | 1930.38 | 8816.52 | 0.33% | 0.0339 | -0.1562 |
| 27 | 520532 | Cotton yarn >85% multiple uncomb 714-232 dtex,not ret | 147.85 | 79.16 | 10293.31 | 8813.72 | 0.33% | 0.0459 | -1.8199 |
| 28 | 170199 | Refined sugar, in solid form, nes, pure sucrose | 0.59 | 3.89 | 322.43 | 8539.35 | 0.32% | 0.0488 | -1.0253 |
| 29 | 740819 | Wire of refined copper < 6mm wide | 14.24 | 19.32 | 2003.39 | 8165.84 | 0.31% | 0.0375 | -0.0389 |
| 30 | 520512 | Cotton yarn >85% single uncombed 714-232 dtex,not ret | 64.70 | 16.55 | 11222.85 | 7680.89 | 0.29% | 0.0471 | -1.7607 |

Table 17 Zambia's top 30 exports sorted in decreasing total exports volume (Exports in thousands of US dollars, sophistication in standard deviations of world's average. See text for explanation for the variables density and sophistication)

| RANK | Product Code (HS-6) | ProductName | RCA 2000 | RCA 2005 | Exports 2000 | Exports 2005 | Export Share 2005 | Density | Sophistication |
|---|---|---|---|---|---|---|---|---|---|
| 1 | 810590 | Cobalt, articles thereof, nes | 1665.33 | 1018.91 | 44084.37 | 146580.67 | 5.54% | 0.0316 | -0.9089 |
| 2 | 740919 | Plate, sheet, strip, refined copper, flat, t > 0.15mm | 623.48 | 754.25 | 49565.74 | 188332.05 | 7.12% | 0.0339 | -0.1000 |
| 3 | 261800 | Granulated slag (slag sand) from iron, steel industry | 0.00 | 409.21 | 0.00 | 36992.24 | 1.40% | 0.0430 | 0.3723 |
| 4 | 740911 | Plate, sheet, strip, refined copper, coil, t > 0.15mm | 0.42 | 284.05 | 59.47 | 146676.20 | 5.55% | 0.0254 | 0.4970 |
| 5 | 710310 | Precious, semi-precious stones unworked, partly worke | 408.77 | 272.34 | 14425.79 | 24939.10 | 0.94% | 0.0524 | -2.3704 |
| 6 | 420229 | Handbags, of vulcanised fibre or paperboard | 0.00 | 268.61 | 0.09 | 22599.98 | 0.85% | 0.0367 | -1.0807 |
| 7 | 252230 | Hydraulic lime | 307.12 | 261.47 | 708.09 | 1552.03 | 0.06% | 0.0411 | -0.7093 |
| 8 | 810510 | Cobalt, unwrought, matte, waste or scrap, powders | 494.40 | 230.00 | 86488.35 | 101319.13 | 3.83% | 0.0410 | -1.3302 |
| 9 | 520533 | Cotton yarn >85% multiple uncomb 232-192 dtex,not ret | 297.24 | 220.47 | 3786.79 | 4511.70 | 0.17% | 0.0458 | -1.0900 |
| 10 | 60240 | Roses | 567.38 | 179.86 | 7224.76 | 7482.75 | 0.28% | 0.0503 | -1.2957 |
| 11 | 740311 | Copper cathodes and sections of cathodes unwrought | 146.87 | 179.74 | 218274.66 | 1160623.22 | 43.90% | 0.0449 | -1.7548 |
| 12 | 240110 | Tobacco, unmanufactured, not stemmed or stripped | 41.76 | 134.43 | 9192.46 | 68413.66 | 2.59% | 0.0485 | -2.4344 |
| 13 | 71090 | Frozen vegetable mixtures, uncooked, boiled or steame | 10.47 | 124.39 | 430.46 | 15878.16 | 0.60% | 0.0426 | -0.7065 |
| 14 | 420340 | Clothing accessories nes, of leather or composition | 0.00 | 122.34 | 0.01 | 3767.95 | 0.14% | 0.0366 | -1.0201 |
| 15 | 70810 | Peas, shelled or unshelled, fresh or chilled | 87.07 | 120.42 | 1540.02 | 7338.21 | 0.28% | 0.0523 | -1.7177 |
| 16 | 120720 | Cotton seeds | 47.98 | 115.77 | 1334.91 | 7467.00 | 0.28% | 0.0623 | -2.0223 |
| 17 | 410320 | Reptile skins, raw | 52.45 | 87.19 | 655.74 | 3348.78 | 0.13% | 0.0577 | -2.1215 |
| 18 | 520532 | Cotton yarn >85% multiple uncomb 714-232 dtex,not ret | 147.85 | 79.16 | 10293.31 | 8813.72 | 0.33% | 0.0459 | -1.8199 |
| 19 | 262030 | Ash or residues containing mainly copper | 8.56 | 48.18 | 119.63 | 2744.05 | 0.10% | 0.0380 | -0.9806 |
| 20 | 240130 | Tobacco refuse | 8.84 | 47.81 | 198.61 | 3094.24 | 0.12% | 0.0580 | -1.9958 |
| 21 | 520100 | Cotton, not carded or combed | 16.82 | 45.36 | 15064.99 | 113100.81 | 4.28% | 0.0427 | -2.4294 |
| 22 | 740120 | Cement copper (precipitated copper) | 10.65 | 42.38 | 27.14 | 450.09 | 0.02% | 0.0336 | -1.1460 |
| 23 | 490700 | Documents of title (bonds etc), unused stamps etc | 11.48 | 39.77 | 1576.59 | 15686.61 | 0.59% | 0.0369 | -1.1621 |
| 24 | 790310 | Zinc dust | 0.74 | 39.15 | 10.97 | 1611.28 | 0.06% | 0.0399 | -0.7469 |
| 25 | 260500 | Cobalt ores and concentrates | 1319.40 | 34.09 | 25389.66 | 3778.67 | 0.14% | 0.0656 | -3.6190 |
| 26 | 846019 | Surface grinding machines accurate to 0.01mm nes | 0.00 | 33.40 | 0.00 | 2152.46 | 0.08% | 0.0280 | 0.0814 |
| 27 | 310229 | Ammonium sulphate-nitrate mix, double salts, pack>10k | 9.03 | 28.93 | 65.30 | 1108.63 | 0.04% | 0.0445 | -0.7019 |
| 28 | 410729 | Reptile leather, other than vegetable pre-tanned | 12.79 | 27.63 | 182.67 | 1339.72 | 0.05% | 0.0372 | -1.8625 |
| 29 | 740811 | Wire of refined copper > 6mm wide | 24.73 | 26.77 | 11622.02 | 61706.57 | 2.33% | 0.0350 | -0.2326 |
| 30 | 170111 | Raw sugar, cane | 38.29 | 26.76 | 20508.14 | 44446.28 | 1.68% | 0.0490 | -2.8201 |

Table 18 Zambia's top 30 exports sorted in decreasing order of RCA (Exports in thousands of US dollars, sophistication in standard deviations of world's average. See text for explanation for the variables density and sophistication)

| RANK | Product Code (HS-6) | ProductName | RCA 2000 | RCA 2005 | Exports 2000 | Exports 2005 | Export Share 2005 | Density | Sophistication |
|---|---|---|---|---|---|---|---|---|---|
| 1 | 120740 | Sesamum seeds | 0.00 | 0.00 | 0.00 | 0.00 | 0.00% | 0.0640043 | -2.81658201 |
| 2 | 80131 | | 0.01 | 0.00 | 0.28 | 0.00 | 0.00% | 0.0609461 | -3.802525587 |
| 3 | 30611 | Rock lobster and other sea crawfish, frozen | 0.00 | 0.00 | 0.00 | 0.00 | 0.00% | 0.0589579 | -2.344820423 |
| 4 | 260600 | Aluminium ores and concentrates | 0.24 | 0.00 | 29.64 | 0.00 | 0.00% | 0.0587167 | -2.091582291 |
| 5 | 30612 | Lobsters (Homarus) frozen | 0.00 | 0.00 | 0.00 | 0.00 | 0.00% | 0.0579248 | -2.507652504 |
| 6 | 130214 | Pyrethrum, roots containing rotenone, extracts | 0.00 | 0.00 | 0.00 | 0.00 | 0.00% | 0.0579144 | -2.639856275 |
| 7 | 180200 | Cocoa shells, husks, skins and waste | 0.00 | 0.00 | 0.00 | 0.00 | 0.00% | 0.0578394 | -3.338703232 |
| 8 | 80450 | Guavas, mangoes and mangosteens, fresh or dried | 0.09 | 0.02 | 5.69 | 5.53 | 0.00% | 0.0570 | -2.4673 |
| 9 | 30613 | Shrimps and prawns, frozen | 0.00 | 0.00 | 0.00 | 1.25 | 0.00% | 0.0561089 | -2.934308359 |
| 10 | 91030 | Turmeric (curcuma) | 0.00 | 0.00 | 0.00 | 0.00 | 0.00% | 0.0555379 | -3.048765612 |
| 11 | 120792 | Shea nuts (karite nuts) | 0.00 | 0.00 | 0.00 | 0.00 | 0.00% | 0.0551498 | -0.997291555 |
| 12 | 271000 | Oils petroleum, bituminous, distillates, except crude | 0.06 | 0.07 | 1311.85 | 6809.32 | 0.26% | 0.0545 | -1.2434 |
| 13 | 10420 | Goats, live | 0.01 | 0.00 | 0.08 | 0.10 | 0.00% | 0.0545201 | -2.099132344 |
| 14 | 261610 | Silver ores and concentrates | 0.20 | 0.00 | 7.24 | 0.00 | 0.00% | 0.0543416 | -1.716492522 |
| 15 | 81090 | Fruits, fresh nes | 1.89 | 0.10 | 137.96 | 23.72 | 0.00% | 0.0543 | -1.6492 |
| 16 | 121190 | Plants & parts, pharmacy, perfume, insecticide use ne | 0.00 | 0.00 | 0.00 | 0.00 | 0.00% | 0.05416 | -2.209191886 |
| 17 | 90220 | Tea, green (unfermented) in packages > 3 kg | 0.93 | 0.02 | 26.46 | 1.22 | 0.00% | 0.0540222 | -2.780154948 |
| 18 | 71490 | Arrowroot, salep, etc fresh or dried and sago pith | 0.00 | 0.02 | 0.00 | 0.98 | 0.00% | 0.0537441 | -2.563694138 |
| 19 | 90411 | Pepper of the genus Piper, whole | 0.48 | 0.00 | 57.75 | 0.00 | 0.00% | 0.0535223 | -2.135676159 |
| 20 | 80590 | Citrus fruits, fresh or dried, nes | 0.16 | 0.01 | 0.98 | 0.06 | 0.00% | 0.0534618 | -1.673994283 |
| 21 | 151311 | Coconut (copra) oil crude | 0.00 | 0.00 | 0.00 | 0.00 | 0.00% | 0.0532363 | -3.20474816 |
| 22 | 80132 | | 0.00 | 0.00 | 0.00 | 0.00 | 0.00% | 0.052865 | -2.936215331 |
| 23 | 261000 | Chromium ores and concentrates | 0.00 | 0.00 | 0.00 | 0.00 | 0.00% | 0.0526065 | -1.881465082 |
| 24 | 70310 | Onions and shallots, fresh or chilled | 0.05 | 0.00 | 4.91 | 0.16 | 0.00% | 0.0525118 | -1.698551416 |
| 25 | 71332 | Beans, small red (Adzuki) dried, shelled | 0.00 | 0.00 | 0.00 | 0.04 | 0.00% | 0.0524677 | -2.34505393 |
| 26 | 30621 | Rock lobster and other sea crawfish not frozen | 0.00 | 0.00 | 0.00 | 0.00 | 0.00% | 0.0522595 | -2.238847251 |
| 27 | 410520 | Sheep or lamb skin leather, nes | 0.00 | 0.00 | 0.00 | 0.00 | 0.00% | 0.0521354 | -1.446402982 |
| 28 | 30379 | Fish nes, frozen, whole | 0.00 | 0.00 | 0.00 | 0.16 | 0.00% | 0.0520813 | -2.110768767 |
| 29 | 71310 | Peas dried, shelled | 0.00 | 0.00 | 0.00 | 0.01 | 0.00% | 0.0518357 | -1.891972888 |
| 30 | 261100 | Tungsten ores and concentrates | 0.00 | 0.00 | 0.00 | 0.00 | 0.00% | 0.0518145 | -2.542717443 |

Table 19 Zambia's top 30 export opportunities: products currently with an RCA<0.1, sorted by density and sophistication larger than the World's average (Exports in thousands of US dollars, sophistication in standard deviations of world's average. See text for explanation for the variables density and sophistication)

| RANK | Product Code (HS-6) | ProductName | RCA 2000 | RCA 2005 | Exports 2000 | Exports 2005 | Export Share 2005 | Density | Sophistication |
|---|---|---|---|---|---|---|---|---|---|
| 1 | 230620 | Linseed oil-cake and other solid residues | 0.00 | 0.00 | 0.00 | 0.00 | 0.00% | 0.0476736 | 0.271428765 |
| 2 | 440392 | Logs, Beech (Fagus spp) | 0.00 | 0.00 | 0.00 | 0.00 | 0.00% | 0.0440978 | 0.226284116 |
| 3 | 210420 | Homogenised composite food preparations | 0.69 | 0.07 | 12.84 | 4.94 | 0.00% | 0.0432 | 0.1135 |
| 4 | 720441 | Waste from the mechanical working of iron or steel ne | 0.04 | 0.01 | 3.06 | 3.14 | 0.00% | 0.0430 | 0.3433 |
| 5 | 310420 | Potassium chloride, in packs >10 kg | 0.00 | 0.00 | 0.00 | 0.00 | 0.00% | 0.0420678 | 0.116769423 |
| 6 | 252330 | Aluminous cement | 0.54 | 0.09 | 12.70 | 6.01 | 0.00% | 0.0418 | 0.1379 |
| 7 | 470200 | Chemical wood pulp, dissolving grades | 0.00 | 0.00 | 0.00 | 0.00 | 0.00% | 0.0413235 | 0.987360637 |
| 8 | 710900 | Base metals or silver, clad with gold, semi-manuf | 349.87 | 0.00 | 1455.64 | 0.00 | 0.00% | 0.0412288 | 0.200754038 |
| 9 | 721050 | Flat rolled iron or non-alloy steel, coated chromium/oxides, w> 600 mm | 0.00 | 0.00 | 0.00 | 0.00 | 0.00% | 0.0410204 | 0.532333705 |
| 10 | 260120 | Roasted iron pyrites | 0.00 | 0.00 | 0.00 | 0.00 | 0.00% | 0.0407831 | 0.33860089 |
| 11 | 810600 | Bismuth, articles thereof, waste or scrap | 0.00 | 0.00 | 0.00 | 0.00 | 0.00% | 0.0397322 | 0.150939253 |
| 12 | 320210 | Synthetic organic tanning substances | 0.45 | 0.00 | 14.09 | 0.00 | 0.00% | 0.0393211 | 0.353817751 |
| 13 | 731413 | | 0.00 | 0.00 | 0.00 | 0.00 | 0.00% | 0.0390303 | 0.685124993 |
| 14 | 740313 | Billets, copper, unwrought | 0.00 | 0.01 | 0.00 | 1.76 | 0.00% | 0.0387 | 0.8747 |
| 15 | 721491 | | 0.00 | 0.00 | 0.00 | 0.06 | 0.00% | 0.0387035 | 0.60429272 |
| 16 | 310240 | Ammonium nitrate limestone etc mixes, pack >10 kg | 0.00 | 0.01 | 0.00 | 3.30 | 0.00% | 0.0386 | 0.4717 |
| 17 | 360690 | Pyrophoric alloys, firelighters, etc | 0.00 | 0.00 | 0.00 | 0.00 | 0.00% | 0.0385066 | 0.321399222 |
| 18 | 720712 | Semi-finished bars, iron or non-alloy steel <0.25%C, rectangular, nes | 0.00 | 0.00 | 0.00 | 0.00 | 0.00% | 0.0385029 | 0.357865202 |
| 19 | 482190 | Paper labels of all kinds, not printed | 0.00 | 0.06 | 0.06 | 11.32 | 0.00% | 0.0384 | 0.1183 |
| 20 | 281830 | Aluminium hydroxide | 0.00 | 0.00 | 0.00 | 0.00 | 0.00% | 0.0383547 | 0.000560869 |
| 21 | 780411 | Lead foil of a thickness <2mm | 0.00 | 0.00 | 0.00 | 0.00 | 0.00% | 0.0383515 | 0.458350965 |
| 22 | 210220 | Yeasts, inactive, dead unicellular organisms nes | 0.00 | 0.00 | 0.00 | 0.02 | 0.00% | 0.0383332 | 0.439865009 |
| 23 | 760429 | Bars, rods and other profiles, aluminium alloyed | 0.00 | 0.00 | 0.00 | 3.13 | 0.00% | 0.0382 | 0.0315 |
| 24 | 210330 | Mustard flour or meal and prepared mustard | 0.00 | 0.00 | 0.00 | 0.00 | 0.00% | 0.0380779 | 0.260103685 |
| 25 | 310250 | Sodium nitrate, in packs >10 kg | 0.07 | 0.00 | 0.61 | 0.00 | 0.00% | 0.038069 | 0.134905119 |
| 26 | 281512 | Sodium hydroxide (caustic soda) in aqueous solution | 0.00 | 0.00 | 0.00 | 0.00 | 0.00% | 0.038004 | 0.057575448 |
| 27 | 310221 | Ammonium sulphate, in packs >10 kg | 0.00 | 0.00 | 0.00 | 0.00 | 0.00% | 0.0379474 | 0.147047473 |
| 28 | 150100 | Lard, other pig fat and poultry fat, rendered | 0.00 | 0.00 | 0.00 | 0.00 | 0.00% | 0.0379043 | 0.674889611 |
| 29 | 720915 | | 0.00 | 0.00 | 0.00 | 0.00 | 0.00% | 0.0378411 | 0.436946174 |
| 30 | 730719 | Pipe fittings of malleable iron or steel, cast | 0.00 | 0.01 | 0.00 | 4.96 | 0.00% | 0.0378 | 0.4623 |

Table 20 Zambia's top 30 export opportunities: products currently with an RCA<0.1, sorted by density and sophistication larger than the World's average (Exports in thousands of US dollars, sophistication in standard deviations of world's average. See text for explanation for the variables density and sophistication)

Table 21 ranking of Products not currently being significantly exported (rca<0.5) which see the largest increases in density after comparing with the productive structure of the combined region.

| Kenya | | Mozambique | | Rwanda | |
|---|---|---|---|---|---|
| | | | | | |
| HS-6 code | Product Name | HS-6 code | Product Name | HS-6 code | Product Name |
| 260500 | Cobalt ores and concentrates | 130214 | Pyrethrum, roots containing rotenone, extracts | 130214 | Pyrethrum, roots containing rotenone, extracts |
| 846911 | N/A | 260500 | Cobalt ores and concentrates | 260500 | Cobalt ores and concentrates |
| 180100 | Cocoa beans, whole or broken, raw or roasted | 320120 | Wattle tanning extract | 440349 | N/A |
| 440349 | N/A | 90111 | Coffee, not roasted, not decaffeinated | 170111 | Raw sugar, cane |
| 710210 | Diamonds, unsorted | 180100 | Cocoa beans, whole or broken, raw or roasted | 180100 | Cocoa beans, whole or broken, raw or roasted |
| 230650 | Coconut or copra oil-cake and other solid residues | 846911 | N/A | 80131 | N/A |
| 440729 | N/A | 260900 | Tin ores and concentrates | 530310 | Jute and other textile bast fibres, raw or retted |
| 90300 | Mate | 60310 | Cut flowers and flower buds for bouquets, etc., fresh | 846911 | N/A |
| 260300 | Copper ores and concentrates | 442010 | Statuettes and other ornaments of wood | 120740 | Sesamum seeds |
| 630510 | Sacks & bags, packing, of jute or other bast fibres | 410612 | Goat or kid skin leather, otherwise pre-tanned | 440724 | N/A |
| 180200 | Cocoa shells, husks, skins and waste | 71320 | Chickpeas, dried, shelled | 151311 | Coconut (copra) oil crude |
| 120720 | Cotton seeds | 90220 | Tea, green (unfermented) in packages > 3 kg | 240110 | Tobacco, unmanufactured, not stemmed or stripped |
| 91020 | Saffron | 91030 | Turmeric (curcuma) | 30611 | Rock lobster and other sea crawfish, frozen |
| 100820 | Millet | 70820 | Beans, shelled or unshelled, fresh or chilled | 71390 | Leguminous vegetables dried, shelled |
| 230610 | Cotton seed oil-cake and other solid residues | 410512 | Sheep or lamb skin leather, otherwise pre-tanned | 71339 | Beans dried, shelled, nes |
| 140420 | Cotton linters | 151110 | Palm oil, crude | 30612 | Lobsters (Homarus) frozen |
| 440399 | Logs, non-coniferous nes | 261100 | Tungsten ores and concentrates | 530410 | Sisal and Agave, raw |
| 270900 | Petroleum oils, oils from bituminous minerals, crude | 80450 | Guavas, mangoes and mangosteens, fresh or dried | 10420 | Goats, live |
| 251020 | Natural calcium phosphates, ground | 271490 | Bitumen and asphalt, asphaltites and asphaltic rocks | 520100 | Cotton, not carded or combed |
| 280522 | Strontium and barium | 160414 | Tuna, skipjack, bonito, prepared/preserved, not mince | 240120 | Tobacco, unmanufactured, stemmed or stripped |
| 710812 | Gold in unwrought forms non-monetary | 30110 | Ornamental fish, live | 71331 | Urd,mung,black or green gram beans dried shelled |
| 50790 | Whalebone, horns, etc unworked or simply prepared nes | 70990 | Vegetables, fresh or chilled nes | 440729 | N/A |
| 50710 | Ivory, unworked or simply prepared, powder and waste | 90700 | Cloves (whole fruit, cloves and stems) | 71320 | Chickpeas, dried, shelled |
| 740311 | Copper cathodes and sections of cathodes unwrought | 630510 | Sacks & bags, packing, of jute or other bast fibres | 170310 | Cane molasses |
| 441850 | Shingles and shakes, of wood | 90190 | N/A | 91030 | Turmeric (curcuma) |
| 970500 | Collections and collectors pieces | 410229 | Sheep or lamb skins, raw, except pickled, no wool | 90220 | Tea, green (unfermented) in packages > 3 kg |
| 151229 | Cotton-seed or fractions simply refined | 60210 | Cuttings and slips, not rooted | 70820 | Beans, shelled or unshelled, fresh or chilled |
| 260200 | Manganese ores, concentrates, iron ores >20% Manganes | 80430 | Pineapples, fresh or dried | 30613 | Shrimps and prawns, frozen |
| 410320 | Reptile skins, raw | 90830 | Cardamoms | 151110 | Palm oil, crude |
| 91010 | Ginger | 30410 | Fish fillet or meat, fresh or chilled, not liver, roe | 30559 | Dried fish, other than cod, not smoked |
| 750511 | Bars, rods and profiles, nickel, not alloyed | 140410 | Raw vegetable materials for dyeing or tanning | 80450 | Guavas, mangoes and mangosteens, fresh or dried |
| 170191 | Refined sugar, in solid form, flavoured or coloured | 90112 | Coffee, not roasted, decaffeinated | 271490 | Bitumen and asphalt, asphaltites and asphaltic rocks |
| 110100 | Wheat or meslin flour | 640220 | Footwear, rubber, plastic, straps fix to sole by plug | 30110 | Ornamental fish, live |
| 440799 | Lumber, non-coniferous nes | 71490 | Arrowroot, salep, etc fresh or dried and sago pith | 270900 | Petroleum oils, oils from bituminous minerals, crude |
| 120210 | Ground-nuts in shell not roasted or cooked | 81090 | Fruits, fresh nes | 30621 | Rock lobster and other sea crawfish not frozen |
| 71490 | Arrowroot, salep, etc fresh or dried and sago pith | 121190 | Plants & parts, pharmacy, perfume, insecticide use ne | 70990 | Vegetables, fresh or chilled nes |
| 80119 | N/A | 270900 | Petroleum oils, oils from bituminous minerals, crude | 160414 | Tuna, skipjack, bonito, prepared/preserved, not mince |
| 261210 | Uranium ores and concentrates | 60390 | Cut flowers and flower buds for bouquets, dried, etc. | 71332 | Beans, small red (Adzuki) dried, shelled |
| 230230 | Wheat bran, sharps, other residues | 410310 | Goat or kid hides and skins, raw, nes | 30410 | Fish fillet or meat, fresh or chilled, not liver, roe |
| 271111 | Natural gas, liquefied | 121299 | Vegetable products nes for human consumption | 230650 | Coconut or copra oil-cake and other solid residues |
| 90920 | Coriander seeds | 80720 | Papaws (papayas), fresh | 240130 | Tobacco refuse |
| 30490 | Fish meat & mince, except liver, roe & fillets, froze | 30379 | Fish nes, frozen, whole | 71490 | Arrowroot, salep, etc fresh or dried and sago pith |
| 120220 | Ground-nuts shelled, not roasted or cooked | 70890 | Legumes except peas & beans, fresh or chilled | 90700 | Cloves (whole fruit, cloves and stems) |
| 120799 | Oil seeds and oleaginous fruits, nes | 410429 | Bovine and equine leather, tanned or retanned, nes | 60210 | Cuttings and slips, not rooted |
| 30219 | Salmonidae, not trout or salmon,fresh or chilled whol | 240290 | Cigars, cheroots, cigarettes, with tobacco substitute | 90190 | N/A |
| 261690 | Precious metal ores and concentrates except silver | 610343 | Mens, boys trousers, shorts, of synthetic fibres, kni | 410229 | Sheep or lamb skins, raw, except pickled, no wool |
| 250621 | Quartzite, crude or roughly trimmed | 610462 | Womens, girls trousers & shorts, of cotton, knit | 100700 | Grain sorghum |
| 30549 | Smoked fish & fillets other than herrings or salmon | 340119 | Soaps for purposes other than toilet soap, solid | 30379 | Fish nes, frozen, whole |
| 110620 | Flour or meal of sago, starchy roots or tubers | 80119 | N/A | 10600 | Animals, live, except farm animals |
| 90411 | Pepper of the genus Piper, whole | 410790 | Leather, of animals nes | 80430 | Pineapples, fresh or dried |
| 100610 | Rice in the husk (paddy or rough) | 611020 | Pullovers, cardigans etc of cotton, knit | 151190 | Palm oil or fractions simply refined |
| 630533 | N/A | 80590 | Citrus fruits, fresh or dried, nes | 121190 | Plants & parts, pharmacy, perfume, insecticide use ne |
| 230120 | Flour or meal, pellet, fish, etc, for animal feed | 510111 | Greasy shorn wool, not carded or combed | 140410 | Raw vegetable materials for dyeing or tanning |
| 110610 | Flour or meal of dried legumes | 610610 | Womens, girls blouses & shirts, of cotton, knit | 71333 | Kidney beans and white pea beans dried shelled |
| 252400 | Asbestos | 410210 | Sheep or lamb skins, raw, wool on, except Persian etc | 70890 | Legumes except peas & beans, fresh or chilled |
| 271112 | Propane, liquefied | 130190 | Natural gum, resin, gum-resin, balsam, not gum arabic | 90830 | Cardamoms |
| 710813 | Gold, semi-manufactured forms, non-monetary | 30420 | Fish fillets, frozen | 80132 | N/A |
| 230220 | Rice bran, sharps, other residues | 610220 | Womens, girls overcoats, etc, of cotton, knit | 80720 | Papaws (papayas), fresh |
| 200891 | Palm hearts, otherwise prepared or preserved | 90300 | Mate | 410429 | Bovine and equine leather, tanned or retanned, nes |
| 80300 | Bananas, including plantains, fresh or dried | 611130 | Babies garments, accessories of synthetic fibres, kni | 440399 | Logs, non-coniferous nes |
| 260111 | Iron ore, concentrate, not iron pyrites,unagglomerate | 630190 | Blankets (except electric) & travel rugs, material ne | 610510 | Mens, boys shirts, of cotton, knit |
| 520210 | Cotton yarn waste (including thread waste) | 611420 | Garments nes, of cotton, knit | 80119 | N/A |
| 842920 | Graders and levellers, self-propelled | 200820 | Pineapples, otherwise prepared or preserved | 560721 | Binder or baler twine, of sisal or agave |
| 710229 | Diamonds industrial, worked | 260300 | Copper ores and concentrates | 340119 | Soaps for purposes other than toilet soap, solid |
| 250629 | Quartzite, slabs etc. | 610452 | Womens, girls skirts, of cotton, knit | 260300 | Copper ores and concentrates |
| 760110 | Aluminium unwrought, not alloyed | 610463 | Womens, girls trousers, shorts, synthetic fibres, kni | 90300 | Mate |
| 30749 | Cuttle fish, squid, frozen, dried, salted or in brine | 60240 | Roses | 30269 | Fish nes, fresh or chilled, whole |
| 521051 | Plain weave cotton, <85% +manmade fibre, <200g print | 91010 | Ginger | 60390 | Cut flowers and flower buds for bouquets, dried, etc. |
| 230810 | Acorns and horse-chestnuts for animal feed | 252329 | Portland cement, other than white cement | 121299 | Vegetable products nes for human consumption |
| 30329 | Salmonidae, nes,frozen, whole | 610910 | T-shirts, singlets and other vests, of cotton, knit | 230610 | Cotton seed oil-cake and other solid residues |
| 30375 | Dogfish and other sharks, frozen, whole | 120999 | Seed, fruits and spores for sowing, nes | 240290 | Cigars, cheroots, cigarettes, with tobacco substitute |
| 100620 | Rice, husked (brown) | 220710 | Undenatured ethyl alcohol > 80% by volume | 630533 | N/A |
| 90820 | Mace | 252010 | Gypsum, anhydride | 80590 | Citrus fruits, fresh or dried, nes |
| 71420 | Sweet potatoes, fresh or dried | 460290 | Basketwork and wickerwork products, non-vegetable | 510111 | Greasy shorn wool, not carded or combed |
| 100590 | Maize except seed corn | 251020 | Natural calcium phosphates, ground | 50800 | Coral,seashell,cuttle bone,etc, unworked,powder,waste |
| 360500 | Matches | 70810 | Peas, shelled or unshelled, fresh or chilled | 610462 | Womens, girls trousers & shorts, of cotton, knit |
| 870520 | Mobile drilling derricks | 620462 | Womens, girls trousers & shorts, of cotton, not knit | 610343 | Mens, boys trousers, shorts, of synthetic fibres, kni |
| 721420 | Bar/rod, iron or non-alloy steel, indented or twisted, nes | 460210 | Basketwork, wickerwork products of vegetable material | 30420 | Fish fillets, frozen |
| 400129 | Natural rubber in other forms | 610230 | Womens, girls overcoats, etc, manmade fibres, knit | 110100 | Wheat or meslin flour |
| 400122 | Technically specified natural rubber (TSNR) | 620791 | Mens, boys dressing gowns, etc cotton, not knit | 90420 | Capsicum or Pimenta, dried, crushed or ground |
| 81290 | Fruits and nuts, provisionally preserved nes | 610520 | Mens, boys shirts, of manmade fibres, knit | 120720 | Cotton seeds |
| 320110 | Quebracho tanning extract | 410110 | Bovine skins, whole, raw | 30799 | Aquatic invertebrates nes, frozen or preserved |
| 440690 | Ties, railway or tramway, impregnated wood | 410129 | Hide sections, bovine, nes, fresh or wet-salted | 91010 | Ginger |
| 440310 | Poles, treated or painted with preservatives | 70310 | Onions and shallots, fresh or chilled | 611020 | Pullovers, cardigans etc of cotton, knit |
| 20442 | Sheep cuts, bone in, frozen | 611120 | Babies garments, accessories of cotton, knit | 130190 | Natural gum, resin, gum-resin, balsam, not gum arabic |
| 30232 | Tuna(yellowfin) fresh or chilled, whole | 620343 | Mens, boys trousers shorts, synthetic fibre, not knit | 610610 | Womens, girls blouses & shirts, of cotton, knit |
| 71410 | Manioc (cassava), fresh or dried | 620530 | Mens, boys shirts, of manmade fibres, not knit | 610220 | Womens, girls overcoats, etc, of cotton, knit |
| 80121 | N/A | 620463 | Womens, girls trousers, shorts, synth fibres, not kni | 120999 | Seed, fruits and spores for sowing, nes |
| 410221 | Sheep or lamb skins, pickled, without wool | 520532 | Cotton yarn >85% multiple uncomb 714-232 dtex,not ret | 220710 | Undenatured ethyl alcohol > 80% by volume |
| 780200 | Lead waste or scrap | 140490 | Vegetable products nes | 71310 | Peas dried, shelled |
| 630590 | Sacks & bags, packing, of materials nes | 620920 | Babies garments, accessories of cotton, not knit | 611130 | Babies garments, accessories of synthetic fibres, kni |
| 220190 | Ice, snow and potable water not sweetened or flavoure | 80300 | Bananas, including plantains, fresh or dried | 610463 | Womens, girls trousers, shorts, synthetic fibres, kni |
| 30530 | Fish fillets, dried, salted or in brine, not smoked | 410121 | Bovine hides, whole, fresh or wet-salted | 50790 | Whalebone, horns, etc unworked or simply prepared nes |
| 520291 | Garnetted stock of cotton | 30339 | Flatfish except halibut, plaice or sole, frozen, whol | 60240 | Roses |
| 90412 | Pepper of the genus Piper, crushed or ground | 610342 | Mens, boys trousers & shorts, of cotton, knit | 30619 | Crustaceans nes, frozen, |
| 310210 | Urea, including aqueous solution in packs >10 kg | 30232 | Tuna(yellowfin) fresh or chilled, whole | 610452 | Womens, girls skirts, of cotton, knit |
| 120890 | Flour or meal of oil seed, fruit, except mustard, soy | 520522 | Cotton yarn >85% single combed 714-232 dtex,not retai | 610910 | T-shirts, singlets and other vests, of cotton, knit |
| 110319 | Cereal groats or meal except wheat, maize, rice, oats | 30549 | Smoked fish & fillets other than herrings or salmon | 611420 | Garments nes, of cotton, knit |
| 520513 | Cotton yarn >85% single uncombed 232-192 dtex,not ret | 70960 | Peppers (Capsicum, Pimenta) fresh or chilled | 252010 | Gypsum, anhydride |
| 30231 | Tuna(albacore,longfin) fresh or chilled, whole | 410390 | Raw hide/skins except bovine/equine/sheep/goat/reptil | 30759 | Octopus, frozen, dried, salted or in brine |
| 152190 | Beeswax, other insect waxes and spermaceti | 620711 | Mens, boys underpants or briefs, of cotton, not knit | 30339 | Flatfish except halibut, plaice or sole, frozen, whol |
| 80111 | N/A | 610130 | Mens, boys overcoats, etc, of manmade fibres, knit | 520299 | Cotton waste, except garnetted stock |
| 560811 | Made up fishing nets, of manmade textile materials | 611030 | Pullovers, cardigans etc of manmade fibres, knit | 30375 | Dogfish and other sharks, frozen, whole |
| 271119 | Petroleum gases & gaseous hydrocarbons nes, liquefied | 620342 | Mens, boys trousers & shorts, of cotton, not knit | 80290 | Nuts edible, fresh or dried, nes |
| 30622 | Lobsters (Homarus), not frozen | 630590 | Sacks & bags, packing, of materials nes | 410129 | Hide sections, bovine, nes, fresh or wet-salted |
| 721410 | Bar/rod, iron or non-alloy steel, forged | 30375 | Dogfish and other sharks, frozen, whole | 70310 | Onions and shallots, fresh or chilled |
| 960110 | Worked ivory, articles of ivory | 610120 | Mens, boys overcoats, etc, of cotton, knit | 30569 | Fish nes, salted or in brine, not dried or smoked |
| 854810 | N/A | 120210 | Ground-nuts in shell not roasted or cooked | 30232 | Tuna(yellowfin) fresh or chilled, whole |
| 401511 | Rubber surgical gloves | 330129 | Essential oils, nes | 70810 | Peas, shelled or unshelled, fresh or chilled |
| 520523 | Cotton yarn >85% single combed 232-192 dtex,not retai | 180200 | Cocoa shells, husks, skins and waste | 251020 | Natural calcium phosphates, ground |
| 260112 | Iron ore, concentrate, not iron pyrites, agglomerated | 120799 | Oil seeds and oleaginous fruits, nes | 200820 | Pineapples, otherwise prepared or preserved |
| 80520 | Mandarin, clementine & citrus hybrids, fresh or dried | 560710 | Twine, cordage, ropes and cables, of jute, bast fibre | 610230 | Womens, girls overcoats, etc, manmade fibres, knit |
| 40229 | Milk and cream powder sweetened < 1.5% fat | 520512 | Cotton yarn >85% single uncombed 714-232 dtex,not ret | 180200 | Cocoa shells, husks, skins and waste |
| 81400 | Peel of citrus fruit or melons | 91099 | Spices nes | 30549 | Smoked fish & fillets other than herrings or salmon |
| 30342 | Tunas(yellowfin) frozen, whole | 610990 | T-shirts, singlets etc, of material nes, knit | 120799 | Oil seeds and oleaginous fruits, nes |
| 40900 | Honey, natural | 170410 | Chewing gum containing sugar, except medicinal | 620343 | Mens, boys trousers shorts, synthetic fibre, not knit |
| 110313 | Maize (corn) groats or meal | 610891 | Womens, girls bathrobe, dressing gowns, of knit cotto | 620462 | Womens, girls trousers & shorts, of cotton, not knit |
| 100830 | Canary seed | 90210 | Tea, green (unfermented) in packages < 3 kg | 120210 | Ground-nuts in shell not roasted or cooked |
| 80510 | Oranges, fresh or dried | 620640 | Womens, girls blouses, shirts, manmade fibre, not kni | 761511 | N/A |
| 252310 | Cement clinkers | 440420 | Poles, piles etc, non-coniferous, pointed, not sawn | 620791 | Mens, boys dressing gowns, etc cotton, not knit |
| 20230 | Bovine cuts boneless, frozen | 100820 | Millet | 410390 | Raw hide/skins except bovine/equine/sheep/goat/reptil |
| 80410 | Dates, fresh or dried | 621142 | Womens, girls garments nes, of cotton, not knit | 560729 | Twine nes, cordage, ropes and cables, of sisal |
| 400121 | Natural rubber in smoked sheets | 640590 | Footwear, nes | 440799 | Lumber, non-coniferous nes |
| 30349 | Tunas nes, frozen, whole | 510129 | Degreased wool nes, not carded, combed or carbonized | 51000 | Ambergris, civet, musk, etc for pharmaceutical use |
| 110311 | Wheat meal | 440200 | Wood charcoal (including shell or nut charcoal) | 70960 | Peppers (Capsicum, Pimenta) fresh or chilled |
| 10599 | Poultry, live except domestic fowls, > 185 grams | 482020 | School, etc, exercise books | 140490 | Vegetable products nes |
| 530390 | Jute and other bast fibres, not spun, nes, tow, waste | 410422 | Bovine leather, otherwise pre-tanned except whole ski | 230230 | Wheat bran, sharps, other residues |
| 440831 | N/A | 140420 | Cotton linters | 610520 | Mens, boys shirts, of manmade fibres, knit |
| 30739 | Mussels, frozen, dried, salted or in brine | 170191 | Refined sugar, in solid form, flavoured or coloured | 90411 | Pepper of the genus Piper, whole |
| 230210 | Maize bran, sharps, other residues | 30342 | Tunas(yellowfin) frozen, whole | 620463 | Womens, girls trousers, shorts, synth fibres, not kni |
| 260600 | Aluminium ores and concentrates | 810510 | Cobalt, unwrought, matte, waste or scrap, powders | 440420 | Poles, piles etc, non-coniferous, pointed, not sawn |
| 261000 | Chromium ores and concentrates | 200940 | Pineapple juice, not fermented or spirited | 611120 | Babies garments, accessories of cotton, knit |
| 440920 | Non-conifer wood continuously shaped along any edges | 230120 | Flour or meal, pellet, fish, etc, for animal feed | 520532 | Cotton yarn >85% multiple uncomb 714-232 dtex,not ret |
| 30614 | Crabs, frozen | 740311 | Copper cathodes and sections of cathodes unwrought | 230120 | Flour or meal, pellet, fish, etc, for animal feed |
| 30239 | Tuna nes, fresh or chilled, whole | 610442 | Womens, girls dresses, of cotton, knit | 620530 | Mens, boys shirts, of manmade fibres, not knit |
| 440839 | N/A | 620799 | Mens, boys dressing gowns, material nes, not knit | 100820 | Millet |
| 261610 | Silver ores and concentrates | 30490 | Fish meat & mince, except liver, roe & fillets, froze | 280522 | Strontium and barium |
| 731300 | Wire for fencing, including barbed wire | 91020 | Saffron | 120220 | Ground-nuts shelled, not roasted or cooked |
| 440610 | Ties, railway or tramway, wood not impregnated | 110620 | Flour or meal of sago, starchy roots or tubers | 620920 | Babies garments, accessories of cotton, not knit |
| 551341 | Woven plain >85% polyester + cotton, <170g/m2 printed | 520210 | Cotton yarn waste (including thread waste) | 620342 | Mens, boys trousers & shorts, of cotton, not knit |
| 200190 | Veg, fruit, nuts nes prepared or preserved by vinegar | 340111 | Soaps, for toilet use, solid | 710812 | Gold in unwrought forms non-monetary |
| 100190 | Wheat except durum wheat, and meslin | 620590 | Mens, boys shirts, of material nes, not knit | 110220 | Maize (corn) flour |
| 710221 | Diamonds, industrial, unworked or simply sawn, cleave | 400110 | Natural rubber latex, including prevulcanised | 610342 | Mens, boys trousers & shorts, of cotton, knit |
| 441214 | N/A | 621143 | Womens, girls garments nes, manmade fibres, not knit | 30342 | Tunas(yellowfin) frozen, whole |
| 251690 | Monumental or building stone nes, porphyry and basalt | 71420 | Sweet potatoes, fresh or dried | 140420 | Cotton linters |
| 510121 | Degreased shorn wool, not carded, combed or carbonize | 710812 | Gold in unwrought forms non-monetary | 520522 | Cotton yarn >85% single combed 714-232 dtex,not retai |
| 200980 | Single fruit, veg juice nes, not fermented or spirite | 81290 | Fruits and nuts, provisionally preserved nes | 30490 | Fish meat & mince, except liver, roe & fillets, froze |
| 271113 | Butanes, liquefied | 610792 | Mens, boys bathrobes, dressing gown manmade fibre kni | 630590 | Sacks & bags, packing, of materials nes |
| 440110 | Fuel wood | 90920 | Coriander seeds | 170410 | Chewing gum containing sugar, except medicinal |
| 630539 | Sacks & bags, packing, of other manmade yarn | 620821 | Womens, girls nightdress, pyjamas, of cotton, not kni | 560710 | Twine, cordage, ropes and cables, of jute, bast fibre |

Table 22 ranking of Products not currently being significantly exported (rca<0.5) which see the largest increases in density after comparing with the productive structure of the combined region.

| Tanzania | | Zambia | |
|---|---|---|---|
| | | | |
| HS-6 code | Product Name | HS-6 code | Product Name |
| 151311 | Coconut (copra) oil crude | 130214 | Pyrethrum, roots containing rotenone, extracts |
| 151110 | Palm oil, crude | 440349 | N/A |
| 271490 | Bitumen and asphalt, asphaltites and asphaltic rocks | 180100 | Cocoa beans, whole or broken, raw or roasted |
| 80430 | Pineapples, fresh or dried | 80131 | N/A |
| 160414 | Tuna, skipjack, bonito, prepared/preserved, not mince | 846911 | N/A |
| 440724 | N/A | 530310 | Jute and other textile bast fibres, raw or retted |
| 80450 | Guavas, mangoes and mangosteens, fresh or dried | 260900 | Tin ores and concentrates |
| 80720 | Papaws (papayas), fresh | 151311 | Coconut (copra) oil crude |
| 720430 | Waste or scrap, of tinned iron or steel | 120740 | Sesamum seeds |
| 200820 | Pineapples, otherwise prepared or preserved | 442010 | Statuettes and other ornaments of wood |
| 610343 | Mens, boys trousers, shorts, of synthetic fibres, kni | 30611 | Rock lobster and other sea crawfish, frozen |
| 610462 | Womens, girls trousers & shorts, of cotton, knit | 530410 | Sisal and Agave, raw |
| 121299 | Vegetable products nes for human consumption | 440724 | N/A |
| 611020 | Pullovers, cardigans etc of cotton, knit | 71390 | Leguminous vegetables dried, shelled |
| 610610 | Womens, girls blouses & shirts, of cotton, knit | 261590 | Niobium, tantalum and vanadium ores and concentrates |
| 81090 | Fruits, fresh nes | 10420 | Goats, live |
| 610220 | Womens, girls overcoats, etc, of cotton, knit | 30612 | Lobsters (Homarus) frozen |
| 611130 | Babies garments, accessories of synthetic fibres, kni | 410612 | Goat or kid skin leather, otherwise pre-tanned |
| 611420 | Garments nes, of cotton, knit | 90220 | Tea, green (unfermented) in packages > 3 kg |
| 610452 | Womens, girls skirts, of cotton, knit | 440729 | N/A |
| 510111 | Greasy shorn wool, not carded or combed | 151110 | Palm oil, crude |
| 71490 | Arrowroot, salep, etc fresh or dried and sago pith | 91030 | Turmeric (curcuma) |
| 610463 | Womens, girls trousers, shorts, synthetic fibres, kni | 230650 | Coconut or copra oil-cake and other solid residues |
| 620462 | Womens, girls trousers & shorts, of cotton, not knit | 90700 | Cloves (whole fruit, cloves and stems) |
| 270900 | Petroleum oils, oils from bituminous minerals, crude | 30621 | Rock lobster and other sea crawfish not frozen |
| 220710 | Undenatured ethyl alcohol > 80% by volume | 160414 | Tuna, skipjack, bonito, prepared/preserved, not mince |
| 620791 | Mens, boys dressing gowns, etc cotton, not knit | 30410 | Fish fillet or meat, fresh or chilled, not liver, roe |
| 610520 | Mens, boys shirts, of manmade fibres, knit | 30613 | Shrimps and prawns, frozen |
| 620463 | Womens, girls trousers, shorts, synth fibres, not kni | 261100 | Tungsten ores and concentrates |
| 610342 | Mens, boys trousers & shorts, of cotton, knit | 80450 | Guavas, mangoes and mangosteens, fresh or dried |
| 620343 | Mens, boys trousers shorts, synthetic fibre, not knit | 410512 | Sheep or lamb skin leather, otherwise pre-tanned |
| 620530 | Mens, boys shirts, of manmade fibres, not knit | 71490 | Arrowroot, salep, etc fresh or dried and sago pith |
| 460210 | Basketwork, wickerwork products of vegetable material | 71332 | Beans, small red (Adzuki) dried, shelled |
| 611120 | Babies garments, accessories of cotton, knit | 80430 | Pineapples, fresh or dried |
| 610230 | Womens, girls overcoats, etc, manmade fibres, knit | 151190 | Palm oil or fractions simply refined |
| 611030 | Pullovers, cardigans etc of manmade fibres, knit | 720430 | Waste or scrap, of tinned iron or steel |
| 620920 | Babies garments, accessories of cotton, not knit | 271490 | Bitumen and asphalt, asphaltites and asphaltic rocks |
| 620342 | Mens, boys trousers & shorts, of cotton, not knit | 30379 | Fish nes, frozen, whole |
| 460290 | Basketwork and wickerwork products, non-vegetable | 60210 | Cuttings and slips, not rooted |
| 610120 | Mens, boys overcoats, etc, of cotton, knit | 270900 | Petroleum oils, oils from bituminous minerals, crude |
| 610130 | Mens, boys overcoats, etc, of manmade fibres, knit | 90830 | Cardamoms |
| 90230 | Tea, black (fermented or partly) in packages < 3 kg | 630510 | Sacks & bags, packing, of jute or other bast fibres |
| 620711 | Mens, boys underpants or briefs, of cotton, not knit | 80132 | N/A |
| 80300 | Bananas, including plantains, fresh or dried | 100700 | Grain sorghum |
| 620640 | Womens, girls blouses, shirts, manmade fibre, not kni | 121190 | Plants & parts, pharmacy, perfume, insecticide use ne |
| 620520 | Mens, boys shirts, of cotton, not knit | 610510 | Mens, boys shirts, of cotton, knit |
| 91099 | Spices nes | 121299 | Vegetable products nes for human consumption |
| 610891 | Womens, girls bathrobe, dressing gowns, of knit cotto | 80119 | N/A |
| 140490 | Vegetable products nes | 81090 | Fruits, fresh nes |
| 70930 | Aubergines(egg-plants), fresh or chilled | 30269 | Fish nes, fresh or chilled, whole |
| 510129 | Degreased wool nes, not carded, combed or carbonized | 140410 | Raw vegetable materials for dyeing or tanning |
| 640590 | Footwear, nes | 71333 | Kidney beans and white pea beans dried shelled |
| 200940 | Pineapple juice, not fermented or spirited | 560721 | Binder or baler twine, of sisal or agave |
| 611212 | Track suits, synthetic fibres, knit | 440399 | Logs, non-coniferous nes |
| 91040 | Thyme and bay leaves | 60390 | Cut flowers and flower buds for bouquets, dried, etc. |
| 620799 | Mens, boys dressing gowns, material nes, not knit | 30420 | Fish fillets, frozen |
| 200899 | Fruit, edible plants nes otherwise prepared/preserved | 50800 | Coral,seashell,cuttle bone,etc, unworked,powder,waste |
| 621143 | Womens, girls garments nes, manmade fibres, not knit | 610343 | Mens, boys trousers, shorts, of synthetic fibres, kni |
| 151329 | Palm kernel & babassu oil, fractions, simply refined | 510111 | Greasy shorn wool, not carded or combed |
| 621142 | Womens, girls garments nes, of cotton, not knit | 91010 | Ginger |
| 400110 | Natural rubber latex, including prevulcanised | 340119 | Soaps for purposes other than toilet soap, solid |
| 510119 | Greasy wool (other than shorn) not carded or combed | 30619 | Crustaceans nes, frozen, |
| 610442 | Womens, girls dresses, of cotton, knit | 610462 | Womens, girls trousers & shorts, of cotton, knit |
| 440200 | Wood charcoal (including shell or nut charcoal) | 30799 | Aquatic invertebrates nes, frozen or preserved |
| 220840 | Rum and tafia | 611020 | Pullovers, cardigans etc of cotton, knit |
| 610791 | Mens, boys bathrobes, dressing gowns etc cotton, knit | 610610 | Womens, girls blouses & shirts, of cotton, knit |
| 610349 | Mens, boys trousers & shorts, of material nes, knit | 610220 | Womens, girls overcoats, etc, of cotton, knit |
| 410422 | Bovine leather, otherwise pre-tanned except whole ski | 130190 | Natural gum, resin, gum-resin, balsam, not gum arabic |
| 121292 | Sugar cane | 460210 | Basketwork, wickerwork products of vegetable material |
| 310520 | Nitrogen-phosphorus-potassium fertilizers, pack >10kg | 410210 | Sheep or lamb skins, raw, wool on, except Persian etc |
| 610590 | Mens, boys shirts, of materials nes, knit | 90300 | Mate |
| 620452 | Womens, girls skirts, of cotton, not knit | 30339 | Flatfish except halibut, plaice or sole, frozen, whol |
| 620721 | Mens, boys nightshirts or pyjamas, cotton, not knit | 611130 | Babies garments, accessories of synthetic fibres, kni |
| 611430 | Garments nes, of manmade fibres, knit | 80590 | Citrus fruits, fresh or dried, nes |
| 330190 | Essential oils, terpenic by-products etc., nes | 200820 | Pineapples, otherwise prepared or preserved |
| 30232 | Tuna(yellowfin) fresh or chilled, whole | 610452 | Womens, girls skirts, of cotton, knit |
| 91050 | Curry | 610463 | Womens, girls trousers, shorts, synthetic fibres, kni |
| 460120 | Mats, matting and screens, vegetable plaiting materia | 611420 | Garments nes, of cotton, knit |
| 610792 | Mens, boys bathrobes, dressing gown manmade fibre kni | 30569 | Fish nes, salted or in brine, not dried or smoked |
| 620590 | Mens, boys shirts, of material nes, not knit | 71310 | Peas dried, shelled |
| 621490 | Shawls, scarves, etc, of material nes, not knit | 610910 | T-shirts, singlets and other vests, of cotton, knit |
| 840710 | Aircraft engines, spark-ignition | 30759 | Octopus, frozen, dried, salted or in brine |
| 610620 | Womens, girls blouses & shirts, manmade fibre, knit | 180200 | Cocoa shells, husks, skins and waste |
| 610832 | Womens, girls nightdress or pyjama manmade fibre, kni | 30232 | Tuna(yellowfin) fresh or chilled, whole |
| 151790 | Edible mix & preparations of animal & veg fat, oil ne | 610230 | Womens, girls overcoats, etc, manmade fibres, knit |
| 170199 | Refined sugar, in solid form, nes, pure sucrose | 80300 | Bananas, including plantains, fresh or dried |
| 701093 | N/A | 80290 | Nuts edible, fresh or dried, nes |
| 261790 | ores and concentrates nes | 220710 | Undenatured ethyl alcohol > 80% by volume |
| 30375 | Dogfish and other sharks, frozen, whole | 410129 | Hide sections, bovine, nes, fresh or wet-salted |
| 151620 | Veg fats, oils or fractions hydrogenated, esterified | 30549 | Smoked fish & fillets other than herrings or salmon |
| 30342 | Tunas(yellowfin) frozen, whole | 620462 | Womens, girls trousers & shorts, of cotton, not knit |
| 71290 | Vegetables nes & mixtures, dried, not further prepare | 620463 | Womens, girls trousers, shorts, synth fibres, not kni |
| 71420 | Sweet potatoes, fresh or dried | 30375 | Dogfish and other sharks, frozen, whole |
| 620821 | Womens, girls nightdress, pyjamas, of cotton, not kni | 140490 | Vegetable products nes |
| 620349 | Mens, boys trousers & shorts, material nes, not knit | 610520 | Mens, boys shirts, of manmade fibres, knit |
| 460199 | Products of non-vegetable plaiting materials | 620343 | Mens, boys trousers shorts, synthetic fibre, not knit |
| 151710 | Margarine, except liquid margarine | 761511 | N/A |
| 721420 | Bar/rod, iron or non-alloy steel, indented or twisted, nes | 620791 | Mens, boys dressing gowns, etc cotton, not knit |
| 610892 | Women/girl bathrobe, dressing gown, knit manmade fibr | 230120 | Flour or meal, pellet, fish, etc, for animal feed |
| 481930 | Sacks and bags, of paper, having a width > 40 cm | 620530 | Mens, boys shirts, of manmade fibres, not knit |
| 680229 | Cut or sawn slabs of stone nes | 252010 | Gypsum, anhydride |
| 610469 | Womens, girls trousers & shorts, material nes, knit | 410121 | Bovine hides, whole, fresh or wet-salted |
| 611190 | Babies garments, accessories of material nes, knit | 611120 | Babies garments, accessories of cotton, knit |
| 611211 | Track suits, of cotton, knit | 70310 | Onions and shallots, fresh or chilled |
| 620453 | Womens, girls skirts, synthetic fibres, not knit | 51000 | Ambergris, civet, musk, etc for pharmaceutical use |
| 620292 | Womens, girls anoraks etc of cotton, not knit | 620920 | Babies garments, accessories of cotton, not knit |
| 610453 | Womens, girls skirts, synthetic fibres, knit | 90210 | Tea, green (unfermented) in packages < 3 kg |
| 620469 | Womens, girls trousers, shorts, material nes, not kni | 90230 | Tea, black (fermented or partly) in packages < 3 kg |
| 970600 | Antiques older than one hundred years | 610130 | Mens, boys overcoats, etc, of manmade fibres, knit |
| 280440 | Oxygen | 330129 | Essential oils, nes |
| 80711 | N/A | 610342 | Mens, boys trousers & shorts, of cotton, knit |
| 620333 | Mens, boys jackets, blazers, synthetic fibre, not kni | 30342 | Tunas(yellowfin) frozen, whole |
| 620891 | Womens, girls panties, bathrobes etc, cotton, not kni | 560710 | Twine, cordage, ropes and cables, of jute, bast fibre |
| 220210 | Beverage waters, sweetened or flavoured | 251020 | Natural calcium phosphates, ground |
| 30239 | Tuna nes, fresh or chilled, whole | 610120 | Mens, boys overcoats, etc, of cotton, knit |
| 200980 | Single fruit, veg juice nes, not fermented or spirite | 611030 | Pullovers, cardigans etc of manmade fibres, knit |
| 680299 | Worked monumental or building stone nes | 620342 | Mens, boys trousers & shorts, of cotton, not knit |
| 200990 | Mixtures of juices not fermented or spirited | 30490 | Fish meat & mince, except liver, roe & fillets, froze |
| 610831 | Womens, girls nightdress or pyjamas, of cotton, knit | 120799 | Oil seeds and oleaginous fruits, nes |
| 91091 | Mixtures of spices | 620711 | Mens, boys underpants or briefs, of cotton, not knit |
| 621132 | Mens, boys garments nes, of cotton, not knit | 620640 | Womens, girls blouses, shirts, manmade fibre, not kni |
| 610432 | Womens, girls jackets & blazers, of cotton, knit | 91099 | Spices nes |
| 610711 | Mens, boys underpants or briefs, of cotton, knit | 90411 | Pepper of the genus Piper, whole |
| 630120 | Blankets (non-electric) & travelling rug, wool or hai | 610990 | T-shirts, singlets etc, of material nes, knit |
| 281121 | Carbon dioxide | 230230 | Wheat bran, sharps, other residues |
| 620341 | Mens, boys trousers & shorts, wool or hair, not knit | 220840 | Rum and tafia |
| 30219 | Salmonidae, not trout or salmon,fresh or chilled whol | 120210 | Ground-nuts in shell not roasted or cooked |
| 620690 | Womens, girls blouses & shirts, material nes, not kni | 621142 | Womens, girls garments nes, of cotton, not knit |
| 271112 | Propane, liquefied | 110620 | Flour or meal of sago, starchy roots or tubers |
| 251690 | Monumental or building stone nes, porphyry and basalt | 630590 | Sacks & bags, packing, of materials nes |
| 380810 | Insecticides, packaged for retail sale | 720410 | Waste or scrap, of cast iron |
| 620630 | Womens, girls blouses & shirts, of cotton, not knit | 620520 | Mens, boys shirts, of cotton, not knit |
| 252922 | Fluorspar, >97% calcium fluoride | 71410 | Manioc (cassava), fresh or dried |
| 610821 | Womens, girls briefs or panties, of cotton, knit | 440200 | Wood charcoal (including shell or nut charcoal) |
| 730690 | Tube/pipe/hollow profile, iron/steel,riveted/open sea | 30239 | Tuna nes, fresh or chilled, whole |
| 253090 | Mineral substances, nes | 621143 | Womens, girls garments nes, manmade fibres, not knit |
| 731300 | Wire for fencing, including barbed wire | 30749 | Cuttle fish, squid, frozen, dried, salted or in brine |
| 310390 | Phosphatic fertilizers, mixes, nes, pack >10kg | 620799 | Mens, boys dressing gowns, material nes, not knit |
| 40229 | Milk and cream powder sweetened < 1.5% fat | 400129 | Natural rubber in other forms |
| 230120 | Flour or meal, pellet, fish, etc, for animal feed | 482020 | School, etc, exercise books |
| 251512 | Marble and travertine in blocks etc. | 620590 | Mens, boys shirts, of material nes, not knit |
| 490700 | Documents of title (bonds etc), unused stamps etc | 610891 | Womens, girls bathrobe, dressing gowns, of knit cotto |
| 610422 | Womens, girls ensembles, of cotton, knit | 610442 | Womens, girls dresses, of cotton, knit |
| 200190 | Veg, fruit, nuts nes prepared or preserved by vinegar | 100820 | Millet |
| 360500 | Matches | 710812 | Gold in unwrought forms non-monetary |
| 250900 | Chalk | 200940 | Pineapple juice, not fermented or spirited |
| 610443 | Womens, girls dresses, of synthetic fibres, knit | 51191 | Fish, shellfish and crustaceans (non-food) |
| 262019 | Ash or residues containing mainly zinc (not spelter) | 151590 | Veg fats, oils nes, fractions, not chemically modifie |
| 620990 | Babies garments, accessories of material nes, not kni | 91020 | Saffron |
| 630291 | Toilet or kitchen linen, of cotton, nes | 710210 | Diamonds, unsorted |